\documentclass[12pt,letterpaper]{article}
\usepackage{fullpage,indentfirst,rotating}
\usepackage{float}
\usepackage{color}

\usepackage{epsfig}
\usepackage{rotating}
\usepackage{subfigure}
\usepackage{amsfonts}
\usepackage{latexsym}
\usepackage{amsmath}
\usepackage{amssymb}
\usepackage{amsthm}
\usepackage{amstext}
\usepackage[utf8]{inputenc}
\usepackage[english]{babel}
\usepackage{mathbbol}
\usepackage{bigints}
\usepackage{longtable,array}
\usepackage{longtable}
\usepackage{mathtools}
\usepackage{adjustbox}
\usepackage{bbm}
\usepackage{bm}
\usepackage{booktabs}
\usepackage{setspace}
\usepackage{xcolor}
\usepackage{blindtext}
\usepackage{sectsty}
\usepackage{todonotes}
\usepackage{bigints}
\usepackage{multicol}
\usepackage{multirow}
\usepackage{comment}
\usepackage{subfigure}
\usepackage{algorithm2e}
\usepackage{authblk}
\usepackage[flushleft]{threeparttable}
\usepackage{enumitem}
\usepackage{relsize}
\sectionfont{\centering}
\subsectionfont{\centering}
\subsubsectionfont{\centering}
\usepackage{dsfont}
\usepackage{booktabs}
\usepackage{multirow}
\usepackage{siunitx}
\usepackage{bbold}
\usepackage{array}
\newcounter{alphasect}
\def\alphainsection{0}

\let\oldsection=\section
\def\section{%
  \ifnum\alphainsection=1%
    \addtocounter{alphasect}{1}
  \fi%
\oldsection}%

\renewcommand\thesection{%
  \ifnum\alphainsection=1%
    \Alph{alphasect}
  \else%
    \arabic{section}
  \fi%
}%

\newenvironment{alphasection}{%
  \ifnum\alphainsection=1%
    \errhelp={Let other blocks end at the beginning of the next block.}
    \errmessage{Nested Alpha section not allowed}
  \fi%
  \setcounter{alphasect}{0}
  \def\alphainsection{1}
}{%
  \setcounter{alphasect}{0}
  \def\alphainsection{0}
}%

\usepackage[colorlinks=true, linkcolor=blue, citecolor=blue]{hyperref}
\usepackage{soul}
\usepackage[authoryear]{natbib}
\usepackage[top=1 in, bottom=1 in, left=1 in , right=1 in]{geometry}
\usepackage{cleveref}

\allowdisplaybreaks

\newcommand{\ben}{\begin{enumerate}}
\newcommand{\een}{\end{enumerate}}
\newcommand{\bit}{\begin{itemize}}
\newcommand{\eit}{\end{itemize}}

\newcommand{\indep}[0]{\ensuremath{\perp \! \! \! \perp}}

\providecommand{\keywords}[1]{\textit{Keywords:} #1}
\providecommand{\codes}[1]{\textit{JEL Classification:} #1}

\newtheorem{assum}{Assumption}
\newtheorem{cor}{Corollary}
\newtheorem{prop}{Proposition}
\newtheorem{theorem}{Theorem}
\newtheorem{lemma}{Lemma}
\newtheorem{exmp}{Example}

\newtheorem{definition}{Definition}
\theoremstyle{definition}

\newcommand{\blind}{0}

\title{Randomization Inference of Heterogeneous Treatment Effects  under Network Interference} 
\if0\blind
{
\author{
\noindent  Julius Owusu  \thanks{This paper has benefited from my time at the School of Economics, University of Bristol, and the valuable comments of Jeffrey S. Racine, Youngki Shin, Michael Veall, Saraswata Chaudhuri, Liang Zhong, Thomas M. Russell, Sukjin Han, David Pacini, and three anonymous referees. I also thank participants at the 2024 Midwest Econometrics Group Conference, the 2024 Canadian Econometrics Study Group Conference, the University of Exeter Econometrics Seminar, and the 57th CEA Annual Conference for helpful feedback. } \\
Department of Economics, Concordia University
\vspace{-0.3cm}
}
}\fi

\if1\blind
{
\author{
\bigskip

}
}\fi

\renewcommand{\thesubsection}{\thesection\!\!.\arabic{subsection}}
\begin{document}
\onehalfspacing

\maketitle
\vspace{-0.5cm}
\begin{abstract}
\thispagestyle{empty}
We develop randomization-based tests for heterogeneous treatment effects in the presence of network interference. Leveraging the exposure mapping framework, we study a broad class of null hypotheses that represent various forms of constant treatment effects in networked populations. These null hypotheses, unlike the classical Fisher sharp null, are not sharp due to unknown parameters and multiple potential outcomes. Existing conditional randomization procedures either fail to control size or suffer from low statistical power in this setting.  We propose a testing procedure that constructs a data-dependent focal assignment set and permits variation in focal units across focal assignments.  These features complicate both estimation and inference, necessitating new technical developments. We establish the asymptotic validity of the proposed procedure under general conditions on the test statistic and characterize the asymptotic size distortion in terms of observable quantities. The procedure is applied to experimental network data and evaluated via Monte Carlo simulations.   
\newline
\newline
\\
 \codes{C12, C14, C46}
\end{abstract}

\setlength{\parindent}{3ex}
\newpage

\setcounter{page}{1}
\doublespacing

\section{Introduction}
The no-interference assumption is a fundamental premise in causal inference, particularly in experiments where individuals are randomly assigned to treatments. It posits that an individual's treatment assignment does not affect the outcomes of others \citep{cox1958planning}. However, this assumption is often unrealistic in settings where individuals interact, such as social networks, markets, or community-based interventions. When treatment effects propagate through such connections, \textit{interference} arises, complicating causal inference. For instance, in a public health intervention, the treatment received by one individual may influence the health behaviors or outcomes of their peers. Accounting for interference is, therefore, essential to accurately characterize the data-generating process (DGP) and obtain credible causal estimates.

    This paper adds to the growing literature on causal inference under network interference.\footnote{See, for example, \cite{halloran1995causal,hudgens2008toward, tchetgen2012causal, manski2013identification, aronow2017estimating, leung2020treatment, vazquez2023identification}.} Our distinct focus is on developing statistical methods for testing heterogeneous treatment effects (HTEs) and zero treatment effects in experimental network data. Specifically, we propose a randomization testing procedure that applies to a broad class of non-sharp null hypotheses of constant treatment effects (CTEs) in the presence of interference.

We adopt a randomization-based inference approach for two reasons. First, the interconnected nature of social networks violates the assumption of cross-sectional independence, making classical asymptotic approximation methods challenging to apply. Second, randomization tests are nonparametric and yield exact p-values for \textit{sharp null hypotheses}, where all potential outcomes can be imputed.

However, unlike the classical  Fisher null hypothesis of zero treatment effect under no interference  \citep{fisher1925statistical}, the null hypotheses in this paper are \textit{not} sharp for two fundamental reasons. First, they may depend on unknown parameters, making it possible to impute only a subset of potential outcomes under the null. Second, network interference leads to many potential outcomes, rendering the complete imputation of potential outcomes under the null hypothesis infeasible. As a result, the classical Fisher randomization testing procedure is not directly applicable.

Several papers in statistics have studied conditional randomization procedures, where inference is based on a subset of treatment assignment vectors and units, to handle non-sharp null hypotheses \citep{aronow2012general, athey2018exact, basse2019randomization, puelz2022graph}.\footnote{See Section 8 of the Supplementary Material for an overview of the literature on conditional randomization tests.} The key challenge in designing such procedures lies in selecting the subset of treatment assignment vectors and units, often referred to as \textit{focal assignments and units}. If focal units depend on the functions of the observed treatment assignment, they tend to vary across focal assignments, complicating randomization-based inference. Consequently, \cite{athey2018exact} cautions against selecting focal units based on observed treatment assignment. However, many null hypotheses, such as those concerning constant treatment effects in the presence of interference, inherently depend on treatment assignment functions. This necessitates the development of randomization testing procedures that condition on functions of observed treatment assignment while maintaining desirable finite-sample properties. 
This paper makes three main contributions.

 First, we introduce a broad class of null hypotheses that formalize various notions of constant treatment effects (CTEs) under network interference. These hypotheses encompass distinct forms of homogeneity and allow researchers to test for different dimensions of treatment effect constancy. When tested individually, they enable inference on specific homogeneity conditions; when tested jointly, they facilitate the identification of sources of heterogeneity. To the best of our knowledge, this is the first formal analysis of such a class of null hypotheses. Related work by \cite{owusu2024nonparametric} proposes asymptotic tests for covariate-defined heterogeneous treatment effects (HTEs) in clustered networks.

Second, we develop a new randomization testing procedure for non-sharp null hypotheses. In contrast to existing approaches, our procedure selects focal units and assignments as functions of the observed treatment assignment.  We establish three main theoretical results: (i) for a class of non-negative test statistics satisfying a pairwise stochastic dominance condition, we prove that the proposed test is unconditionally asymptotically valid for any nominal level $\alpha\in (0,1),$ (ii) for a broader class of test statistics, we establish asymptotic validity at some nominal levels $\alpha\in (0,1),$ and, (iii) we characterize the asymptotic size distortion of the proposed testing procedure in terms of observables.  

Third, we show that our randomization procedure can be integrated with the confidence interval method of \cite{berger1994p} to account for nuisance parameters under the null. We prove the asymptotic validity of the resulting procedure, further broadening the applicability of our approach.

The remainder of the paper is organized as follows. Section~\ref{framework}  presents the framework and formulates the hypothesis testing problem. Section \ref{sec 3}  introduces the proposed testing procedure and establishes the main results.  Section \ref{guidelines} \!\! provides implementation guidelines. Section \ref{Monte Carlo Simulation} \!\! reports results from a Monte Carlo study. In Section \ref{application}, we illustrate the practical application of the proposed method using the experimental data of \cite{cai2015social}, testing for heterogeneous treatment effects. Section \ref{conclusion} \!\!   concludes. Proofs of the main results are presented in the Appendix. Extensions and proofs of auxiliary results are provided in the Supplementary Material.

\paragraph*{Notation} For any positive integer $N,$ let $[N]=\{1,2, \dots N\}.$ For any random variables/vectors $W$ and $V,$ $\Pr_{_V}$ denotes the probability with respect to $V,$  $\Pr_{_{V|W}}$ denotes the conditional probability with respect to $V$ given $W.$  We use an analogous notation for the expectation of random variables/vectors. For any set $\mathcal{B},$ $|\mathcal{B}|$ denotes the cardinality of the set. Finally,  $\mathbbm{I}(\cdot)$ represents the indicator function.

\section{Framework}\label{framework}
\subsection{Setup}
Consistent with the framework in \cite{athey2018exact}, we consider the following setting. Suppose we have a population of $N$ units (with $i$ indexing the units) connected through a single acyclic exogenous network denoted by a $N\times N$  adjacency matrix $\mathbf{A} \in \mathbbm{A},$ where $\mathbbm{A}$ is the space of possible adjacency matrices. The $(i,j)$-th element of the adjacency matrix, $A_{ij}$,  equals one if units $i$ and $j$ interact, and zero otherwise. Hereafter, we refer to units $i$ and $j$  as neighbors if $A_{ij} = 1$.

Consider an experimental setting in which each unit \( i \in [N] \) is randomly assigned to one of two treatments: \( T_i = 0 \) (control) or \( T_i = 1 \) (treatment). Let \( N_0 \) and \( N_1 \) denote, respectively, the number of control and treated units, so that \( N_0 + N_1 = N \). The vector of treatment indicators denoted by \( \mathbf{T} \in \{0,1\}^N \) follows an assignment mechanism \( p : \{0,1\}^N \to [0,1] \), where \( p(\mathbf{t}) \) represents the probability that \( \mathbf{T} = \mathbf{t} \). We define the support of \( \mathbf{T} \) as $\mathcal{T}_0 = \left\{ \mathbf{t} \in \{0,1\}^N : p(\mathbf{t}) > 0 \right\}.$
For any assignment \( \mathbf{T} \in \mathcal{T}_0 \), let \( \Lambda(\mathbf{T}) \) denote a permutation of \( \mathbf{T} \) such that \( p(\Lambda(\mathbf{T})) > 0 \). Similarly, \( \Lambda(T_i) \) refers to the treatment status of unit \( i \) under the permuted assignment \( \Lambda(\mathbf{T}) \).

We adopt the Rubin--Neyman potential outcomes framework to formalize the causal setting. Specifically,  there exists a mapping of potential outcomes \( \mathbf{Y} : \mathcal{T}_0 \to \mathbbm{Y}^N \subset \mathbbm{R}^N \), where \( \mathbf{Y}(\mathbf{t}) \) denotes the vector of potential outcomes corresponding to assignment \( \mathbf{t} \in \mathcal{T}_0 \). The \( i^{\text{th}} \) element of \( \mathbf{Y}(\mathbf{t}) \) is written as \( Y_i(\mathbf{t}) \), representing the potential outcome of unit \( i \) under assignment \( \mathbf{T} = \mathbf{t} \).
The notation  \( Y_i(\mathbf{t}) \) allows for the possibility that the
potential outcome for unit $i$ may depend on the treatment assignment of other units in the population, thereby violating the classical Stable Unit Treatment Value Assumption (SUTVA) as formulated by \citet{cox1958planning}.

The observed treatment assignment vector  denoted by \( \mathbf{T}^{\mathrm{obs}} \) is randomly drawn from \( p(\mathbf{T}) \). We let  \( \mathbf{Y}^{\mathrm{obs}} =\mathbf{Y}(\mathbf{T}^{\mathrm{obs}}) \) represent the observed outcome vector. Analogously, for each unit \( i\in [N] \), \( Y_i = Y_i(\mathbf{T}^{\mathrm{obs}}) \) denotes the observed outcome.

In addition to observed outcomes and treatment variables, we assume that for each unit \( i \in [N] \), the researcher observes a vector of $L$ pretreatment covariates \( X_i \in \mathbbm{X} \subset \mathbbm{R}^L \), where $\mathbbm{X}$ is a finite set. Let \( \mathbf{X} \) denote the \( N \times L \) matrix collecting the covariates of all units. Thus, the observed data consist the quadruple \( (\mathbf{Y}^{\mathrm{obs}}, \mathbf{T}^{\mathrm{obs}}, \mathbf{A}, \mathbf{X}) \). We adopt a design-based approach where treatment assignment is the sole source of randomness.

A defining feature of our setting is the presence of interference, whereby the potential outcome of a given unit may depend not only on its treatment status but also on the treatment assignments of other units. Interference fundamentally exacerbates the missing potential outcomes problem: without any restrictions on the nature of the interference, each unit has \( 2^N \) potential outcomes, yet only one is observed. Consequently, studying causal effects under interference typically requires additional structure to render the analysis tractable.

A common approach is to impose an exposure mapping, as proposed by \citet{aronow2017estimating}. Under this framework, the dependence of each unit's outcome on the treatment of others is summarized by a low-dimensional exposure variable. For instance, \citet{leung2020treatment} considered settings where the fraction of treated neighbors is a sufficient statistic for the influence of network peers' treatments on a unit's outcome.

In this paper, we adopt the exposure mapping approach and formally define the network exposure mapping  as
\begin{equation}\label{exposure}
    \pi: [N]\times \mathcal{T}_0\times \mathbbm{A}  \mapsto \mathbf{\Pi},
\end{equation}
where $ \mathbf{\Pi}$ is an arbitrary finite set of exposure values.  We assume that the exposure mapping is the \textit{same} for all units. However, since the network structure \( \mathbf{A} \) is fixed (exogenous), we simplify notation and write \( \pi(i, \mathbf{T}, \mathbf{A}) = \pi_i(\mathbf{T}) \), unless otherwise necessary. Moreover, we let $\Pi_i \in \mathbf{\Pi}$ denote the resulting random variable, with $\Pi_i=\pi_i(\mathbf{T}).$  Hereon, we refer to \( \Pi_i \) as the \textit{network exposure variable}. Borrowing the terminology of \citet{manski2013identification}, the pair \( (T_i, \Pi_i) \in \{0,1\} \times \mathbf{\Pi} \) is called the \textit{effective treatment} of unit \( i \). The effective treatment jointly captures a unit's treatment status and exposure to others' treatment assignments, as determined by the network structure.

The following assumptions formally characterize the network structure and the role of exposure mapping in the framework.
\begin{assum}[No Second and Higher-Order Spillovers]\label{direct neighbors}
Let $\mathcal{M}(i, j)$ denote the shortest distance between units $i$ and $j$. In cases with no path between $i$ and $j$, $\mathcal{M}(i, j)=\infty.$  If $t_j = t'_j$ holds for all units $j\neq i$ where $\mathcal{M}(i, j) < 2$, then $Y_i(\mathbf{t}') = Y_i(\mathbf{t})$ for all $i$, given any pair of assignment vectors $\mathbf{t}$ and $\mathbf{t}' \in \mathcal{T}_0$.
\end{assum}

\begin{assum}[Correctly Specified Exposure Mapping]\label{exp map}
For any two treatment assignment vectors $\mathbf{t} \neq \mathbf{t}'$ with $\mathbf{t} = (t_i, \mathbf{t}_{-i}) \in \mathcal{T}_0$ and $\mathbf{t}' = (t_i, \mathbf{t}'_{-i}) \in \mathcal{T}_0$, there exists an exposure mapping $\pi(\cdot, \cdot, \cdot)$ such that, for all units $i$, \(
Y_i(\mathbf{t}) = Y_i(\mathbf{t}') \,\, \text{whenever}\,\, \pi(i, \mathbf{t}, \mathbf{A}) = \pi(i, \mathbf{t}', \mathbf{A})\), for a given adjacency matrix $\mathbf{A}$  and for all $i\in[N].$
\end{assum}

Assumption \ref{direct neighbors} restricts interference to first-order spillovers. Specifically, it allows the treatment status of a unit’s immediate neighbors to affect its potential outcome but assumes that treatments assigned to neighbors-of-neighbors have no effect. This restriction is convenient and empirically testable, as discussed in \cite{athey2018exact}. Moreover, it implies that the observed adjacency indicator, $A_{ij}$, coincides with the \textit{interference dependence variable} defined in \cite{savje2021average}. While we adopt this formulation for concreteness, all theoretical results in this paper extend to alternative definitions of the interference dependence variable.

Assumption~\ref{exp map} requires that the researcher correctly specifies the exposure mapping given the network structure. This assumption is standard in the literature on causality under interference; see, for example, \cite{aronow2017estimating} and \cite{leung2020treatment}. Although the assumption is strong, recent methodological contributions---such as \cite{hoshino2023randomization}---develop procedures for assessing or selecting plausible exposure mappings in applied settings. We examine the consequences of exposure mapping misspecification in Section 5 of the  Supplementary Material.

\subsection{The Testing Problem}\label{testing pro}
Given an arbitrary exposure mapping, we consider a general class of non-sharp null hypotheses defined by
  \vspace{-0.2cm}
\begin{align} \label{Gen null}
H^{G}_{0}:
   &Y_i(t, \pi) - Y_i(t', \pi')= \tau(t, t', \pi, \pi', X_i) \,\,\text{for some function} \,\, \tau(\cdot), \nonumber\\
   &\quad \quad  \quad  \quad \text{where}\,\, t,t'\in \{0,1\},\,\, \pi, \pi' \in \mathbf{\Pi},\,\,X_i \in \mathbbm{X} \,\, \text{and} \,\,\,\, i\in [N] .
\end{align}
These hypotheses assert that differences in potential outcomes across effective treatment realizations are governed by a function of the effective treatments and unit-level covariates. They possess several distinct features. First, the function \(\tau(\cdot)\) is typically unknown and unobserved in practice.  Second, the restrictions imposed by the null do not generally permit the imputation of all potential outcomes in the population. Third, because exposure values depend explicitly on the assigned treatment vector, the set of treatment assignments consistent with the null varies across units. We elaborate on these features in Section \ref{sec 3.1}.

The hypothesis \( H^{G}_{0} \) encompasses several null hypotheses of substantive interest, which may be broadly classified into three categories: (i) those asserting \emph{no/zero treatment effects under network interference}; (ii) those characterizing various forms of \emph{constant direct treatment effects under network interference}; and (iii) those asserting \emph{constant indirect (spillover) effects under network interference}. For brevity, we defer detailed discussion of each class to Section 7 of the Supplementary Material.

Throughout the paper, we focus on a representative null hypothesis, \( H_0 \), that illustrates the core challenges associated with the broader class \( H^{G}_{0} \). The testing procedures we develop for \( H_0 \) extend directly to other hypotheses encompassed by \( H^{G}_{0} \). Specifically, we consider
  \vspace{-0.1cm}
$$H_0: Y_i(1, \pi_k) - Y_i(0, \pi_k) = \tau(\pi_k) \quad \text{for some function } \tau(\cdot), \text{ and for some } \pi_k \in \boldsymbol{\Pi}, \forall\, i\, \in [N].
  \vspace{-0.1cm}
  $$
  
In other words, $H_0$  posits that individual-level direct effects depend solely on exposure values. Testing this hypothesis is of direct relevance for designing treatment assignment policies aimed at maximizing direct welfare;  see Section 7 of the Supplementary Material for more discussion.

In contrast to the sharp null of no treatment effects considered by \cite{fisher1925statistical}, the hypothesis $H_0 $ and its variants do not permit the imputation of all potential outcomes under alternative treatment assignments for any given experimental design. That is, $H_0$ is not sharp. The following example illustrates the nature of this non-sharpness.
\begin{exmp}[Non-sharpness of $H_0$]\label{eg1} 
Consider an undirected network with $N=10$ units depicted in Figure \ref{graph}.
Given the data $(Y_i^{obs}, T_i^{obs}, \Pi_i^{obs})_{i=1}^N,$ where $T_i^{obs}$ denotes the $i^{th}$ element of $\mathbf{T}^{obs}$ and $\Pi_i^{obs}= \pi_i(\mathbf{T}^{obs}):= \mathbbm{I}(\sum_{j=1}^NT^{obs}_jA_{ij}/ \sum_{j=1}^NA_{ij}\geq 0.5).$ Suppose we want to test the null
$
H_0^1: Y_i(1, 1) - Y_i(0, 1)= \tau(1) \,\,\text{for some function}\,\, \tau(\cdot),\,\,  \forall \,\,i\in [N].
$

Table \ref{sciencetable} shows that under $H_0^1,$ no potential outcome can be imputed.
Specifically, each unit has three out of four potential outcomes missing, indicated by question marks and exclamation marks. There are two reasons why potential outcomes are not imputable under $H_0^1,$ distinguished by the question and exclamation marks. First, under interference, there are four potential outcomes for each unit $\{Y_i(1,0),$ $Y_i(0,0)$ $Y_i(1,1),$ $Y_i(0,1)\}.$ Yet $H_0$ only restricts $\{Y_i(1, 1), Y_i(0, 1)\}$; as a result, we have no information on $\{Y_i(1,0),$ $Y_i(0,0)\}$ under $H_0.$ The missing potential outcomes resulting from multiple potential outcomes are denoted as the question marks $(\textcolor{red}{?})$ in Table \ref{sciencetable}. Second, in general, we do not know the specification of the function $\tau(\cdot)$ in the null. Thus, $H_0^1$ only ``partially'' restricts $\{Y_i(1, 1), Y_i(0, 1)\}.$ The missing potential outcomes resulting from the unknown functional form of $\tau(\cdot)$ are denoted as the exclamation marks $(\textcolor{blue}{!})$ in Table \ref{sciencetable}.
For instance, for unit 1, the potential outcomes $\{Y_1(1,0),$ $Y_1(0,0)\}$ are missing  due to the multiplicity of potential outcomes and  $Y_1(1,1)$ is missing due to the unknown parameter $\tau.$ 
 
To illustrate concretely the non-imputability of the potential outcomes under $H_0^1,$  consider the randomized treatment assignment vector $\Lambda(\mathbf{T}^{obs})=\Tilde{\mathbf{t}}=(1,1,1,1,1,0,0,0,0,0),$  with corresponding exposure values $(1, 1, 1, 1, 1, 0, 0,0,0,0).$ Under $H^1_0$ and given   $\Tilde{\mathbf{t}},$  any two-sample test statistic $z(\cdot)$  \textit{cannot} be computed.

\begin{figure}[ht]
    \centering
    \includegraphics[scale=0.3]{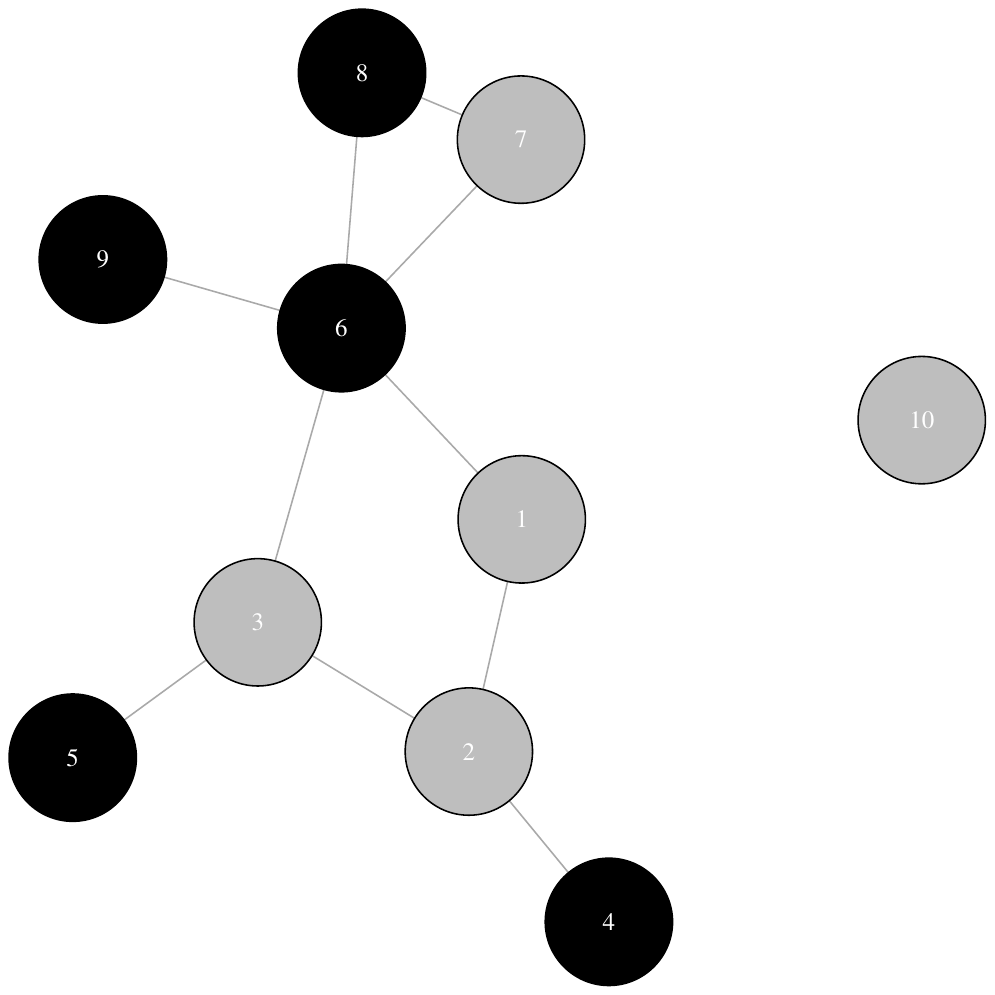}
    \vspace{-2cm}
    \caption{An undirected social network. (Grey nodes are the control units, and black  nodes are the treated units)}
    \label{graph}
\end{figure}
   \vspace{-0.5cm}
   \renewcommand{\arraystretch}{0.8} 
\begin{table}[ht]
\centering
\caption{A Science Table under $H^{\tau}_0$ using Example \ref{eg1}. NB: $Y^{P}_i$ represents the new realized outcome of unit $i$ under $H^{\tau}_0$ for the new treatment vector $\Tilde{\mathbf{t}}$.}
\resizebox{\textwidth}{!}{%
\begin{tabular}{cclllllllll}
  \multicolumn{1}{c}{\underline{Units} }  &
 \multicolumn{3}{c}{\underline{Observed Variables} } &
      \multicolumn{4}{c}{\underline{Counterfactual Outcomes}} & \multicolumn{3}{c}{\underline{Permuted Variables}} \\
\hline
$i$ & $T_i$& $\Pi_i$&  $Y^{obs}_i$ & $Y_i(1,0)$ & $Y_i(0,0)$ & $Y_i(1,1)$ & $Y_i(0,1)$   & $\Lambda(T_i)$ & $\pi_i(\Tilde{\mathbf{t}})$ & $Y^{P}_i$   \\
\hline
 1& 0 & 1& $y_1$ & \textcolor{red}{?} & \textcolor{red}{?} & $y_1+\textcolor{blue}{!}$ & $y_1$  & 1& 1& $y_1+\textcolor{blue}{!}$\\
 2& 0 & 0& $y_2$ & \textcolor{red}{?} & \textcolor{red}{?} & \textcolor{red}{?} & \textcolor{red}{?}  &1 & 1& \textcolor{red}{?}\\
 3& 0 & 1& $y_3$ & \textcolor{red}{?} & \textcolor{red}{?} & $y_3+\textcolor{blue}{!}$ & $y_3$ & 1& 1 & $y_3+\textcolor{blue}{!}$\\
 4& 1 & 0& $y_4$ & \textcolor{red}{?} & \textcolor{red}{?} & \textcolor{red}{?}& \textcolor{red}{?} &1 &1 &\textcolor{red}{?} \\
 5& 1 & 0& $y_5$ & \textcolor{red}{?} & \textcolor{red}{?} & \textcolor{red}{?} & \textcolor{red}{?}  &1 & 1&\textcolor{red}{?} \\
 6& 1& 0& $y_6$ & \textcolor{red}{?} & \textcolor{red}{?} & \textcolor{red}{?} & \textcolor{red}{?} & 0& 0&\textcolor{red}{?}\\
 7& 0& 1& $y_7$ & \textcolor{red}{?} & \textcolor{red}{?} & $y_7+\textcolor{blue}{!}$ & $y_7$  &0 &0&\textcolor{red}{?}\\
 8& 1& 1& $y_8$ & \textcolor{red}{?} & \textcolor{red}{?} & $y_8$ & $y_8-\textcolor{blue}{!}$  & 0 & 0&\textcolor{red}{?}\\
 9& 1& 1& $y_9$ & \textcolor{red}{?} & \textcolor{red}{?} & $y_9$ & $y_9-\textcolor{blue}{!}$  & 0 &0&\textcolor{red}{?} \\
 10& 0& 0& $y_{10}$ & \textcolor{red}{?} & \textcolor{red}{?} & \textcolor{red}{?}& \textcolor{red}{?} & 0 &0&\textcolor{red}{?} \\
 \hline
\end{tabular}%
}
\label{sciencetable}
\end{table}
\end{exmp}
The non-sharpness exhibited in Example~\ref{eg1} is a general feature of the null hypotheses encompassed by $H_0^G.$ These hypotheses belong to a broader class referred to as partial nulls by \cite{zhang2023randomization}. Because potential outcomes cannot be fully imputed under such nulls, the classical unconditional Fisher randomization test is not applicable. Moreover, as we demonstrate in the next section, existing conditional Fisher randomization procedures may exhibit low statistical power, even when applicable. These limitations motivate the development of specialized testing procedures tailored to the null hypotheses considered in this paper.

\section{The Testing Procedure: Randomization Inference} \label{sec 3}
A fundamental component of any testing procedure is the choice of test statistic. Throughout, we assume the existence of a predetermined two-sample-type statistic $z(\mathbf{Y}, \mathbf{T}, \mathbf{X}, \mathbf{A})$  appropriate for the null hypothesis under consideration. Under Assumptions~\ref{direct neighbors} and~\ref{exp map}, this statistic simplifies to \( z(\mathbf{Y}, \mathbf{T}, \boldsymbol{\pi}(\mathbf{T}), \mathbf{X}) \). The main theoretical results in this paper are derived under general conditions that allow for a broad class of test statistics. Nevertheless, the choice of statistics plays a central role in determining the power of the test.

Another essential component of randomization-based inference is the treatment assignment mechanism. We assume this mechanism is known and may follow an arbitrary but pre-specified design. The results in this paper accommodate a broad class of designs, including complete randomization, Bernoulli, paired, and stratified designs; see \cite{imbens2015causal}.\footnote{Under complete randomization,  a sparsity condition on the network is required to ensure that dependencies among individual treatments induced by design do not obscure the underlying network structure \citep{savje2021average}.} We impose the following condition.

\begin{assum}[Strict Overlap]\label{Overlap}
For any $\mathbf{t}=(t_1,\cdots,t_N)  \in \mathcal{T}_0,$ and for all  $x \in \mathcal{X},$  $\pi \in \boldsymbol{\Pi},$ there exist $\zeta, \eta,$ $1>\zeta>\eta> 0,$ such that  
\vspace{-0.3cm}
\begin{align}
    &\zeta <\frac{\sum_{i=1}^N \mathbbm{I}(t_i=1,  X_i=x)}{\sum_{i=1}^N\mathbbm{I}(X_i=x)}< 1-\zeta, \,\,\,\label{ovl 1} 
\end{align}
\vspace{-1cm}
\begin{align}
    &\eta <\frac{\sum_{i=1}^N \mathbbm{I}(t_i=1, \pi_i(\mathbf{t})=\pi, X_i=x)}{\sum_{i=1}^N\mathbbm{I}(X_i=x)}< \frac{\sum_{i=1}^N \mathbbm{I}(t_i=1,  X_i=x)}{\sum_{i=1}^N\mathbbm{I}(X_i=x)}-\eta. \,\,\,\label{ovl 2} 
\end{align}

\end{assum}
Assumption~\ref{Overlap} strengthens the standard overlap condition frequently imposed in causal inference; see \cite{imbens2015causal}. It requires that, for each assignment vector generated by the experimental design, a strictly positive fraction of units within each effective treatment and covariate-defined subgroup is available to compute the test statistic. In the case of the null hypothesis $H_0$, which concerns a single exposure value $\pi_k,$ the condition is only required to hold for that value.

As shown in the main results, Assumption~\ref{Overlap} is sufficient to guarantee the unconditional validity of the proposed testing procedures. However, in contrast to the standard overlap condition, Assumption~\ref{Overlap} may fail under common experimental designs, particularly when the exposure mapping is multi-valued. One approach to address this issue is to restrict attention to treatment assignments that satisfy conditions~\eqref{ovl 1} and~\eqref{ovl 2}. Consequently, the proposed procedures remain conditionally valid, even in traditional designs where Assumption~\ref{Overlap} does not hold globally.

Example~\ref{eg1} shows that the null hypothesis \( H_0 \) is not sharp for two distinct reasons. For clarity of exposition, the next subsection focuses on testing procedures for null hypotheses that do not involve unknown parameters. Subsection~\ref{solution to np} then extends the analysis to nulls that include unknown parameters.

\subsection{Null Hypotheses with Known Parameters}\label{sec 3.1}
In this subsection, we test the null hypotheses in which all parameters are known a priori. Specifically, we consider a variant of $H_0:$
\begin{equation} \label{representative null}
H^{\tau}_{0}:
   Y_i(1, \pi_k) - Y_i(0, \pi_k)= \tau \,\,\text{for some known}\,\, \tau \in \mathbbm{R},\,\,  \text{and for} \,\,  \pi_k \in \mathbf{\Pi},\,\, \forall \,\,i\in [N].  
\end{equation}
The null $H^{\tau}_{0}$ is similar to the CTE null hypothesis in settings without interference; see \cite{ding2016randomization} and  \cite{chung2021permutation}.
 A formal comparison between the testing problem considered in this paper and those in the aforementioned references is provided in Section 6 of the Supplementary Material.

Revisiting Example~\ref{eg1}, observe that under \( H^{\tau}_{0} \), all missing potential outcomes in Table~\ref{sciencetable} that were previously unobserved due to the unknown functional form of \( \tau(\cdot) \) become known. In particular, the entries marked \textcolor{blue}{!} in Table~\ref{sciencetable} are equal to \( \tau \) under \( H^{\tau}_{0} \). Consequently, the potential outcomes of some units are observed under the randomized treatment assignment \( \Lambda(\mathbf{T}) \). This feature---referred to as ``partial imputability''---motivates the conditional randomization inference (CRI) procedures commonly employed in the literature, where inference is conducted using focal units and focal assignments. Applying most existing CRI procedures presents two key technical challenges. 

First, the subset of treatment assignment vectors that guarantees non-empty groups within each effective treatment arm---required to compute a two-sample test statistic---depends on the observed assignment through the realized exposure values. These exposure values determine which randomized assignments yield non-empty treatment and control groups within the relevant subpopulation. For example, in Table~\ref{sciencetable}, the observed exposure values imply that only units 1, 3, 7, 8, and 9 are relevant for testing the null \( H_0^1 \) with \( \tau(1) = \tau \). A randomized assignment qualifies as a focal assignment if and only if at least two of these units are assigned different treatments while maintaining their exposure status. Under this criterion, the assignment vector \( \tilde{\mathbf{t}} \) in Table~\ref{sciencetable} does not qualify as a focal assignment for \( H_0^1 \) under the observed treatment, although it may qualify under a different observed assignment.
Consequently, it is challenging to construct focal assignment selection rules that are both universal and independent of the observed assignment. Moreover, when focal assignment sets are defined based on exposure values, they may overlap across different observed assignments, violating the non-overlapping partition condition required for unconditional validity in CRI procedures.

Second, the subset of units for which potential outcomes can be imputed under \( H^{\tau}_0 \) may vary across both observed and randomized treatment assignments. This is because exposure values, which determine the potential outcomes that can be imputed under the null, are functions of the treatment assignment vector. As a result, the set of focal units is inherently assignment-dependent, complicating their selection in traditional CRI procedures. For example, Table~\ref{sciencetab 2}, which builds on Example~\ref{eg1}, displays two randomized treatment vectors. Under the first, only units 1 and 7 have observable outcomes under the null, while under the second, units 1, 3, 7, 8, and 9 do. Thus, the definition of focal units varies across randomized assignments. However, \cite{athey2018exact} recommends against selecting focal units based on functions of the assignment vector, creating a tension between practical feasibility and theoretical validity.

\renewcommand{\arraystretch}{1.8} 

\begin{table}[ht]
    \centering
        \caption{Note: This table builds on Example \ref{eg1}. We consider two new randomized treatment vectors (in column 1), corresponding exposure values (in column 2), and potential outcomes under $H_0^\tau$ (in column 3).}
    \resizebox{1\textwidth}{!}{%
    \begin{tabular}{>{\centering\arraybackslash}p{5cm}>{\centering\arraybackslash}p{7cm}>{\centering\arraybackslash}p{7cm}}
    \hline
    $\Lambda(\mathbf{T})=(\Lambda(T_1),\dots, \Lambda(T_{10}))$ & 
    $(\pi_1(\Lambda(\mathbf{T})),\dots, \pi_{10}(\Lambda(\mathbf{T})))$ & 
    $(Y_1(\Lambda(\mathbf{T})), \dots,Y_{10}(\Lambda(\mathbf{T})))$ \\
    \hline
    (1,1,1, 0,0,0,0,1,0,1) & 
    (1,1,0,1,1,1,1,0,0,0) & 
    ($y_1+\tau$, \textcolor{red}{$?$}, \textcolor{red}{$?$}, \textcolor{red}{$?$}, \textcolor{red}{$?$}, \textcolor{red}{$?$}, $y_7$, \textcolor{red}{$?$}, \textcolor{red}{$?$}, \textcolor{red}{$?$}) \\
  
    (0,0,0,0,1,1,1,1,1,0) & 
    (1,0,1,0,0,0,1,1,1,0) & 
    ($y_1$, \textcolor{red}{$?$}, $y_3$, \textcolor{red}{$?$}, \textcolor{red}{$?$}, \textcolor{red}{$?$}, $y_7+\tau$, $y_8$, $y_9$, \textcolor{red}{$?$}) \\
    \hline
    \end{tabular}
    }
    \label{sciencetab 2}
\end{table}

Three existing CRI procedures may, in principle, be applicable to testing \( H_0^\tau \).

The first, which we refer to as the \textit{naive method}, defines the focal units as those with observed exposure equal to \( \pi_k \). The focal assignments are then restricted to treatment vectors under which these units retain their observed exposure value \( \pi_k \). While straightforward to implement, this approach may yield an empty set of focal assignments, particularly in settings with complex network structures and multi-valued exposure mappings.

The second feasible procedure, known as the \textit{intersection method} \citep{zhang2023randomization}, partitions the space of possible treatment assignments into non-overlapping subsets, with the subset containing the observed assignment designated as the focal assignment set. The focal units are defined as the intersection of units that, across all focal assignments, maintain a fixed exposure value \( \pi_k \). This approach suffers from two main limitations. First, constructing a reasonable partition of the treatment space is non-trivial and inherently hypothesis-specific. Second, by design, the resulting set of focal units is often ``near empty'' or empty, leading to a test with little or no power \citep[p.~2936]{zhang2023randomization}.

The third feasible procedure is the graph-theoretic approach proposed by \cite{puelz2022graph}, commonly referred to as the \textit{biclique method}. This is the state-of-the-art method that constructs a bipartite graph---termed the null exposure graph---by linking units and treatment assignments that are jointly consistent with the null hypothesis. The selection of focal units and focal assignments is then reduced to identifying large bicliques within this graph. While conceptually appealing, the method suffers from a critical limitation: in complex networks with multi-valued exposure mappings, the largest bicliques are often small, resulting in tests with trivial statistical power.

In this paper, we propose a procedure that addresses the challenges outlined above while overcoming the limitations of existing feasible methods. The approach is grounded in the idea of conditional randomization but is designed to deliver unconditional validity without requiring a partition of the treatment assignment space. We begin by presenting an abstract formulation of the procedure and introducing key definitions to facilitate comparison with conventional CRI methods. The exposition then specializes to the case of \( H_0^\tau \), allowing for clearer interpretation and implementation.

Following the notation of \cite{basse2019randomization}, our procedure begins by defining an event space \( \mathbbm{C} = \{\mathcal{C} = (\mathcal{S}, \mathcal{U}, \mathcal{T}) : \mathcal{S} \in \mathbbm{U}, \mathcal{U} \in \mathbbm{U}, \mathcal{T} \in \mathbbm{T} \} \), where \( \mathcal{S} \subseteq [N] \) denotes the set of \textit{super-focal units}---the pool from which focal units are drawn---\( \mathcal{U} \subseteq \mathcal{S} \) denotes the set of \textit{focal units},\footnote{Focal units are referred to as \textit{imputable units} in \cite{zhang2023randomization}.} and \( \mathcal{T} \subseteq \mathcal{T}_0 \) denotes the set of \textit{focal assignments}. Next, we specify a conditioning mechanism \( m(\mathcal{C} \mid \mathbf{T}^{\text{obs}}) \) over the event space. The mechanism is chosen such that, conditional on \( \mathcal{C} \), the predetermined test statistic \( z(\cdot) \) is well-defined and computable under the null hypothesis. This framework provides a principled basis for selecting focal units and assignments.

In its general form, the proposed conditioning mechanism  decomposes as 
  \vspace{-0.2cm}
\begin{equation}\label{owusu cm}
   m(\mathcal{C}|\mathbf{T}^{obs}) = \Tilde{f}(\mathcal{S}|\mathbf{T}^{obs}) \cdot \Tilde{g}(\mathcal{T}|\mathcal{S},  \mathbf{T}^{obs} ) \cdot \Tilde{h}(\mathcal{U}|\mathcal{S},\mathbf{T}^{obs}, \mathcal{T}).
     \vspace{-0.2cm}
\end{equation}
where  $\Tilde{f}$ and $\Tilde{h}$ are distributions over $\mathbbm{U}, $ and $\tilde{g}$ is a distribution over $\mathbbm{T}$.
This decomposition implies that the proposed conditioning process involves the following steps:
\begin{enumerate}
    \item \textbf{Selection of Super-Focal Units}: Identify the set of super-focal units as a function of the observed treatment assignment, specifically via the observed exposure variable.
    \item \textbf{Selection  of Focal Assignments}: Define the set of focal assignments $\mathcal{T},$ given the selected super-focal units. Since super-focal units are chosen based on observed assignments, the selection of
    $\mathcal{T}$ is implicitly informed by the observed treatment vector.
    \item \textbf{Selection of Focal Units}: Determine the focal units given the chosen super-focal units and the focal treatment assignments. Consequently, the observed treatment assignment indirectly influences the selection of focal units.
\end{enumerate}

We now sequentially apply the three-step conditioning procedure described above to the null hypothesis \(H_0^\tau\), thereby establishing the technical details of our approach.

\paragraph*{\textbf{Step 1}: Selection of Super-Focal Units} 
\indent As \( H_0^{\tau} \) pertains only to the potential outcomes of units with exposure value \( \pi_k \), the units relevant for testing the null are those whose observed exposure equals \( \pi_k \). Formally, given an arbitrary observed assignment \( \mathbf{t}^{\text{obs}} \), we define the set of super-focal units as
\vspace{-0.25cm}
\begin{equation}\label{super-focal units}
\mathcal{S}(\mathbf{t}^{obs}):=\{i \in [N]: \pi_i(\mathbf{t}^{obs})=\pi_k \}.
\vspace{-0.25cm}
\end{equation}
Assumption \ref{Overlap} ensures that $|\mathcal{S}(\mathbf{t}^{obs})|$ is sufficiently large, thereby enabling the computation of the observed test statistic for any $\mathbf{t}^{obs} \in \mathcal{T}_0$. Note that for unit $i$ in the population, the probability of being a super-focal unit is $\Pr{}_{\mathbf{T}^{obs}}(\pi_i(\mathbf{T}^{obs})=\pi_k)= \sum_{\mathbf{t} \in \mathcal{T}_0}\mathbbm{I}\{\pi_i(\mathbf{t})=\pi_k\}\cdot \Pr{}_{\mathbf{T} }(\mathbf{T}=\mathbf{t}).$
Based on Assumption \ref{Overlap}, these inclusion probabilities are strictly positive for all units.

\paragraph*{\textbf{Step 2}: Selection  of Focal Assignments} 
\indent Exposure values vary across treatment assignments; as a result, some randomized assignments may alter the exposure status of all units in \( \mathcal{S}(\mathbf{t}^{\text{obs}}) \), while others may leave these exposures unchanged. Assignment vectors that preserve the exposure status of all super-focal units constitute the focal assignment set in the naive method. However, as previously noted, such sets tend to have limited cardinality in complex networks with rich exposure mappings, resulting in low statistical power and increased computational burden.

Between these extremes, certain randomized assignments may change the exposure of only a subset of units in \( \mathcal{S}(\mathbf{t}^{\text{obs}}) \), while others retain exposure \( \pi_k \). Our proposed conditioning procedure selects subsets of assignments \( \mathcal{T} \subseteq \mathbbm{T} \) for which the number of super-focal units retaining their exposure status exceeds a pre-specified threshold determined by a tuning parameter \( \epsilon \).

Formally, for an arbitrary observed assignment, $\mathbf{t}^{obs},$ we define the focal assignment set for $H_0^\tau$ as:
\vspace{-0.3cm}
\begin{equation}\label{focal assignments}
  \mathcal{T}_\epsilon(\mathbf{t}^{obs}):=  \{\mathbf{t}' \in \mathcal{T}_0:  \hat{R}(0,\mathbf{t}';\mathcal{S}(\mathbf{t}^{obs})) \in  \mathcal{I}_{0,\epsilon}\,\, \text{and}\,\, \hat{R}(1,\mathbf{t}';\mathcal{S}(\mathbf{t}^{obs})) \in  \mathcal{I}_{1,\epsilon}\}, 
  \vspace{-0.3cm}
\end{equation}
where
\vspace{-0.3cm}
 $$ \hat{R}(t,\mathbf{t}';\mathcal{S}(\mathbf{t}^{obs}))\coloneqq{\sum_{i\in \mathcal{S}(\mathbf{t}^{obs})} \mathbbm{I}\{t'_i=t,\pi_i(\mathbf{t}')=\pi_k \}},\,\, \text{for}\,\, t=0, 1,
  \vspace{-0.1cm}
 $$
 defines the count of  super-focal units assigned to treatment $t$ and exposure value $\pi_k$ under a randomized assignment $\mathbf{t}'.$ For $t\in \{0,1\},$  $\mathcal{I}_{t,\epsilon}$  denotes closed intervals whose widths are governed by the tuning parameter $\epsilon\geq 0.$  These intervals ensure the computability of the test statistics at the focal assignments.  For instance, $\mathcal{I}_{t,\epsilon}$ may represent the exact \textit{central interval} of the distribution of  $\hat{R}(t,\mathbf{t}';\mathcal{S}(\mathbf{t}^{obs}))$ excluding extreme values  $0$ and $ |\mathcal{S}(\mathbf{t}^{obs})|.$ Thus, unlike the naive method, this selection rule does not limit focal assignments to those where focal units are fixed and equal to the super-focal units.

The tuning parameter $\epsilon$ controls the balance of treated and untreated units for each focal assignment, analogous to the role of the minimum biclique size tuning parameter in the biclique method.  
For any  $\mathbf{t} \in \mathcal{T}_0$, 
$\hat R(1,\mathbf{t};\mathcal{S}(\mathbf{t}^{obs}))+ \hat R(0,\mathbf{t};\mathcal{S}(\mathbf{t}^{obs}))\leq|\mathcal{S}(\mathbf{t}^{obs})|.$  Hence, extreme high values of $\hat R(0,\mathbf{t};\mathcal{S}(\mathbf{t}^{obs}))$ can lead to small complementary values of $\hat R(1,\mathbf{t};\mathcal{S}(\mathbf{t}^{obs})),$ and vice versa. 
To mitigate such imbalanced sample sizes---which can render test statistics non-computable or substantially reduce their precision---we select assignments where $\hat R(0,\mathbf{t};\mathcal{S}(\mathbf{t}^{obs}))$ and $R(1,\mathbf{t};\mathcal{S}(\mathbf{t}^{obs}))$ lie within meaningful intervals, ensuring adequate sample size balance.

Larger values of $\epsilon$---corresponding to narrower intervals $\mathcal{I}_{0,\epsilon}$ and $\mathcal{I}_{1,\epsilon}$---help mitigate sample size imbalances between treatment arms. This improves the precision of the test statistic and reduces the sensitivity of the inference to variability in effective sample sizes.
However, narrower intervals also restrict the focal assignment set, thereby reducing statistical power and increasing computational burden. The choice of \( \epsilon \) thus entails a fundamental trade-off between estimation precision and statistical power. Practical guidance for selecting \( \mathcal{I}_{t, \epsilon} \) is provided in Section 2 of the Supplementary Material.

It is worth noting that for any \( \mathbf{t} \in \mathcal{T}_0 \) and \( t \in \{0,1\} \), the probability that the realized sample sizes \( \hat{R}(0, \mathbf{t}; \mathcal{S}(\mathbf{T}^{\text{obs}})) \) and \( \hat{R}(1, \mathbf{t}; \mathcal{S}(\mathbf{T}^{\text{obs}})) \) fall within the respective intervals \( I_{0,\epsilon} \) and \( I_{1,\epsilon} \) is defined by
\vspace{-0.15cm}
\begin{equation}\label{assigment inclusion prob}
\resizebox{0.92\textwidth}{!}{$\phi_{\mathbf{t}}^\epsilon := \sum_{\mathbf{t}^{\text{obs}} \in \mathcal{T}_0}
  \mathbbm{I}\left\{
    \hat{R}(0, \mathbf{t}; \mathcal{S}(\mathbf{t}^{\text{obs}})) \in I_{0,\epsilon}
    \,\, \text{and} \,\,
    \hat{R}(1, \mathbf{t}; \mathcal{S}(\mathbf{t}^{\text{obs}})) \in I_{1,\epsilon}
  \right\}
  \cdot \Pr{}_{\mathbf{T}^{\text{obs}}}(\mathbf{T}^{\text{obs}} = \mathbf{t}^{\text{obs}}).$}
  \vspace{-0.15cm}
\end{equation}
We assume that all assignments in the design have a positive probability of selection as focal assignments; that is, \( 0 < \phi_{\mathbf{t}}^\epsilon < 1 \) for all \( \mathbf{t} \in \mathcal{T}_0 \). If \( \phi_{\mathbf{t}}^\epsilon = 0 \) for some \( \mathbf{t} \), the focal assignment set must exclude those assignments, or the intervals \( I_{t,\epsilon} \) must be redefined accordingly.

\paragraph*{\textbf{Step 3}: Selection of Focal Units} 
\indent Given the focal assignment set defined in~\eqref{focal assignments}, the subset of units in \( \mathcal{S}(\mathbf{t}^{\text{obs}}) \) for which \( \pi_i(\mathbf{t}) = \pi_k \) will, in general, vary across assignments \( \mathbf{t} \in \mathcal{T}_\epsilon(\mathbf{t}^{\text{obs}}) \). As a result, the proposed focal unit selection rule implies that focal units are inherently \emph{assignment-dependent}. This assignment dependence facilitates more flexible and efficient data use relative to traditional CRI methods, potentially enhancing statistical power.

Formally, for a given observed assignment \( \mathbf{t}^{\text{obs}} \), the set of focal units under any \( \mathbf{t} \in \mathcal{T}_\epsilon(\mathbf{t}^{\text{obs}}) \) is defined by
\vspace{-0.25cm}
\begin{equation}\label{focal units}
  \mathcal{U}(\mathbf{t}^{obs},\mathbf{t}) :=\{i\in \mathcal{S}(\mathbf{t}^{obs}): \pi_i(\mathbf{t})= \pi_k   \}.
 \vspace{-0.25cm}
\end{equation}

Accordingly, the probability that unit \( i \in [N] \) is selected as a focal unit can be expressed as $\Pr(\pi_i(\mathbf{T}^{obs})=\pi_k, \pi_i(\mathbf{T})=\pi_k)= \Pr{}_{\mathbf{T}}(\pi_i(\mathbf{T}^{obs})=\pi_k)\cdot Pr{}_{\mathbf{T}|\mathcal{T}_\epsilon(\mathbf{t}^{obs}) }(\pi_i(\mathbf{T})=\pi_k),$ where
 $\Pr{}_{\mathbf{T}|\mathcal{T}_\epsilon(\mathbf{t}^{obs}) }(\pi_i(\mathbf{T})=\pi_k)=\sum_{\mathbf{t} \in \mathcal{T}_\epsilon(\mathbf{t}^{obs})}\mathbbm{I}\{\pi_i(\mathbf{t})=\pi_k\}\cdot \Pr{}_{\mathbf{T}|\mathcal{T}_\epsilon(\mathbf{t}^{obs}) }(\mathbf{T}=\mathbf{t}).$ Under Assumption~\ref{Overlap}, these inclusion probabilities are strictly positive for all \( i \in [N] \).

Applying the selection rules outlined above, we introduce a novel randomization testing procedure for \( H_0^\tau \), summarized in Procedure~\ref{alg:unadjusted CDE}. This procedure serves to illustrate the unique challenges associated with the proposed conditioning framework and provides a foundation for the methodological developments that follow.

\RestyleAlgo{ruled}
\SetKwComment{Comment}{/* }{ */}
\renewcommand{\algorithmcfname}{Procedure}
\begin{algorithm} [ht]
\caption{Randomization Test of $H^{\tau}_{0}$}\label{alg:unadjusted CDE}
\KwData{$(\mathbf{Y}^{obs}, \mathbf{t}^{obs}:=(t_1^{\text{obs}}, \dots, t_N^{\text{obs}}), \boldsymbol{\pi}(\mathbf{t}^{obs}):=(\pi_i(\mathbf{t}^{\text{obs}}), \dots\pi_N(\mathbf{t}^{\text{obs}}) ))$}
\KwResult{the estimated p-values: $pval_k(\mathbf{Y}^{obs},\mathbf{t}^{obs},\boldsymbol{\pi}(\mathbf{t}^{obs});\mathcal{C}(\mathbf{t}^{obs})).$ }

\vspace{0.2cm}

1. Earmark the  super-focal units as $\mathcal{S}(\mathbf{t}^{obs})=\{i \in [N] : \pi_i(\mathbf{t}^{obs})=\pi_k \}.$ 
\\
\vspace{0.2cm}
2. Select the  intervals $\mathcal{I}_{t, \epsilon} \subset \mathbbm{Z}^{+},$ e.g., $\mathcal{I}_{t,2}=[2,\,\, |\mathcal{S}(\mathbf{t}^{obs})|-2]. $\\
\vspace{0.2cm}
3. Find the set of focal assignments  $\mathcal{T}_\epsilon(\mathbf{t}^{obs})$ using the rule in \eqref{focal assignments}. \\  
\vspace{0.2cm}
4. \For{$\{b=1 \,\,\text{to}\,\,|\mathcal{T}_\epsilon(\mathbf{t}^{obs})|\}$}{
 i. Draw $\mathbf{t}^{(b)}:=(t_1^{(b)}, \dots, t_N^{(b)})$  from $\mathcal{T}_\epsilon(\mathbf{t}^{obs}).$\\
 \vspace{0.2cm}
  ii. Choose  the focal units, $\mathcal{U}(\mathbf{t}^{\text{obs}},\mathbf{t}^{(b)})=\{i\in \mathcal{S}(\mathbf{t}^{obs}) :  \pi_i(\mathbf{t}^{(b)})=\pi_k \}.$\\
  \vspace{0.2cm}
  iii. Impute the potential outcomes for the focal units under the null restriction,\\ \,\,\,\,\,\,\, $Y_i(\mathbf{t}^{(b)})=Y_i^{obs}\cdot \mathbbm{I}\{{t}_i^{(b)}={t}_i^{obs} \,\,\text{or}\,\, i \notin \mathcal{U}(\mathbf{t}^{obs}, \mathbf{t}^{(b)})\} + (Y_i^{obs}-\tau)\cdot \mathbbm{I}\{{t}_i^{(b)}=0,$ \\ \,\,\,\,\,\,\,\, ${t}_i^{obs}=1 \,\,\text{and}\,\, i \in \mathcal{U}(\mathbf{t}^{obs},\mathbf{t}^{(b)})\} + (Y_i^{obs}+\tau)\cdot \mathbbm{I}\{{t}_i^{(b)}=1, {t}_i^{obs}=0 \,\,\text{and}$\,\, \\ \,\,\,\,\,\,\,\,\,\,$ i \in \mathcal{U}(\mathbf{t}^{obs}, \mathbf{t}^{(b)})\}.$\\
  \vspace{0.2cm}
iv. Compute and store the imputed test statistic using $[\mathbf{t}^{(b)}, \boldsymbol{\pi}(\mathbf{t}^{(b)}), \mathbf{Y}(\mathbf{t}^{(b)})]$ and \\\,\,\,\,\,\,\,\,  the focal units, i.e., $z( \mathbf{Y}(\mathbf{t}^{(b)}), \mathbf{t}^{(b)}, \boldsymbol{\pi}(\mathbf{t}^{(b)});\mathcal{U}(\mathbf{t}^{obs},\mathbf{t}^{(b)}) ).$\\
 }
 \vspace{0.2cm}
 5. Compute the \textit{observed test statistic} using the observed variables of the  units in \\\,\,\,\,\,  $\mathcal{S}(\mathbf{t}^{obs})$ 
   denoted as $z^{\text{obs}}:=z( \mathbf{Y}^{obs}, \mathbf{t}^{obs}  , \boldsymbol{\pi}(\mathbf{t}^{obs}); \mathcal{S}(\mathbf{t}^{obs})).$ \\
 \vspace{0.2cm}
 6. Compute the p-value,\\ \,\,\, \,\,$pval_k(\mathbf{Y}^{\text{obs}}, \mathbf{t}^{\text{obs}}, \boldsymbol{\pi}(\mathbf{t}^{obs});\mathcal{C}(\mathbf{t}^{obs}))= \mathbbm{E}[ \mathbbm{I}\{z( \mathbf{Y}(\mathbf{T}), \mathbf{T}, \boldsymbol{\pi}(\mathbf{T});\mathcal{U}(\mathbf{t}^{obs}, \mathbf{T}) )\geq$\\$\,\,\,\,\,\,\, z^{\text{obs}} \}|H^{\tau}_{0} ],$ where the expectation is with respect to the conditional randomization \\\,\,\,\,\, distribution  $\Pr_{\mathbf{T}| \mathcal{T}_\epsilon(\mathbf{t}^{obs})}(\mathbf{T}).$
\end{algorithm}
Several features of Procedure~\ref{alg:unadjusted CDE} depart deliberately from conventional CRI methods, giving rise to nontrivial technical challenges that must be addressed to ensure control of the Type I error rate. First, the set of focal units varies across focal assignments. Thus, even under a true null hypothesis, the distribution of the observed test statistic computed with the super-focal units may differ from that of the imputed test statistics. Second, under Assumption~\ref{Overlap}, the observed test statistic is well-defined under all assignments in the design. By contrast, the imputed test statistics are computable only over the subset of focal assignments, which are not independent and identical (i.i.d) draws from the full assignment space. This may introduce a systematic discrepancy between the observed and randomized reference distributions. Third, the proposed procedure does not induce a partition of \( \mathcal{T}_0 \) into mutually exclusive focal assignment sets. In particular, a single assignment may belong to multiple focal assignment sets. As shown by \citet{hennessy2016conditional} and \citet{zhang2023randomization}, such a partitioning property is central to establishing the unconditional validity of CRI procedures via conditional arguments.

We address the first two challenges by employing inverse probability weighted (IPW) estimators for both test statistics and p-values \citep{horvitz1952generalization, hajek1971comment}. First, observe that super-focal and focal units are sampled without replacement from the finite population, with unequal inclusion probabilities given by $\Pr{}_{\mathbf{T}}(\pi_i(\mathbf{T}^{obs})=\pi_k)$ and $\Pr{}_{\mathbf{T}}(\pi_i(\mathbf{T}^{obs})=\pi_k)\cdot \Pr{}_{\mathbf{T}|\mathcal{T}_\epsilon(\mathbf{t}^{obs}) }(\pi_i(\mathbf{T})=\pi_k)$ respectively, for each $i \in [N].$ Incorporating these probabilities yields consistent (and unbiased) estimators of population parameters. Similarly, focal assignments are drawn without replacement from the finite population of assignments induced by the experimental design, again with unequal inclusion probabilities \( \phi_{\mathbf{t}}^\epsilon \), defined in \eqref{assigment inclusion prob}, for all \( \mathbf{t} \in \mathcal{T}_0 \). Hence, IPW estimators can be used to construct consistent (and unbiased) estimators of the p-value.

To address the third challenge, we propose test statistics that ensure unconditional validity directly without relying on conditional validity as an intermediate step. Proposition~\ref{Validity of the individual unadjusted tests} provides a general sufficient condition for finite-sample unconditional validity. This condition takes the form of a pairwise pointwise dominance requirement and does not require the standard exchangeability between observed and imputed test statistics typically invoked to justify the validity of randomization tests.\footnote{Following the release of an earlier version of this paper on \texttt{arXiv}, \citet{zhong2024unconditional} pursued this direction by proposing test statistics specifically constructed to satisfy the exchangeability condition required for finite-sample validity. However, to attain a test of size $\alpha\in (0,1)$, his procedure requires a target nominal level of $\alpha/2$. This not only complicates interpretation but also undermines the conventional theoretical justification for level-$\alpha$ tests, as the procedure no longer guarantees control at the nominal level specified by the researcher.
 }

\begin{prop}{\!\!(Finite Sample Validity of the Randomization Test for  $H^{\tau}_{0}$)} \label{Validity of the individual unadjusted tests}\\
Suppose Assumptions \ref{direct neighbors}--\ref{Overlap} hold. Assume $z(\cdot)$ is an unbiased test statistic\footnote{Unbiasness means that the expectation of the test statistic equals its population counterpart. For instance, the Horvitz--Thompson difference-in-means test statistic is an unbiased estimator of the average treatment effect estimand in this setting.} with the condition  that 
\vspace{-0.5cm}
\begin{align}\label{sto dom main}
   \Pr{}&_{\mathbf{T}|\mathcal{T}_\epsilon(\mathbf{t}^{obs})}\left( z( \mathbf{Y}(\mathbf{T}), \mathbf{T} , \boldsymbol{\pi}(\mathbf{T});\mathcal{U}(\mathbf{t}^{obs},\mathbf{T}) )\leq z^{obs}\Big|\mathcal{C}(\mathbf{t}^{obs}), H^{\tau}_{0} \right)\leq \nonumber\\
  &\Pr{}_{\mathbf{T}^{obs}}\left( z(\mathbf{Y}^{obs}, \mathbf{T}^{obs} , \boldsymbol{\pi}(\mathbf{T}^{obs});\mathcal{S}(\mathbf{T}^{obs}) )\leq z^{obs}\Big|
  H^{\tau}_{0} \right),
\end{align}
for all $\mathbf{t}^{obs}\in \mathcal{T}_0,$ where $z^{\text{obs}}:=z( \mathbf{Y}^{obs}, \mathbf{t}^{obs}  , \boldsymbol{\pi}(\mathbf{t}^{obs}); \mathcal{S}(\mathbf{t}^{obs})).$
Then,  the randomization testing procedure in Procedure \ref{alg:unadjusted CDE} based on an unbiased p-value estimator is unconditionally valid  at any significant level $\alpha,$ i.e.,   
\vspace{-0.3cm}
\begin{equation} \label{size 1 main}
    \Pr{}_{\mathbf{T}^{obs}}(pval_{_k}(\mathbf{Y}^{obs}, \mathbf{T}^{obs}, \boldsymbol{\pi}(\mathbf{T}^{obs});\mathcal{C}(\mathbf{T}^{obs}))\leq \alpha|H^{\tau}_{0})\leq \alpha.
\end{equation} 
\end{prop}

The Appendix contains proofs of all main results; all remaining proofs are given in the Supplementary Material.

Proposition~\ref{Validity of the individual unadjusted tests} establishes that Procedure~\ref{alg:unadjusted CDE}, when implemented with unbiased estimators of test statistics and p-values, achieves finite-sample unconditional validity under the null hypothesis. The key requirement is that the conditional null distribution of the imputed test statistic must not exceed the observed distribution at any realized value of the observed test statistic, as expressed in inequality~\eqref{sto dom main}. This condition is strictly weaker than pairwise first-order stochastic dominance: while first-order dominance implies~\eqref{sto dom main}, the converse does not hold. Indeed, \eqref{sto dom main} may be satisfied even when the null and observed distributions cross, allowing for some forms of second-order dominance. Thus, Proposition~\ref{Validity of the individual unadjusted tests} provides a flexible criterion that broadens the class of valid testing procedures beyond those justified by stronger distributional assumptions.

Identifying test statistics that satisfy condition~\eqref{sto dom main} in finite samples is generally difficult, as the relevant null conditional probabilities typically lack closed-form representations. To circumvent this challenge, we focus on constructing test statistics for which condition~\eqref{sto dom main} holds asymptotically.  Thus, we aim to establish validity in the limit as $N \to \infty$, in the spirit of \citet{wu2021randomization}. Our asymptotic framework follows the classical sequence-of-populations approach initiated by \citet{brewer1979class} and extended by \citet{isaki1982survey}. Specifically, we consider a sequence of nested finite populations indexed by size \( N \), where the treatment assignment mechanism is independently reapplied to each population in the sequence. The following theorem shows that for a broad class of test statistics, condition~\eqref{sto dom main} is satisfied in the limit as \( N \to \infty \).
\begin{theorem}{\!\!(Asymptotic Validity of the Randomization Test for  $H^{\tau}_{0}$)} \label{Asymptotic Validity}\\
    Suppose Assumptions \ref{direct neighbors}--\ref{Overlap} hold. Let $z(\cdot)$ denote a consistent non-negative test statistic, with support $[\underline{z},\infty).$ If $z(\mathbf{Y}^{obs}, \mathbf{T}^{obs} , \boldsymbol{\pi}(\mathbf{T}^{obs});\mathcal{S}(\mathbf{T}^{obs}) )=\underline{z}+o_{p}(1)$ under $H^{\tau}_{0},$ then for all $z^{obs}\in [\underline{z},\infty),$ 
 \begin{align}\label{asysto dom main}
   \Pr{}&_{\mathbf{T}|\mathcal{T}_\epsilon(\mathbf{t}^{obs})}\left( z( \mathbf{Y}(\mathbf{T}), \mathbf{T} , \boldsymbol{\pi}(\mathbf{T});\mathcal{U}(\mathbf{t}^{obs},\mathbf{T}) )\leq z^{obs}\Big|\mathcal{C}(\mathbf{t}^{obs}), H^{\tau}_{0} \right)\leq \nonumber\\
  &\Pr{}_{\mathbf{T}^{obs}}\left( z(\mathbf{Y}^{obs}, \mathbf{T}^{obs} , \boldsymbol{\pi}(\mathbf{T}^{obs});\mathcal{S}(\mathbf{T}^{obs}) )\leq z^{obs}\Big|
  H^{\tau}_{0} \right)\,\,\, \text{as}\,\,\, N\to \infty.
\end{align}
Consequently, Procedure  \ref{alg:unadjusted CDE} is unconditionally valid at any significant level $\alpha\in(0,1)$ as $N\to\infty.$
\end{theorem}

Theorem \ref{Asymptotic Validity} asserts that any consistent non-negative test statistic that converges in probability under the null to the minimum value in its support satisfies the pairwise dominance condition in the limit as \( N \to \infty \). 
Thus, for this class of test statistics, the randomization test described in Procedure \ref{alg:unadjusted CDE} is unconditionally valid at any significant level $\alpha\in(0,1)$ as $N\to\infty.$ 

Next, we introduce a set of primitive statistics that serve as building blocks for the construction of test statistics satisfying the sufficient conditions of Theorem~\ref{Asymptotic Validity}. Let \( \bar{y}_{t}(\pi_k) \) denote the Horvitz--Thompson (HT) estimator of the mean potential outcome under effective treatment \( (t, \pi_k) \), defined as $\bar{y}_{t}(\pi_k):=N^{-1}\sum_{i=1}^NY_i\cdot D_i/\Pr{}_{\mathbf{T}}(D_i=1)$ where $D_i= \mathbbm{I}\{T_i=t,\Pi_i=\pi_k\}.$  Also, let $\hat{\sigma}^2_{t}(\pi_k)$ denote the Yates-Grundy consistent and unbiased estimator of the population variance of the potential outcome under effective treatment $(t, \pi_k),$ i.e.,  $\hat{\sigma}^2_{t}(\pi_k):=(N(N-1))^{-1}\sum_{i=1}^N\sum_{j>i}D_iD_j (Y_i-Y_j)^2/\Pr{}_{\mathbf{T}}(D_i=1, D_j=1)$ \citep{yates1953selection}. Finally, let $\hat{F}_{t,\pi_k}(\cdot)$ denote the  HT estimator of the empirical distribution of the potential outcome under effective treatment $(t, \pi_k),$ i.e., $\hat{F}_{t,\pi_k}(y):=N^{-1}\sum_{i=1}^N D_i\mathbbm{I}\{Y_i\leq y\}/\Pr{}_{\mathbf{T}}(D_i=1).$

We consider three feasible statistics to test for $H_0^\tau.$
Specifically,  we let
 $$z_{_{AR}}(\mathbf{Y}^{obs}, \mathbf{T}^{obs} , \boldsymbol{\pi}(\mathbf{T}^{obs});\mathcal{S}(\mathbf{T}^{obs}) ):=\left|\frac{\hat{\sigma}^2_1(\pi_k)- \hat{\sigma}^2_0(\pi_k) }{ \hat{\sigma}^2_0(\pi_k)}\right|$$ denote the ``absolute ratio of variance''(AR)  statistic,
$$z_{_{MR}}(\mathbf{Y}^{obs}, \mathbf{T}^{obs} , \boldsymbol{\pi}(\mathbf{T}^{obs});\mathcal{S}(\mathbf{T}^{obs}) ):=\max\Bigg\{\frac{\hat{\sigma}^2_1(\pi_k)}{ \hat{\sigma}^2_0(\pi_k)}, \frac{\hat{\sigma}^2_0(\pi_k)}{ \hat{\sigma}^2_1(\pi_k)} \Bigg\}$$ denote the ``maximum ratio''(MR) statistic, and 
   $$z_{_{SK}}(\mathbf{Y}^{obs}, \mathbf{T}^{obs} , \boldsymbol{\pi}(\mathbf{T}^{obs});\mathcal{S}(\mathbf{T}^{obs}) ):=\max_{y}|\hat{F}_{0,\pi_k}(y)-\hat{F}_{1,\pi_k}(y+\tau)|$$  denote the shifted Kolmogorov--Smirnov statistic.

 To analytically justify that the foregoing statistics satisfy the sufficient conditions of Theorem \ref{Asymptotic Validity}---particularly the convergence in probability condition---we introduce the following regularity conditions.
 \vspace{-0.1cm}
 \begin{assum}{\!\!(Boundedness of potential outcomes and exposure probabilities)}\label{bounded}
\begin{enumerate}
    \item[(i)]  For all $i\in[N], t\in\{0,1\}$ and $\pi\in \boldsymbol{\Pi},$  $|y_i(t, \pi)|\leq c_1<\infty.$ 
    \item[(ii)]  For all $i, j\in[N], t\in\{0,1\}$ and $\pi\in \boldsymbol{\Pi},$  $|1/\Pr_{_\mathbf{T}}(T_i=t,\pi_i(\mathbf{T})=\pi, T_j=t,\pi_j(\mathbf{T})=\pi)|\leq c_2<\infty.$ \item [(iii)] For $t\in\{0,1\}$ and $\pi\in \boldsymbol{\Pi},$ the variance of the population value of potential outcomes  $y_1(t,\pi)\cdots y_N(t,\pi)$---denoted as ${\sigma}^2_{t}(\pi_k)= (N-1)^{-1}\sum_{i=1}^N(y_i(t,\pi)-N^{-1}\sum_{i=1}^Ny_i(t,\pi))^2$---is bounded away from zero and infinity, $0<\sigma^2_{t}(\pi)<c_3<\infty.$ 
\end{enumerate}
\end{assum}

\begin{assum}{\!\!(Restriction of the dependency between exposures)}\label{dependency graph}
 Let $h_{ijkl}$ be a dependency indicator such that if $h_{ijkl}=0,$ then $(T_i, \pi_i(\mathbf{T}))\not\!\perp\!\!\!\perp(T_j, \pi_j(\mathbf{T})),$ $(T_k,\pi_k(\mathbf{T}))\not\!\perp\!\!\!\perp (T_l,\pi_l(\mathbf{T})),$  $(T_i, \pi_i(\mathbf{T}))\indep \left((T_k,\pi_k(\mathbf{T})), (T_l, \pi_l(\mathbf{T}))\right)$ and $(T_j,\pi_j(\mathbf{T}))$\\ $\indep ((T_k,\pi_k(\mathbf{T})), (T_l, \pi_l(\mathbf{T}))).$ Then \quad
    $\sum_{i=1}^N\sum_{j=1}^N\sum_{k=1}^N\sum_{l=1}^Nh_{ijkl}=o((N(N-1))^2).$
\end{assum}
Assumption \ref{bounded} imposes standard boundedness conditions on potential outcomes and exposure probabilities. Assumption~\ref{dependency graph} restricts the extent of dependence induced by the design and the exposure mapping across quadruples of units. In particular, it ensures that, although individual pairs of units may exhibit nontrivial clustering in their exposures, the joint dependence between distinct pairs becomes asymptotically negligible as the population size grows. This condition is less restrictive than the \textit{local dependence} condition required for standard asymptotic-based inference under network interference; see \cite{aronow2017estimating, liu2014large, leung2020treatment} and \cite{leung2022causal}.

\begin{theorem}\label{test statistics}
   If Assumptions \ref{direct neighbors}--\ref{dependency graph} hold, then the statistics $z_{_{AR}}(\cdot),$ $z_{_{MR}}(\cdot),$ and $z_{_{SK}}(\cdot)$ satisfy the asymptotic pairwise condition in \eqref{asysto dom main} under $H^{\tau}_{0}.$  Hence, using these statistics,  $$\lim_{N\to \infty}\Pr{}_{\mathbf{T}^{obs}}(pval_{_k}(\mathbf{Y}^{obs}, \mathbf{T}^{obs}, \boldsymbol{\pi}(\mathbf{T}^{obs});\mathcal{C}(\mathbf{T}^{obs}))\leq \alpha|H^{\tau}_{0})\leq \alpha,$$ for all $\alpha\in (0,1).$ 
\end{theorem}

Theorem \ref{test statistics} identifies three test statistics\footnote{The list is not exhaustive; additional statistics satisfy the sufficient conditions of Theorem~\ref{Asymptotic Validity} under \( H_0^\tau \) and related null hypotheses. For instance, under no treatment effect nulls, the absolute difference-in-means satisfies these conditions.} for $H_0^\tau$ that guarantees asymptotic validity of Procedure \ref{alg:unadjusted CDE} at all nominal levels. We numerically assess the finite sample performance of these test statistics in Section \ref{Monte Carlo Simulation}.

The class of test statistics that satisfy the sufficient conditions in Theorem~\ref{Asymptotic Validity} may nonetheless be restrictive. In the next theoretical result, we establish the asymptotic validity of Procedure~\ref{alg:unadjusted CDE} for a broader class of test statistics. Specifically, we show that validity holds asymptotically over a restricted set of significance levels, which may depend on the choice of statistic. While this result does not guarantee uniform validity over the entire unit interval, it remains practically relevant since standard nominal levels used in empirical applications typically lie in the interval $(0, 0.1]$.

\begin{cor}{\!\!(Resticted Asymptotic Validity of the Randomization Test for  $H^{\tau}_{0}$)} \label{Gen Asymptotic Validity}\\
    Suppose Assumptions \ref{direct neighbors}--\ref{Overlap} hold. Let $z(\cdot)$ denote a consistent test statistic with support $I$ that may be bounded or unbounded.  Let $z(\mathbf{Y}^{obs}, \mathbf{T}^{obs}, \boldsymbol{\pi}(\mathbf{T}^{obs});\mathcal{S}(\mathbf{T}^{obs}) )=z^{*}+o_{p}(1)$ under $H^{\tau}_{0}$ with $z^{*}<\sup I$ and $z(\mathbf{Y}^{obs}, \mathbf{T}^{obs}, \boldsymbol{\pi}(\mathbf{T}^{obs});\mathcal{S}(\mathbf{T}^{obs}) )\overset{d}{\to}F$  under $H^{\tau}_{0},$ where $F$ is non-degenerate. If $\alpha \in (0, 1-F(z^{*})),$ then 
    \vspace{-0.3cm}
    $$\Pr{}_{\mathbf{T}^{obs}}(pval_{_k}(\mathbf{Y}^{obs}, \mathbf{T}^{obs}, \boldsymbol{\pi}(\mathbf{T}^{obs});\mathcal{C}(\mathbf{T}^{obs}))\leq \alpha|H^{\tau}_{0})\leq \alpha
\,\,\, \text{as}\,\,\, N\to \infty.$$
\end{cor}
Corollary~\ref{Gen Asymptotic Validity} states that if the test statistic is consistent and its probability limit lies strictly below the supremum of its support, then Procedure~\ref{alg:unadjusted CDE} is asymptotically valid for nominal levels approximately between zero and the limiting survival function evaluated at the point of convergence. For example, if the asymptotic distribution is symmetric with a mean value equal to the limit \( z^* \), then the procedure is valid for all significance levels \( \alpha \) in the interval \( (0, 0.5] \). Corollary~\ref{Gen Asymptotic Validity} generalizes Theorem~\ref{Asymptotic Validity}: when \( z^* = \inf I \) (the infimum of the support), we recover the earlier result; when \( z^* = \sup I \), the procedure fails to control size for any \( \alpha \in (0,1) \).

Most appropriately scaled test statistics satisfy the sufficient conditions stated in Corollary~\ref{Gen Asymptotic Validity}. In contrast to Theorem~\ref{Asymptotic Validity}, however, verifying these conditions analytically---particularly the range of significance levels for which asymptotic validity holds---requires deriving the limiting distribution of the test statistic. This task typically demands stronger assumptions on the network structure, such as the widely used local dependence condition.
Similar to the result in Theorem \ref{test statistics}, under local dependence and additional regularity conditions, IPW versions of classical statistics for testing equality of variances---such as the variance ratio, Pitman--Morgan \citep{pitman1939note, morgan1939test}, and Levene’s test \citep{levene1960robust}---can be shown to satisfy the conditions in Corollary~\ref{Gen Asymptotic Validity}. 

 It is worth emphasizing that, unlike the classical Fisher randomization test---where the p-value distribution stochastically dominates the uniform distribution solely due to the discreteness of the assignment mechanism---the test statistics that satisfies the sufficient conditions of in Theorem~\ref{Asymptotic Validity} and Corollary~\ref{Gen Asymptotic Validity} may yield even greater divergence from uniformity under the null. As a result, Procedure~\ref{alg:unadjusted CDE} is valid in a conservative sense: the Type I error rate may fall strictly below the nominal level~$\alpha$. This is akin to the validity results of weak nulls in \citet{wu2021randomization}.
 
\subsubsection{Asymptotic size distortion}
In this subsection, we characterize the asymptotic size distortion of Procedure~\ref{alg:unadjusted CDE}. Our objective is to derive informative bounds on the limiting size distortion associated with various test statistics. These bounds, expressed in terms of observable data features, serve to guide practitioners in identifying settings where the proposed procedure performs optimally or exhibits diminished control over Type I errors. We begin by introducing the following definition.

\begin{definition}[Wasserstein Metric] For two probability measures $\mu$ and $\nu,$ the Wasserstein metric (distance) is defined as  $$d_W(\mu, \nu):=\sup\left\{\left|\int \ell(x)d\mu(x)-\int \ell(x)d\nu(x)\right|: \forall x,y,  \ell(x)- \ell(y)|\leq |x-y| \right\}.$$  
\end{definition}

We introduce additional notation to formalize the size distortion of Procedure~\ref{alg:unadjusted CDE}. Let
$
F_{Z|\mathcal{T}_\epsilon(\mathbf{t}^{obs})}(z^{obs}) = \Pr{}_{\mathbf{T}|\mathcal{T}_\epsilon(\mathbf{t}^{obs})}( z\big( \mathbf{Y}(\mathbf{T}), \mathbf{T}, \boldsymbol{\pi}(\mathbf{T}); \mathcal{U}(\mathbf{t}^{obs}, \mathbf{T}) \big) \leq z^{obs} \,\big|\, \mathcal{C}(\mathbf{t}^{obs}), H_0^{\tau} )
$
denote the conditional distribution of the imputed test statistic evaluated at an arbitrary observed value \( z^{obs} \), under the null hypothesis \( H_0^{\tau} \). Similarly, define
$
F_{Z^{obs}}(z^{obs}) = \Pr{}_{\mathbf{T}^{obs}}( z( \mathbf{Y}^{obs},\mathbf{T}^{obs}, \boldsymbol{\pi}(\mathbf{T}^{obs}); \mathcal{S}(\mathbf{T}^{obs})) \leq z^{obs} \,\big|\, H_0^{\tau} )
$
as the marginal distribution of the observed test statistic under \( H_0^{\tau} \), taken over the assignment mechanism. Finally, define the mixture null distribution by averaging the conditional distributions of imputed test statistics across all possible treatment assignments:
$
F_{\mathrm{mix}}(z^{obs}) = \sum_{\mathbf{t}^{obs} \in \mathcal{T}_0} F_{Z|\mathcal{T}_\epsilon(\mathbf{t}^{obs})}(z^{obs})\cdot \Pr{}_{\mathbf{T}^{obs}}(\mathbf{T}^{obs}=\mathbf{t}^{obs}).$

In the presence of multiple imputed distributions, a key challenge is to identify an appropriate reference distribution against which the distribution of the observed test statistic can be meaningfully compared. For test statistics that satisfy the pairwise dominance condition, the mixture distribution \( F_{\text{mix}}(\cdot) \) serves as a suitable reference distribution for this purpose. See Section 4.1 of the Supplementary Material for numerical justification. The degree of size distortion of the proposed procedure can then be measured by the discrepancy between \( F_{\text{mix}}(\cdot) \) and \( F_{Z^{\text{obs}}}(\cdot) \); the smaller this discrepancy, the better the size control.

The next theorem provides a general bound on the size distortion of the proposed procedure, using the mixture randomization distribution, \( F_{\mathrm{mix}}(\cdot), \) as the benchmark distribution for the imputed test statistics.

\begin{theorem}\label{general size bound}
     Suppose the sufficient conditions of Theorem \ref{Asymptotic Validity} hold,  then   
    \vspace{-0.4cm}
    \begin{align}
        \lim_{N\to \infty}|\Pr{}&_{\mathbf{T}^{obs}}(1- F_{\mathrm{mix}}(z(\mathbf{Y}^{obs}, \mathbf{T}^{obs} , \boldsymbol{\pi}(\mathbf{T}^{obs});\mathcal{S}(\mathbf{T}^{obs}))\leq \alpha)-\alpha| \nonumber \\
        \leq& |(1-\alpha)- F_{\mathrm{mix}}(F^{-1}_{\mathrm{mix}}(1-\alpha))| + \lim_{N\to \infty}\sqrt{2\kappa\cdot d_W(F_{\mathrm{mix}}, F_{Z^{obs}})}  \label{size d b main},
        \vspace{-1cm}
    \end{align}
 
    where $\kappa$ is the upper bound of the Lebesgue density of $z(\mathbf{Y}^{obs}, \mathbf{T}^{obs} , \boldsymbol{\pi}(\mathbf{T}^{obs});\mathcal{S}(\mathbf{T}^{obs})).$ 
  
\end{theorem}

Theorem~\ref{general size bound} establishes a general uniform bound on the asymptotic size distortion of the proposed testing procedure in Procedure~\ref{alg:unadjusted CDE}. The bound consists of two components. The first term in~\eqref{size d b main} reflects the conventional size distortion of randomization tests arising from the discreteness of the treatment assignment. The second term captures the contribution of the pairwise dominance condition satisfied by the test statistics under the sufficient conditions of Theorem~\ref{Asymptotic Validity}. This term is determined by the limiting distribution of the test statistic and the Wasserstein distance between the mixture randomization distribution and the limiting distribution of the observed test statistic as \( N \to \infty \).

If the observed test statistic converges weakly to a non-degenerate distribution, such as the normal, gamma, or beta distribution, then Stein's method \citep{stein1972bound} can be employed to derive an informative upper bound on the second term, $\lim_{N \to \infty} \{2\kappa \cdot d_W(F_{\mathrm{mix}}, F_{Z^{obs}})\}^{1/2}$. In the following corollary, we apply this technique to the case where the test statistic is asymptotically standard normal. The resulting bound depends on observable features of the data, thereby offering practical insight into the conditions under which the proposed testing procedure achieves better size control.

\begin{cor}\label{size bound for normals}
   Suppose the sufficient conditions of Corollary \ref{Gen Asymptotic Validity}  hold and $F=N(0,1),$ where $N(0,1)$ denotes the standard normal distribution. Then for all $\alpha \in (0,1)$
    \vspace{-0.1cm}
    \begin{align}
        \lim_{N\to \infty}|\Pr{}_{\mathbf{T}^{obs}}(&1- F_{\mathrm{mix}}(z(\mathbf{Y}^{obs}, \mathbf{T}^{obs}, \boldsymbol{\pi}(\mathbf{T}^{obs});\mathcal{S}(\mathbf{T}^{obs}))\leq \alpha)-\alpha|\nonumber \\
        \leq&  |(1-\alpha)- F_{\mathrm{mix}}(F^{-1}_{\mathrm{mix}}(1-\alpha))| + \left(\frac{2}{\pi}\right)^\frac{1}{4}\cdot \sqrt{d_W( F_{\mathrm{mix}}, \Phi)}.
    \end{align}
   
   Moreover, if the test statistic can be expressed as the sum of random variables (say $\sum_{i=1}^N W_i/\sqrt{Var(\sum_{i=1}^N W_i})$ where $W_i =N^{-1}\cdot(Y_i\cdot \mathbbm{I}\{T_i=1, \Pi_i=\pi_k\}/\Pr{}_{_\mathbf{T}}(T_i=1, \Pi_i=\pi_k)-Y_i\cdot\mathbbm{I}\{T_i=0, \Pi_i=\pi_k\}/\Pr{}_{_\mathbf{T}}(T_i=0, \Pi_i=\pi_k)),$ then 
    \begin{align}
         \sqrt{d_W(F_{\mathrm{mix}}, \Phi)}
         \leq \left\{\frac{A_{\mathrm{max}}^2}{Var(\sum_{i=1}^N W_i)^{\frac{3}{2}}}\sum_{i=1}^N\mathbbm{E}[|W_i|^3] + \frac{\sqrt{28}A_{\mathrm{max}}^{\frac{3}{2}}}{\sqrt{\pi}Var(\sum_{i=1}^N W_i)}\sqrt{\sum_{i=1}^N\mathbbm{E}|W_i^4|} \right\}^{1/2},
        \vspace{-1cm}
    \end{align}
 \normalsize   
with $A_{\mathrm{max}}:=\max_{i\in[N]}\sum_{j=1}^NA_{ij}$ is the maximal degree of the network.
\end{cor}

Corollary~\ref{size bound for normals} states that for test statistics expressible as sums of random variables and converging under the null to the standard normal distribution, the asymptotic size distortion depends on the distribution of focal units through $ W_i$. In particular, the distortion decreases as the number of focal units increases. Furthermore, the size distortion worsens as network density increases, as captured by  $A_{\mathrm{max}},$ the maximum degree in the network.

 \subsection{{Null Hypotheses with Unknown Parameters}} \label{solution to np}
In many applications, the functional form of $\tau(t, t', \pi, \pi', X_i)$ in $H_0^G$ is unknown a priori. A natural approach to testing CTEs is to posit $\tau(\cdot)$ as the conditional average treatment effect (CATE) function. However, this introduces a nuisance parameter into the testing problem, as the CATE function is typically unknown and must be estimated from the data.

A naive strategy involves replacing the CATE function with its sample analog and proceeding with the proposed randomization inference procedure. However, as noted by \citet{ding2016randomization}, such plug-in approaches typically lack theoretical guarantees for valid inference, particularly regarding the control of Type I error in both finite samples and asymptotic regimes.

We propose a tractable and easily implementable procedure that jointly addresses the two sources of non-sharpness: the multiplicity of potential outcomes and the presence of nuisance parameters. The procedure combines the conditioning strategy outlined in Procedure~\ref{alg:unadjusted CDE} with the confidence interval (CI) method of \citet{berger1994p}, as adapted to the randomization testing framework by \citet{ding2016randomization}.

The core idea of the CI method is to compute the p-value corresponding to the least favorable value of the nuisance parameter within a prespecified confidence region, thereby ensuring uniformly valid inference over all values in the region.

Before presenting the validity results associated with applying the CI method to the proposed randomization testing procedure---referred to hereafter as the \textit{CRI-CI procedure}---we formally introduce the method in the context of $H_0^{\tau}$, where $\tau = \text{ATE}(\pi_k)$ denotes the average direct treatment effect at exposure value $\pi_k$.

Let $\mathrm{CI}_\gamma$ denote the $(1-\gamma)$ confidence interval  for the nuisance parameter $\tau,$ where $\gamma \in (0,1).$  The p-value under the CRI-CI procedure is then defined as:
\vspace{-0.5cm}
\begin{equation*} \label{pvalues}
    pval_{k,\gamma}( \mathbf{Y}^{obs},\mathbf{t}^{obs}, \boldsymbol{\pi}(\mathbf{t}^{obs});\mathcal{C}(\mathbf{t}^{obs})):= \sup_{\tau' \in \mathrm{CI}_\gamma} pval_k( \mathbf{Y}^{obs},\mathbf{t}^{obs}, \boldsymbol{\pi}(\mathbf{t}^{obs});\mathcal{C}(\mathbf{t}^{obs}), \tau') +\gamma,
\vspace{-0.5cm}
\end{equation*}
\normalsize
where $pval_k( \cdot , \tau')$ denotes the p-value when $\tau =\tau'.$\footnote{In practice, computing p-values for every possible value of \( \tau \) within the confidence interval is computationally infeasible. Following \citet{ding2016randomization}, we implement the resulting procedure by evaluating p-values over a finite uniform grid within the estimated confidence interval. This approximation preserves the theoretical validity of the method.
}  
Procedure \ref{Alg: Ag2} summarizes the CRI-CI procedure for testing $H_0^{\tau}$ in the presence of the nuisance parameter $\tau.$

 \begin{algorithm}[hbt!] \label{Alg: Ag2}
\caption{CRI-CI procedure to test for $H^{\tau}_{0}$ with  nuisance parameters}\label{alg:three}
\KwData{$(\mathbf{Y}^{obs}, \mathbf{t}^{obs}, \boldsymbol{\pi}(\mathbf{t}^{obs})).$}
\KwResult{the estimated p-values: $pval_{k,\gamma}( \mathbf{Y}^{obs},\mathbf{t}^{obs}, \boldsymbol{\pi}(\mathbf{t}^{obs});\mathcal{C}(\mathbf{t}^{obs})).$ }

\vspace{0.2cm}
1. Select $\gamma \in (0, 1),$ e.g. $\gamma= 0.001.$ \\
\vspace{0.2cm}
2. Estimate the confidence interval of the nuisance parameter and denote it as\\\quad \,$\widehat{CI}_\gamma(\tau).$
\vspace{0.2cm}

\vspace{0.2cm}
3. \For{\{$\tau' \in \widehat{CI}_\gamma(\tau)$ \}}{
\vspace{0.2cm}
 (i) \,Execute steps 1 to 6 of Procedure \ref{alg:unadjusted CDE}.\\
 \vspace{0.2cm}
 (ii) Compute and store the p-value \\
 \,\,\,\,\,\,\,\,\,\,\,$pval_k( \mathbf{Y}^{obs},\mathbf{t}^{obs}, \boldsymbol{\pi}(\mathbf{t}^{obs});\mathcal{C}(\mathbf{t}^{obs})), \tau')=$\\
 \,\,\,\,\,\,\,\,\,\,\, \,\,\,\,\,\,\,\,\,\,\, \,\,\,\,\,\,\,\,\,\,\, \,\,\,\,\,\,\,\,$\mathbbm{E}[ \mathbbm{I}\{z( \mathbf{Y}(\mathbf{T}), \mathbf{T}, \mathbf{A};\mathcal{U}(\mathbf{t}^{obs}, \mathbf{T}),\tau' )\geq z^{\text{obs}} \}|\mathcal{C}(\mathbf{t}^{obs}),H^{\tau}_{0} ].$
 }
 \vspace{0.2cm}
4. Obtain the maximum p-value across $\widehat{CI}_\gamma(\tau)$ and add $\gamma,$
i.e,\\\,\,\,\,\,\,\, $pval_{k,\gamma}( \mathbf{Y}^{obs},\mathbf{t}^{obs}, \boldsymbol{\pi}(\mathbf{t}^{obs});\mathcal{C}(\mathbf{t}^{obs})):=$\\$\,\,\,\,\,\,\,\,\,\,\,\,\,\,\,\,\,\,\,\,\,\,\,\, \,\,\,\,\,\,\,\,\,\,\, \,\,\,\,\,\,\,\,\,\,\, \,\,\,\,\,\sup_{\tau' \in \widehat{CI}_\gamma(\tau)} pval_k( \mathbf{Y}^{obs},\mathbf{t}^{obs}, \boldsymbol{\pi}(\mathbf{t}^{obs});\mathcal{C}(\mathbf{t}^{obs}), \tau') +\gamma.$

\end{algorithm}
The resulting p-values from the CRI-CI procedure represent \textit{worst-case} values over the confidence region and are therefore conservative by construction \citep{ding2016randomization}. Developing methods that mitigate the associated power loss remains an open question and is beyond the scope of this paper. Nonetheless, Theorem~\ref{Asymptotic validity of the CI method on CRI using the individual tests} establishes that the CRI-CI procedure is asymptotically valid.

\begin{theorem}{\!\!(Asymptotic Validity of the  CRI-CI Method for  $H^{\tau}_{0}$\!).} \label{Asymptotic validity of the CI method on CRI using the individual tests}\\ 
Suppose the sufficient conditions of either Theorem \ref{Asymptotic Validity} or Corollary \ref{Gen Asymptotic Validity} hold.  Then, the randomization testing procedure in Procedure \ref{Alg: Ag2} is asymptotically valid  at some significant level $\alpha \in (0,1)$ i.e., 
 \vspace{-0.2cm}
\begin{equation} \label{size 3 main}
   \lim_{N\to \infty}   \Pr{}_{\mathbf{T}^{obs}}(pval_{k,\gamma}( \mathbf{Y}^{obs},\mathbf{T}^{obs}, \boldsymbol{\pi}(\mathbf{T}^{obs});\mathcal{C}(\mathbf{T}^{obs}))\leq \alpha|H^{\tau}_{0})\leq \alpha. 
\end{equation}

\end{theorem}

\section{Implementation Guidelines}\label{guidelines}

In this section, we provide practical guidelines for implementing the testing procedures developed in the preceding sections using Monte Carlo methods. In particular, we focus on estimating test statistics and p-values via the Monte Carlo approach introduced by \citet{dwass1957modified}. For a comprehensive review of Monte Carlo p-value estimation and its applications in econometrics, see \citet{dufour2001monte}. To save space, we relegate other implementation issues to Section 2 of the Supplementary Material.

Computing exact p-values is often infeasible due to the large number of possible focal assignments, even in relatively small populations. This challenge is exacerbated by the fact that unbiased and consistent estimation of the test statistics and p-values in Procedures~\ref{alg:unadjusted CDE} and~\ref{Alg: Ag2} requires knowledge of the inclusion probabilities associated with focal units, super-focal units, and focal assignments.

We consider two estimation strategies: (i) IPW estimators, which are consistent when the relevant inclusion probabilities are known, and (ii) uniformly weighted estimators, which are computationally more tractable but generally biased and inconsistent. We analytically characterize the magnitude and direction of this bias and justify the use of uniformly weighted estimators in settings where the computational burden of IPW estimation is prohibitive.

\paragraph*{Inverse Probability Weighted Estimators}
Based on procedures \ref{alg:unadjusted CDE} and \ref{Alg: Ag2}, recall that super-focal units, focal units and assignments are sampled without replacement from their respective finite populations, typically with unequal probabilities. As discussed in Section \ref{sec 3.1}, IPW estimators remain consistent and unbiased for both the observed and imputed test statistics, as well as for the resulting p-values, under this sampling design. However, computing the exact inclusion probabilities is only feasible when the sample size is small. We propose the following Monte-Carlo estimation procedure that involves:
\begin{enumerate}
    \item Randomly draw and save a moderate number (e.g., 5000) of treatment assignments of the design denoted as $\widehat{\mathcal{T}}_0.$ 
    \item For each $\mathbf{t}\in \widehat{\mathcal{T}}_0,$ compute and store the indicator values $\mathbbm{I}\{\pi_i(\mathbf{t})=\pi_k\}$ in the $N\times |\widehat{\mathcal{T}}_0|$ matrix defined as $\hat{\mathbf{I}}:=[\mathbbm{I}\{\pi_i(\mathbf{t})=\pi_k\}]_{\substack{\mathbf{t}\in \widehat{\mathcal{T}}_0 \\ i\in[N]}}.$ Then, an estimator of the \textit{probability of each unit being a super-focal unit,} $(\Pr(\pi_i(\mathbf{T}^{obs})=\pi_k), i \in [N]),$ is the diagonals of the $N\times N$ matrix
    $(\hat{\mathbf{I}}\hat{\mathbf{I}}'+ 1_N)/(|\widehat{\mathcal{T}}_0|+1),$ where $1_N$ is the $N \times N$ identity matrix that ensures nonzero marginal probabilities. The off-diagonal elements are the joint inclusion probabilities that are relevant in estimating some statistics, like the  Yate-Grundy variance estimator.
    \item For a given observed assignment $\mathbf{t}^{obs},$ store the super-focal unit set as ${\mathcal{S}}(\mathbf{t}^{obs})$ and the focal assignment set from $\widehat{\mathcal{T}}_0$ as $\widehat{\mathcal{T}}_\epsilon(\mathbf{t}^{obs}).$
\item For all $\mathbf{t} \in \widehat{\mathcal{T}}_\epsilon(\mathbf{t}^{obs}),$ compute and store the $N\times |\widehat{\mathcal{T}}_\epsilon(\mathbf{t}^{obs})|$ matrix defined as $\hat{\mathbf{I}}_s:=[\mathbbm{I}\{\pi_i(\mathbf{t})=\pi_k\}]_{\substack{\mathbf{t}\in \widehat{\mathcal{T}}_\epsilon(\mathbf{t}^{obs})\\ i\in[N]}}.$ Then, an estimator of the probabilities $(\Pr{}_{\mathbf{T}|\mathcal{T}_\epsilon(\mathbf{t}^{obs}) }(\pi_i(\mathbf{T})=\pi_k): i \in [N])$ is the diagonals of the $N\times N$ matrix
    $(\hat{\mathbf{I}}_s\hat{\mathbf{I}}_s'+ 1_N)/(|\widehat{\mathcal{T}}_\epsilon(\mathbf{t}^{obs})|+1).$ Thus, the  estimator of the \textit{ probability of each unit being a focal unit} is the diagonal of the matrix $(\hat{\mathbf{I}}\hat{\mathbf{I}}'+ 1_N)/(|\widehat{\mathcal{T}}_0|+1)\cdot (\hat{\mathbf{I}}_s\hat{\mathbf{I}}_s'+ 1_N)/(|\widehat{\mathcal{T}}_\epsilon(\mathbf{t}^{obs})|+1).$

 \item For all  $\mathbf{t}^{obs}\in \widehat{\mathcal{T}}_0$ and $\mathbf{t} \in \widehat{\mathcal{T}}_\epsilon(\mathbf{t}^{obs}),$ compute and store a $|\widehat{\mathcal{T}}_0|\times |\widehat{\mathcal{T}}_\epsilon(\mathbf{t}^{obs})|$ matrix defined as $\hat{\mathbf{I}}_r:=[\mathbbm{I}\{\hat{R}(0,\mathbf{t};{\mathcal{S}}(\mathbf{t}^{obs})) \in  \mathcal{I}_{0,\epsilon}\,\, \text{and}\,\, \hat{R}(1,\mathbf{t};{\mathcal{S}}(\mathbf{t}^{obs})) \in  \mathcal{I}_{1,\epsilon}\}]_{\substack{\mathbf{t}^{obs}\in \widehat{\mathcal{T}}_0 \\\mathbf{t}\in \widehat{\mathcal{T}}_\epsilon(\mathbf{t}^{obs})}}.$   Then, an estimator of the \textit{probability of each assignment being a focal assignment,} (denoted $\hat{\phi}_\mathbf{t}^\epsilon$ : $\mathbf{t}\in \widehat{\mathcal{T}}_0),$ is the diagonals of the $|\widehat{\mathcal{T}}_0|\times |\widehat{\mathcal{T}}_0|$ matrix
    $(\hat{\mathbf{I}}_r\hat{\mathbf{I}}_r'+ 1_{_{|\widehat{\mathcal{T}}_0|}})/(|\widehat{\mathcal{T}}_0|+1),$ where $1_{_{|\widehat{\mathcal{T}}_0|}}$ is the $|\widehat{\mathcal{T}}_0|\times |\widehat{\mathcal{T}}_0|$  identity matrix.  
\end{enumerate}
The estimators of the inclusion probabilities mentioned above are consistent; see \cite{fattorini2006applying} and \cite{aronow2017estimating} for formal proofs.

Using the estimated inclusion probabilities of units in steps 2 and 4, we can compute the unbiased IPW estimators of observed and imputed test statistics. In addition, using the inclusion probability of the randomly drawn assignments in step 5, we can compute the unbiased Horvitz--Thompson estimator of the p-value defined as 
\vspace{-0.1cm}
\begin{equation*}
\resizebox{0.999999\textwidth}{!}{$\widehat{pval}^{HT}_k(\mathbf{Y}^{\text{obs}}, \mathbf{t}^{\text{obs}}, \boldsymbol{\pi}(\mathbf{t}^{obs}) ;\mathcal{C}(\mathbf{t}^{obs}))=\frac{1}{|\widehat{\mathcal{T}}_\epsilon(\mathbf{t}^{obs})|} \sum_{\mathbf{t} \in \widehat{\mathcal{T}}_\epsilon(\mathbf{t}^{obs})}\frac{1}{\hat{\phi}_\mathbf{t}^\epsilon}\mathbbm{I}\{z( \mathbf{Y}(\mathbf{t}), \mathbf{t}, \boldsymbol{\pi}(\mathbf{t}) ;\mathcal{U}(\mathbf{t}^{obs}, \mathbf{t}) )\geq z^{\text{obs}} \} 
$}
\vspace{-0.1cm}
\end{equation*}
or the more efficient \citep{sarndal2003model}, biased but consistent H\'ajek ratio estimator
\vspace{-0.1cm}
\begin{equation*}
\resizebox{0.999999\textwidth}{!}{$\widehat{pval}^{HK}_k(\mathbf{Y}^{\text{obs}}, \mathbf{t}^{\text{obs}}, \boldsymbol{\pi}(\mathbf{t}^{obs}) ;\mathcal{C}(\mathbf{t}^{obs}))=\frac{1}{ \hat{S}^{HT}} \sum_{\mathbf{t} \in \widehat{\mathcal{T}}_\epsilon(\mathbf{t}^{obs})}\frac{1}{\hat{\phi}_\mathbf{t}^\epsilon}\mathbbm{I}\{z( \mathbf{Y}(\mathbf{t}), \mathbf{t}, \boldsymbol{\pi}(\mathbf{t}) ;\mathcal{U}(\mathbf{t}^{obs}, \mathbf{t}) )\geq z^{\text{obs}} \},
$}
\vspace{-0.1cm}
\end{equation*}
\normalsize 
where $ \hat{S}^{HT}=\sum_{\mathbf{t} \in \widehat{\mathcal{T}}_\epsilon(\mathbf{t}^{obs})}1/\hat{\phi}_\mathbf{t}^\epsilon$  is the  Horvitz--Thompson estimator of the $ |\widehat{\mathcal{T}}_\epsilon(\mathbf{t}^{obs})| .$

If the focal assignment set, $\widehat{\mathcal{T}}_\epsilon(\mathbf{t}^{obs}),$ is large, a more computationally efficient alternative for computing the p-value is to use a smaller random sample of $B$ i.i.d.  draws---with $B < |\widehat{\mathcal{T}}_\epsilon(\mathbf{t}^{obs})|$---from the focal assignment set to approximate the null distribution. We defer the discussion of the choice of $B$ to Section 2 of the Supplementary Material.

\paragraph*{Uniformly Weighted Estimators}
The computational demands of the IPW estimators can be prohibitive. This motivates the consideration of uniformly weighted estimators of test statistics and p-values, which treat the super-focal units, focal units, and focal assignments as if they were sampled with replacement with equal probability. These estimators are attractive for their computational simplicity, as they do not require knowledge of inclusion probabilities.

It is well known in the survey sampling literature (see, e.g., \citealp{sarndal2003model}) that uniformly weighted estimators are generally biased and inconsistent when sampling is conducted without replacement and with unequal inclusion probabilities. Nevertheless, in some settings and for certain statistics, the bias introduced by this approximation may be sufficiently small to be inconsequential in practice. To clarify when this holds in the context of variance-based statistics, we formally characterize the bias of the uniformly weighted sample variance estimator in the following proposition.

\begin{prop}\label{variance bias}
Define the uniformly weighted sample variance estimator as
$$s^2_{t}(\pi_k):=\frac{1}{(n_{tk} -1)}\sum_{i=1}^ND_i (Y_i-\bar{Y})^2,$$ where $D_i= \mathbbm{I}\{T_i=t,\Pi_i=\pi_k\},$  $\sum_{i=1}^ND_i=n_{tk},$  and $\bar{Y}=\sum_{i=1}^NY_iD_i/n_{tk}.$ The bias is   

\begin{align*}
    \mathbbm{E}[s^2_{t}(\pi_k)-\sigma^2_t(\pi_k)] =& \frac{1}{n_{tk}} \sum_{i=1}^N \left( p_i - \frac{n_{tk}}{N} \right) D_i  Y_i(t, \pi_k)^2 \nonumber\\
    &-\frac{1}{n_{tk}(n_{tk}-1)} \sum_{i \neq j} \left( p_{ij} - \frac{n_{tk}(n_{tk}-1)}{N(N-1)} \right) Y_i(t, \pi_k) Y_j(t, \pi_k),
\end{align*}
where $p_i=\Pr(D_i=1)$ and $p_{ij}=\Pr(D_i=1, D_j=1).$
\end{prop}

Proposition~\ref{variance bias} asserts that the bias of the uniformly weighted sample variance estimator when computed from a non-i.i.d. sample is governed by two principal factors. First, the bias decreases as the variability in inclusion probabilities \( p_i \) declines---that is, as the sampling design approaches uniformity---since the deviations \( p_i - n_{tk}/N \) and \( p_{ij} - n_{tk}(n_{tk}-1)/N(N-1) \) become smaller. Second, because these deviations are mean-zero (zero-sum) over the population, the bias is further mitigated when inclusion probabilities are uncorrelated with the squared outcomes and cross-products of outcomes. In particular, the bias vanishes when the inclusion probabilities are independent of \( Y_i(t, \pi_k) \), holding the sampling fraction fixed.

These observations imply that, in network settings where the degree distribution is approximately constant across units, the bias of the uniformly weighted sample variance estimator is negligible. Moreover, in randomized experiments, treatment assignments are independent of potential outcomes by design, further mitigating the bias introduced by uniform weighting. Consequently, under such conditions, sample variance-based test statistics and their associated p-values can be computed using uniform weights at minimal accuracy cost, providing a computationally efficient alternative to the more intensive IPW estimators.

Similar to the IPW estimators, uniformly weighted p-values are also computationally infeasible if the focal assignment set is large.  In practice, however, one can use the Monte Carlo method to estimate the p-values in Procedures \ref{alg:unadjusted CDE}--\ref{Alg: Ag2} by generating i.i.d. draws from the focal assignment set using rejection or importance sampling methods as described in \citet{branson2019randomization}. Specifically, for uniformly weighted p-values, we recommend using a random sample of focal assignments of size $B < |\mathcal{T}_\epsilon(\mathbf{t}^{obs})|$---that meet the eligibility requirements of the focal assignment set to compute the p-values. The resulting Monte-Carlo p-value is of the form specified in \cite{lehmann2022general}: 
\vspace{-0.1cm}
\begin{equation*}
\resizebox{0.99\textwidth}{!}{$\widehat{pval}^U_k(\mathbf{Y}^{\text{obs}}, \mathbf{t}^{\text{obs}}, \boldsymbol{\pi}(\mathbf{t}^{obs}) ;\mathcal{C}(\mathbf{t}^{obs}))= \frac{1}{B+1} (1+\sum_{b=1}^B \mathbbm{I}\{z( \mathbf{Y}(\mathbf{t}^{(b)}), \mathbf{t}^{(b)}, \boldsymbol{\pi}(\mathbf{t}^{(b)});\mathcal{U}(\mathbf{t}^{obs},\mathbf{t}^{(b)}))\geq z^{\text{obs}} \}).  
$}
\vspace{-0.1cm}
\end{equation*}
Under the assumption that focal units and assignments are approximately i.i.d draws, the approximated p-value is ``approximately'' consistent.\footnote{\cite{hennessy2016conditional} and \cite{hoshino2023randomization} show that when focal units and assignments are i.i.d draws, then the approximated p-value is a consistent estimator of the true p-value.}

\section{Simulation} \label{Monte Carlo Simulation}
In this section, we present the design and results of Monte Carlo experiments that evaluate the finite-sample performance of the proposed testing procedures for $H_0^\tau$. Specifically, we compare the performance of several feasible test statistics in terms of both empirical size and power. In addition, we benchmark the proposed method against the \textit{biclique method}.

Our simulation design builds on that in \citet{ding2016randomization}, with key modifications to incorporate network interference.   For each unit $i\in [N],$ the potential outcomes and treatment effects are defined as:
\vspace{-0.3cm}
\begin{align} \label{MCdesign}
    &Y_i(1,\pi):= Y_i(0,\pi)+ \tau_i(\pi, x),\quad \quad Y_i(0,\pi):=\Psi_i(\pi)+ \varepsilon_i, \nonumber\\
    &\tau_i(\pi, x):=1+\psi_0\pi+\psi_1x + \sigma_\tau\cdot Y_i(0,\pi),
   \,\,\,\, \forall \pi \in \mathbf{\Pi} \,\,\,\,\forall x \in \mathbbm{X} \,\,\text{and}\,\, i \in [N], \nonumber
\end{align}
where $\psi_0,$ $\psi_1$ and $\sigma_\tau$ parameterize different forms of treatment effect heterogeneity. Throughout, we set $\psi_0=\psi_1=0$  and  $N=200.$  

Treatment is assigned in two stages. First, we generate assignments under complete randomization with $N_0=N_1=100.$  Second, we restrict attention to assignments satisfying the overlap condition in Assumption \ref{Overlap}. To estimate the inclusion probabilities, we randomly draw 5000 treatment assignments and follow the steps outlined in Section~\ref{guidelines}.

Throughout, we use a binary-valued exposure mapping defined as:\footnote{Results for a multi-valued exposure mapping defined as the number of treated neighbors are provided in Section 4.2 of the Supplementary Material.} $$\pi_i(\mathbf{T}):= \mathbbm{I}(\sum_{j=1}^NT_jA_{ij}/\sum_{j=1}^NA_{ij} >0.5).$$
Based on the exposure mapping and the population size, we use the intervals $\mathcal{I}_{t,\epsilon}=\mathcal{I}_{t,20} =[20,\,\, |\mathcal{S}(\mathbf{t}^{obs})|-20]$ for $t=0,1$.\footnote{In Section  2.1 of the Supplementary Material, we numerically examine how variations in $\mathcal{I}_{t,\epsilon}$ influence the size and power of the test.} Finally, we draw the errors $\{\varepsilon_i\}_{i\in[N]}$ from a multivariate normal distribution whose covariance structure reflects both individual variability and network-based correlation.

All rejection rates are computed as the proportion of rejections at the 5\% significance level based on 1,000 Monte Carlo replications, yielding a simulation standard error of approximately 0.00689 under the null. 

\subsection{Performance of Test Statistics}
In this subsection, we assess the finite sample performance of some feasible statistics for testing $H_0^\tau$ using Procedure \ref{alg:unadjusted CDE}. Specifically, we compute the empirical rejection probabilities using the \textit{absolute ratio of variance, maximum ratio, shifted Kolmogorov--Smirnov, Levene,} and \textit{Pitman--Morgan statistics}. Recall that the absolute ratio of variance, maximum ratio, and shifted Kolmogorov--Smirnov statistics satisfy the sufficient conditions of Theorem \ref{Asymptotic Validity}. On the other hand, the  Levene and Pitman--Morgan statistics satisfy the sufficient conditions of Corollary \ref{Gen Asymptotic Validity}. See Section 3 of the Supplementary Material for a formal discussion of the Pitman--Morgan statistic, a description of how it is adapted to the present setting, and its limitations.

For the results reported in this subsection, we utilize an adjacency matrix where each unit is connected to at least one other unit. In addition, $\Psi_i(\pi)'$s are i.i.d draws from a right-skewed \textit{Beta distribution} with shape parameters $0.5$ and $2+\pi$ for all $\pi \in \boldsymbol{\Pi}.$

\begin{figure}[h]
\setcounter{subfigure}{0}
\centering
\caption{Rejection rates at the $5\%$ significance level for different statistics.}
\subfigure[Power curve for variance-type test statistics]{
    \includegraphics[width=0.47\columnwidth, keepaspectratio]{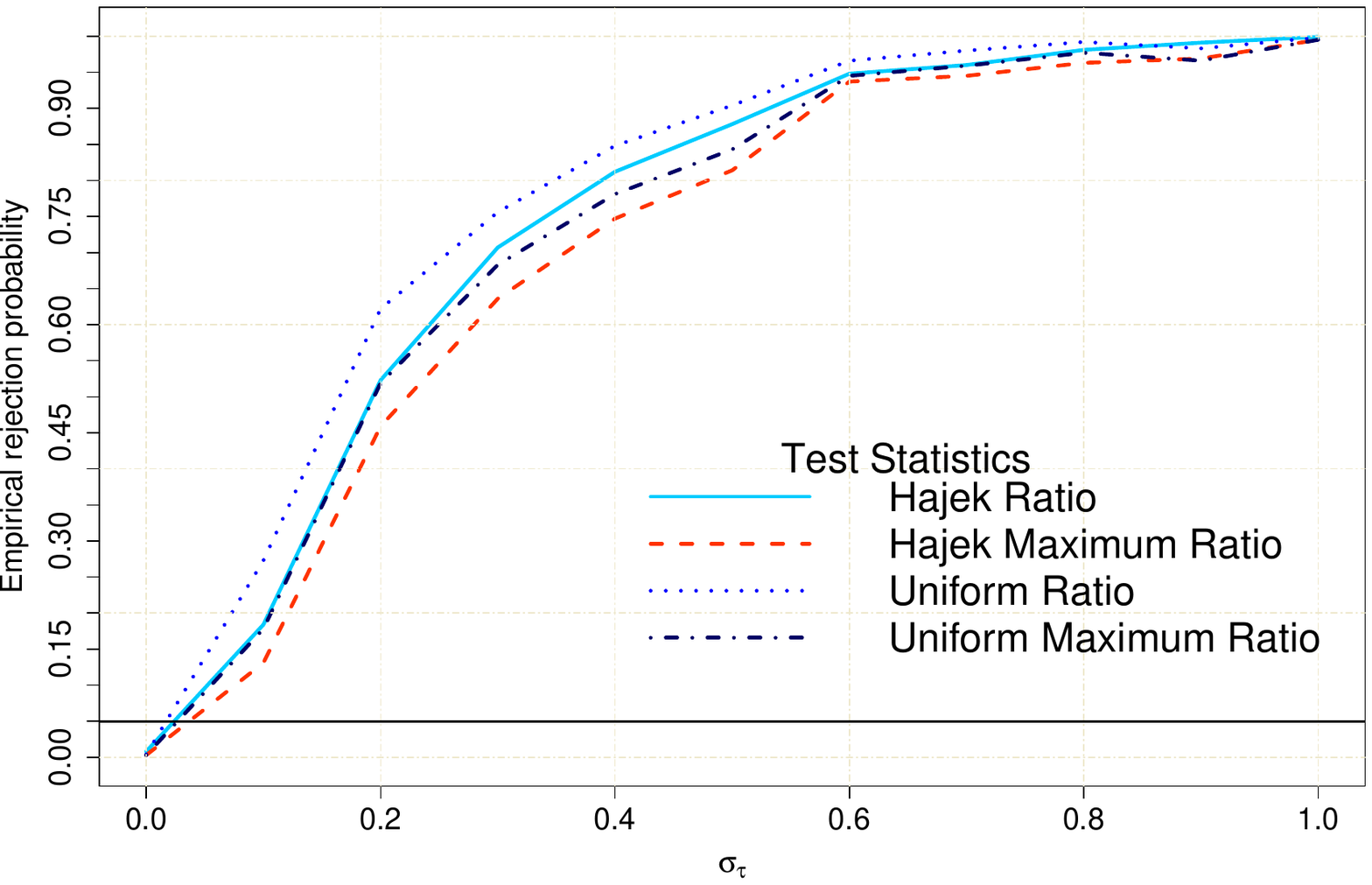}
}
\subfigure[Power curve for uniformly weighted test statistics]{
    \includegraphics[width=0.47\columnwidth, keepaspectratio]{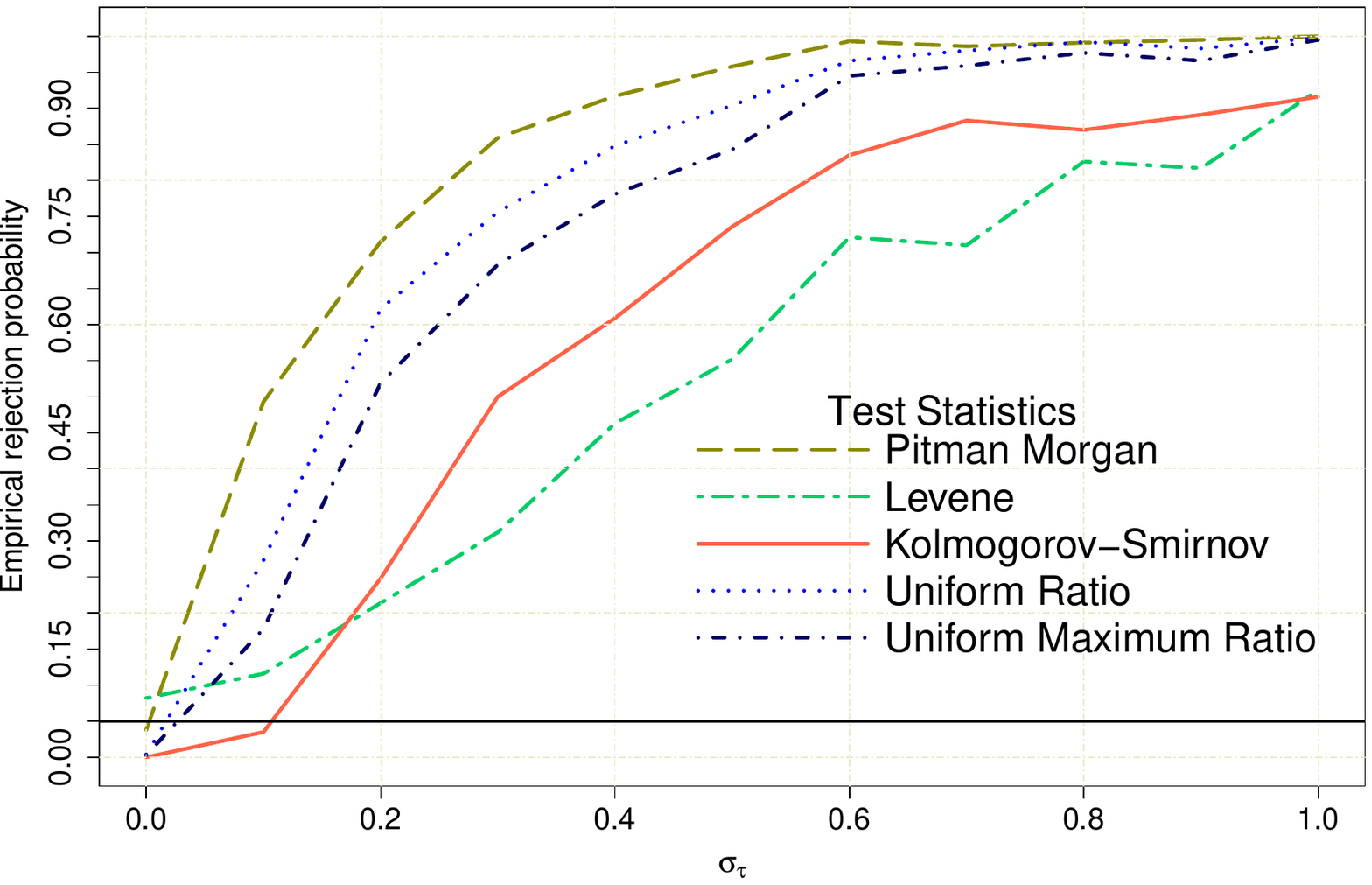}       
}
\label{powercurves-test statistics}
\end{figure}
In Figure \ref{powercurves-test statistics}(a), we compare the performance of the absolute ratio of variance (ratio) and maximum ratio test statistics. Specifically, we report the empirical rejection probabilities by varying $\sigma_\tau$ over the range  $ \{0.00, 0.01, 0.02, \dots, 1.00 \}.$ As expected, the result shows that the statistics tend to under-reject when the null is true. The statistics based on the  H\'ajek estimator tend to exhibit lower rejection rates compared to their uniformly weighted counterparts, although the differences are negligible. The difference is negligible since the exposure mapping is binary, and the variability in inclusion probabilities is small.
 This corroborates the theoretical findings in Section~\ref{guidelines}. 

In Figure \ref{powercurves-test statistics}(b), we compare the rejection rates of all the statistics computed using uniformly weighted estimators. Our results show that the Pitman--Morgan statistics tend to have the lowest size distortion and also outperform the rest in terms of power. This is unsurprising since the Pitman--Morgan statistics is related to the likelihood ratio criterion (see \cite{morgan1939test}); as such, it satisfies the Neyman--Pearson Lemma \citep{neyman1933ix}. 
In both figures, p-values are estimated using the H\'ajek estimator.

\subsection{Performance of Testing Procedures}
In this subsection, we compare the finite sample performance of four procedures: \textit{Biclique-Oracle (B-Oracle), super-focal-Oracle (SF-Oracle), Plug-in, and Confidence Interval (CRI-CI).}\footnote{For CRI-CI, we set $\gamma=0.0001$ and confidence interval grid size of 151.} The first two procedures rely on knowledge of the nuisance parameter and are infeasible in many practical settings.  B-Oracle uses the biclique-based conditioning procedure of \cite{puelz2022graph}, while SF-Oracle is based on Procedure \ref{alg:unadjusted CDE}.  On the other hand, the Plug-in method estimates the ATE and substitutes it directly into the null hypothesis, while the CRI-CI method is based on Procedure \ref{Alg: Ag2}.

Throughout this subsection, we employ an adjacency matrix where each unit is connected to at most five other units to mimic the observed network structure of our empirical application data in the next section. In addition, we consider two distributions of $\Psi_i(\pi)'$s: (i) i.i.d draws from a ``near-symmetric'' \textit{Beta distribution} with shape parameters $10$ and $10+\pi$ for all $\pi \in \boldsymbol{\Pi},$ and (ii) i.i.d draws from a right-skewed \textit{Beta distribution} with shape parameters $0.5$ and $2+\pi$ for all $\pi \in \boldsymbol{\Pi}.$ Finally, we use the  H\'ajek absolute ratio test statistic.

Due to the high computational time requirement of the B-Oracle procedure, we only compare its empirical size to that of SF-Oracle. We set the number of randomizations in both procedures to $ 50$.
\begin{table}[h]
    \centering
    \renewcommand{\arraystretch}{1.1} 
        \caption{Empirical size of the B-Oracle and SF-Oracle procedures}
    \begin{tabular}{cc} 
       \hline
       Procedure & Empirical Size \\
       \hline
       SF-Oracle & 0.004 \\
       B-Oracle  & 0.135 \\
       \hline
    \end{tabular}
    \label{size table}
\end{table}
\vspace{-0.3cm}
The results in Table \ref{size table} show that the Biclique-Oracle procedure overrejects under the null. We conjecture that this is primarily attributable to the small size of the bicliques generated by the algorithm, which limits the amount of information used for inference and thus reduces power. Across 1000 replications, we record an average biclique size (the number of focal units and assignments) of 6.944. In contrast, the SF-Oracle procedure uses 50 focal assignments to estimate the reference null distribution in every iteration. Thus, as expected, the proposed SF-Oracle procedure is valid. In Section 8.1 of the Supplementary Material, we present a formal analytical comparison of the power properties of the B-Oracle and the SF-Oracle. 

Next, we extend our simulation exercise by computing the empirical rejection probabilities of the SF-Oracle, Plug-in, and CRI-CI procedures by varying $\sigma_\tau$ over the range  $ \{0.00, 0.01, 0.02, \dots, 1.00 \}.$ For all the results reported below, we set $B=199.$ 
\begin{figure}[h]
\setcounter{subfigure}{0}
\centering
\caption{Empirical rejection probabilities at the $5\%$ significance level.}
\subfigure[Power curve when  $\Psi_i(\pi)\sim Beta(10,10).$]{
    \includegraphics[width=0.47\columnwidth, keepaspectratio]{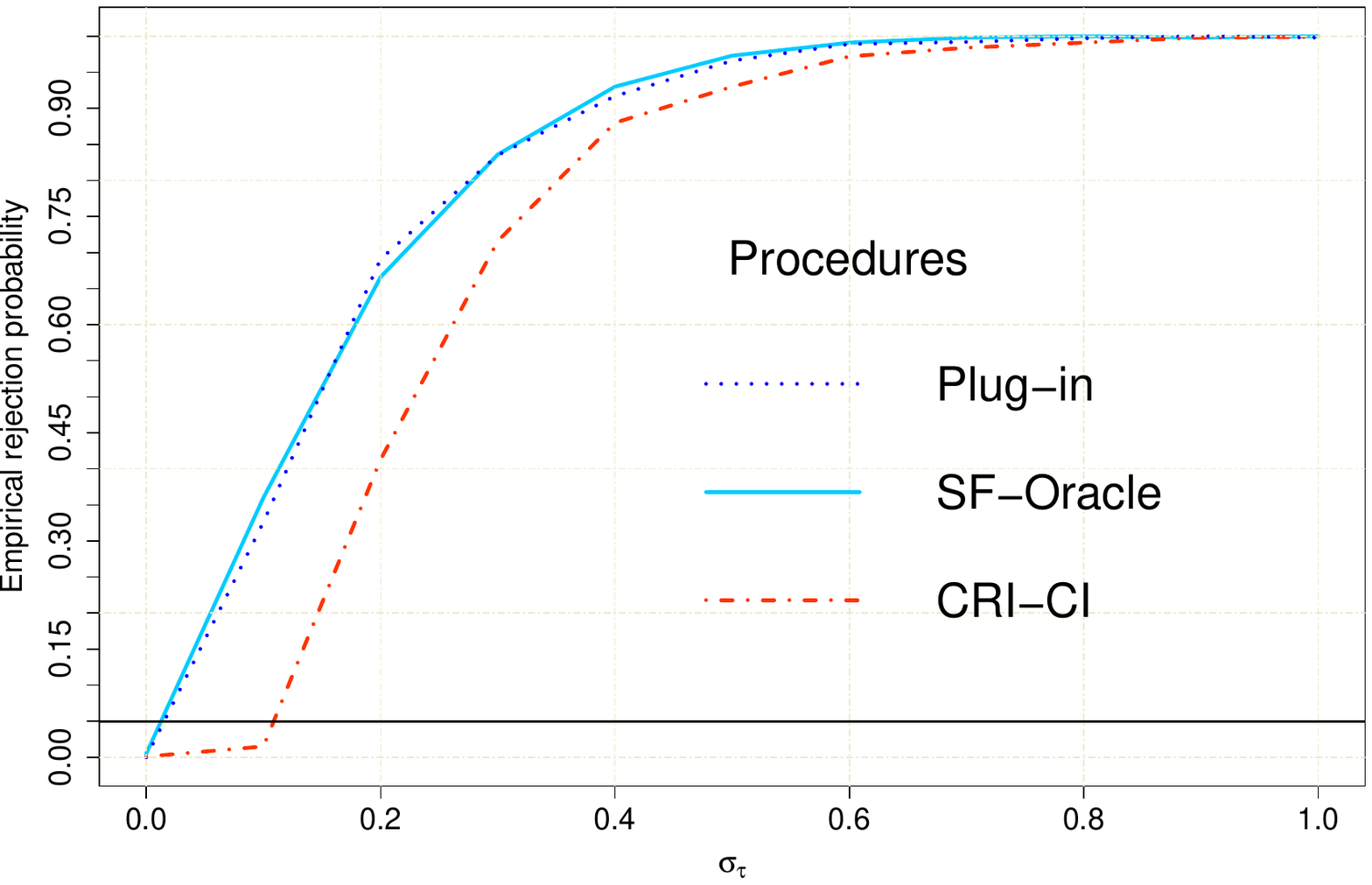}
}
\subfigure[Power curve when $\Psi_i(\pi)\sim Beta(0.5, 2).$]{
    \includegraphics[width=0.47\columnwidth, keepaspectratio]{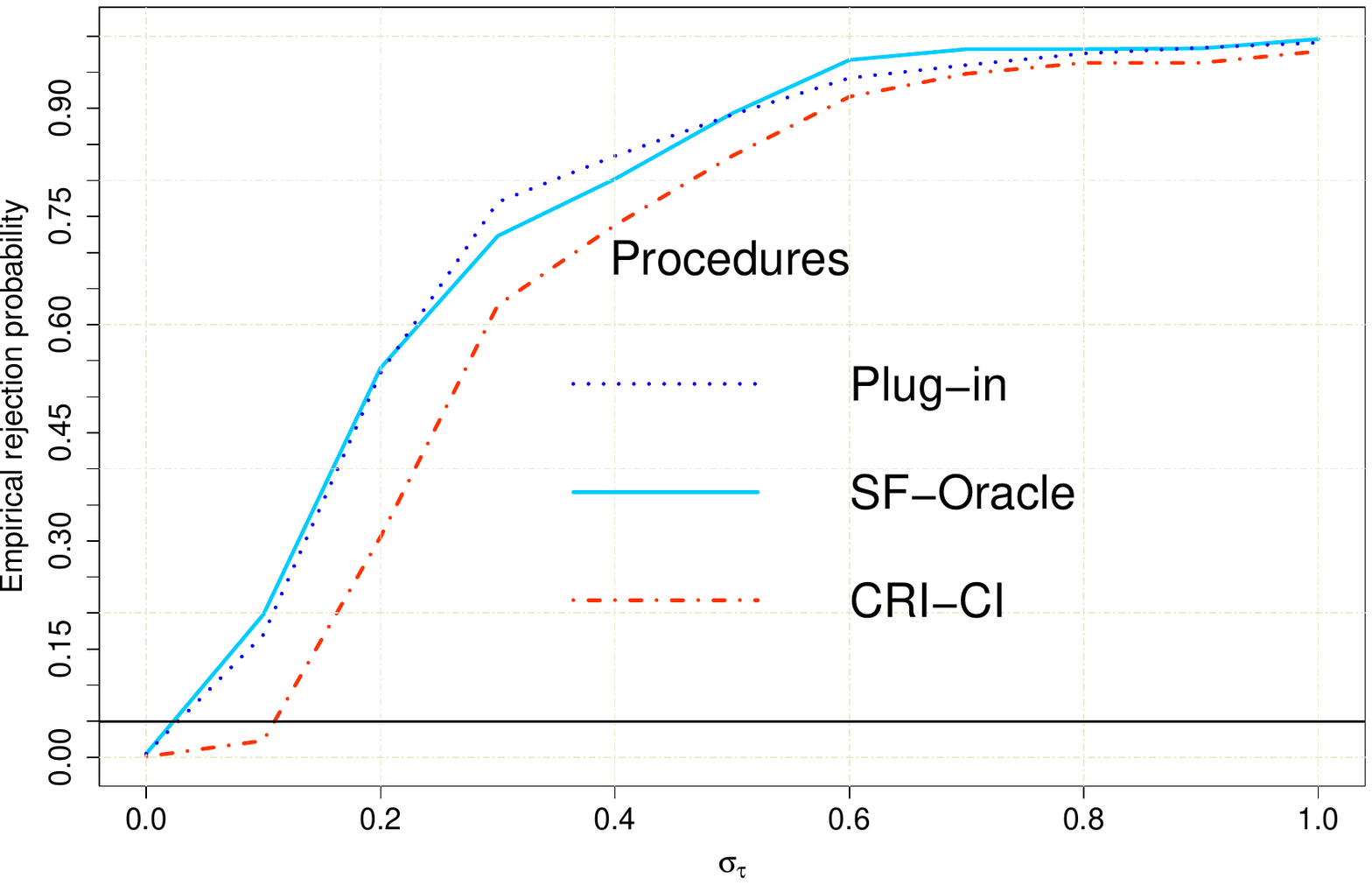}       
}
\label{powercurves}
\end{figure}

Using \textit{symmetrically distributed untreated potential outcomes}, the power curves in Figure \ref{powercurves}(a) show that the \textit{Plugin}, \textit{SF-Oracle}, and \textit{CRI-CI} procedures are all valid, with the \textit{CRI-CI} procedure uniformly exhibiting lower power as expected. Notably, the \textit{Plugin} and the \textit{SF-Oracle} procedures share an identical power function. This equivalence aligns with findings in \cite{ding2016randomization} for symmetric distributions. The \textit{CRI-CI} procedure has low statistical power for parameter values close to the null parameter value $(\sigma_\tau=0)$, corroborating our theory.    In Figure \ref{powercurves}(b), we display the power curves under the \textit{asymmetrically distributed untreated potential outcomes}. In general, the results are similar to those in Figure \ref{powercurves}(a). 

Thus, among procedures that guarantee theoretical control of Type I error, the proposed \textit{SF-Oracle} procedure is most suitable when all parameters are known, whereas the \textit{CRI-CI} procedure is appropriate in the presence of nuisance parameters.

\section{Empirical Application} \label{application}
In this section, we illustrate the application of the proposed testing procedures using real data from the field experiment conducted by \citet{cai2015social}. The study was designed to assess whether farmers’ understanding of a weather insurance policy influences their decision to purchase the product. Specifically, the authors evaluate the effects of two types of information sessions on insurance adoption among 4,902 households residing in 173 small rice-producing villages---nested within 47 administrative villages---across three regions in Jiangxi Province, China. Their findings indicate that the format of the information session not only has a direct impact on participants’ adoption behavior but also exerts significant peer effects on the adoption decisions of their named friends.

The dataset includes detailed social network information at the household level, as well as a range of pre-treatment covariates, including age, gender, rice cultivation area, risk aversion score, and the fraction of household income derived from rice production. The outcome of interest is binary, indicating whether the household purchased the insurance policy or not. The resulting social network is sparse because households could nominate up to five friends.

In each village, the experimental design consisted of two rounds of information sessions introducing the insurance product. During each round, two sessions were conducted \textit{simultaneously}: one providing basic information (\textit{the simple session}) and the other offering more detailed information (\textit{the intensive session}). The second round of sessions was administered three days after the first. This delay was sufficiently long to allow participants to share information with their direct friends but not enough to fully diffuse shared information throughout the broader network via friends-of-friends; see  \citet{cai2015social}. Consequently, Assumption \ref{direct neighbors} is plausibly satisfied.

To illustrate the proposed  \textit{CRI-CI} testing procedure, we restrict attention to three of the largest villages in the study---Dukou, Yazhou, and Yongfeng---comprising 502 households. Because the overall social network is clustered, we can analyze a subset of villages and their corresponding sub-networks in isolation.

A specification test of the interference structure by \citet{hoshino2023randomization} suggests that the correct network exposure mapping for this dataset is a threshold function of the number of neighbors who attended the first-round intensive sessions. This is formally defined as  
$
\pi_i(\mathbf{T}):= \mathbbm{I}(\sum_{j=1}^NT_jP_jA_{ij}/ \sum_{j=1}^NP_jA_{ij}\geq 0),
$
where for $j\in [N],$ $T_j$ and $P_j$ are the treatment indicators: $T_j = 1$ if household $j$ attended the intensive session (and zero otherwise), and $P_j = 1$ if household $j$ attended the first-round sessions (and zero otherwise).\footnote{We do not account for selection in the test, as post-selection inference is beyond the scope of this paper.} 

We test three null hypotheses: (i) $H_0,$ (ii) $H^{\Pi}_{0}:
    Y_i(1, \pi) - Y_i(0, \pi)= \tau (\pi)   \,\,\text{for some function}\,\, $\\$ \tau(\cdot), \,\, \forall\,\, \pi \in \mathbf{\Pi},\,\, \text{and}\,\,\forall \,\,i\in [N],$ and
(ii) $H^{X,\Pi}_{0}:
    Y_{i}(1, \pi) - Y_{i}(0, \pi)= \tau(\pi, X_i)  \, \text{for some }\, \tau(\cdot, \cdot),\, $\\$\forall\pi \in \mathbf{\Pi},\,\forall\, X_i \in \mathbbm{X}\,\text{and} \,\forall\,i\in [N];$ see Section 7 of the Supplementary Material for discussions of these nulls.
For $H^{X,\Pi}_{0}$, we use the binary covariate \textit{insurance\_repay}, which equals one if a household previously received a payout from an insurance policy and zero otherwise.

As described in \citet{cai2015social}, the treatment pair $(T, P)$ was assigned to households using a stratified randomization design using \textit{household size} and \textit{area of rice production per capita}. We treat $T_i$ as the sole source of randomness, conditioning on the observed values of  $P_i,$ which are held fixed and included as covariates for all $i\in [N].$ In particular, we define the null hypotheses with respect to units for which $P_i=0$, i.e., we focus on the subset of households that attended the second-round sessions.

We use the focal assignment set defined as  $\mathcal{T}_\epsilon(\mathbf{t}^{obs}):= \{\mathbf{t}' \in \mathcal{T}_0,\,\, \epsilon \leq \hat{R}(0,\mathbf{t}';\mathcal{S}(\mathbf{t}^{obs}))\leq|\mathcal{S}(\mathbf{t}^{obs})|-\epsilon,\,\, \text{and}\,\, \epsilon \leq \hat{R}(0,\mathbf{t}';\mathcal{S}(\mathbf{t}^{obs}))\leq|\mathcal{S}(\mathbf{t}^{obs})|-\epsilon \},$ where $\epsilon$ is varied between 20 to 50, for robustness check. 

Tables \ref{apply1}–\ref{apply3} report p-values for tests of the null hypotheses evaluated at each exposure value, using the H\'ajek absolute ratio of variance statistic and the \textit{CRI-CI} procedure with $\gamma=0.0001, B=399$ and confidence interval grid size of 151.  To conduct joint inference across exposure values, one may apply the standard multiple-testing procedures (MTPs) that control either the family-wise error rate or the false discovery rate; see \citet{romano2010multiple} for an overview of MTPs.

\begin{table}[h]
\centering
    \centering
   \caption{Test of $H^{\tau}_0$ at  $\alpha=0.05$}
           \renewcommand{\arraystretch}{1}
    \label{apply1}
\begin{adjustbox}{width=9cm}
    \begin{tabular}{p{2.7cm} p{2.7cm} p{2.7cm}}
     \hline
    $\epsilon$ & $\widehat{pval}_{0}$ & $\widehat{pval}_{1}$ \\
     \hline
     20 & 0.952 & 0.687\\
     30 & 0.732 & 0.351\\
     40 & 0.975 & 0.689\\
     50 & 0.998 & 0.549\\
     \hline
    \end{tabular}
    \end{adjustbox}
\end{table}

\vspace{-0.5cm}
\begin{table}[h]
    \centering
    \caption{Test of $H^\Pi_0$ at  $\alpha=0.05$}
         \renewcommand{\arraystretch}{1}
    \label{apply2}
  \begin{adjustbox}{width=9cm}
    \begin{tabular}{p{2.7cm} p{2.7cm} p{2.7cm}}
     \hline
     $\epsilon$ & $\widehat{pval}_{0}$ & $\widehat{pval}_{1}$ \\
     \hline
   20& 0.952 & 0.797\\
   30& 0.744 & 0.544\\
   40& 0.975 & 0.797\\
   50& 0.998 & 0.665\\
     \hline
    \end{tabular}
    \end{adjustbox}
\end{table}

\vspace{-0.5cm}
\begin{table}[h]
\caption{Test of $H^{X,\Pi}_0$ at  $\alpha=0.05$.  }
 \renewcommand{\arraystretch}{1.3}
\label{apply3}
\centering
\begin{adjustbox}{width=13cm}
  \begin{tabular}{ p{2.5cm} p{2.5cm} p{2.5cm} p{2.5cm} p{2.5cm} p{2.5cm}}
    \hline
     {$\epsilon$} &{$\widehat{pval}_{00}$}  & {$\widehat{pval}_{10}$} & {$\widehat{pval}_{01}$} & {$\widehat{pval}_{11}$}   \\
       \hline
  10 & 0.998 & 0.995 &  0.730 & 0.975\\
  15 & 0.933 & 1.000  & 0.448 & 0.998\\
  20 & 0.631 & 0.998  & 0.430 & 0.995\\
    \hline
  \end{tabular}
  \end{adjustbox}
\end{table}
Based on the p-values in Table \ref{apply1}, we fail to reject the null hypothesis of constant treatment effect across the population. As a result, there may be no heterogeneous effect on the decision to purchase weather insurance among participants in the second round. Based on the decision from  the test of $H^{\tau}_0,$ we must also fail to reject $H^{\Pi}_0$ and $H^{X,\Pi}_0.$ The p-values in  Tables \ref{apply2}--\ref{apply3} corroborate this assertion.

\section{Conclusion}\label{conclusion} 
This paper develops randomization-based inference procedures for testing heterogeneous treatment effects in the presence of network interference. Within the exposure mapping framework, we formulate a general class of non-sharp null hypotheses that encompass various notions of constant treatment effects in networked populations. These hypotheses depend on unknown functions that may act as nuisance parameters, and the restrictions they impose do not permit full imputation of unobserved potential outcomes. Existing conditional randomization procedures either lack statistical power, are invalid, or are inapplicable in this setting. 

We propose a novel randomization testing procedure that constructs a data-dependent focal assignment set tailored to the observed treatment assignment and exposure configuration. In contrast to conventional approaches that fix focal units and assignment sets ex-ante, our method allows the set of focal units to vary across focal assignments in accordance with the restrictions imposed by the null hypothesis. These adaptive features introduce technical complications that render the use of uniformly weighted estimators of test statistics and p-values invalid. To overcome this, we propose consistent and unbiased inverse probability-weighted estimators. Under general conditions on the test statistic, we establish the asymptotic validity of the procedure and characterize the limiting size distortion in terms of observable quantities.

We illustrate the proposed procedure using data from the field experiment of \citet{cai2015social}, which investigates how information influences the adoption of weather insurance among rice farmers in rural China. Finally, we present results from an extensive Monte Carlo study that corroborate the theoretical findings.

The randomization testing procedure developed in this paper is broadly applicable to a wide class of partial null hypotheses, including those arising in settings without interference. Its flexibility in accommodating non-sharp nulls while preserving validity makes it a promising foundation for addressing more challenging testing problems, such as Neyman's weak null hypothesis \citep{neyman1923application}. An important direction for future research is the development of refined procedures that mitigate the finite-sample size distortion observed under the proposed approach, potentially through improved conditioning schemes or alternative test statistics.

\bibliographystyle{chicago} 
\bibliography{reference}

\pagebreak
\begin{center}
\textbf{\Huge Supplementary Material}\label{supp mat} 
\end{center}
\appendix

\setcounter{section}{0}
\setcounter{equation}{0}
\setcounter{figure}{0}
\setcounter{table}{0}
\setcounter{page}{1}
\makeatletter
\renewcommand{\theequation}{S\arabic{equation}}
\renewcommand{\thefigure}{S\arabic{figure}}
\renewcommand{\thetable}{S\arabic{table}}
\renewcommand{\bibnumfmt}[1]{[S#1]}
\renewcommand{\citenumfont}[1]{S#1}
\renewcommand{\thesubsection}{\thesection.\arabic{subsection}}

\begin{alphasection}

\numberwithin{equation}{section}
\section{Practical Guidelines}
\subsection{Choice of the Intervals $I_{t,\epsilon}$}\label{choose interval}
The choice of the interval $\mathcal{I}_{t,\epsilon}$ depends on the population size, the exposure mapping, and the particular test statistic employed. In general, we recommend selecting intervals of the form

\begin{equation} \label{interval}
    \mathcal{I}_{t,\epsilon} =[\epsilon,\,\, |\mathcal{S}(\mathbf{t}^{obs})|-\epsilon] \subset \mathbbm{Z}^{+},
\end{equation}
for some $\epsilon\in \mathbbm{Z}^{+},$\ where $\epsilon$ reflects design-specific features and properties of the test statistic.

In most applications, the choice $\epsilon=1$ yields the widest possible intervals $\mathcal{I}_{t,1}$, ensuring that each effective treatment arm contains at least one unit, which is sufficient to compute the test statistic. However, since such intervals accommodate small effective sample sizes, it may result in imprecise null distributions.

When the population is sufficiently large and the number of exposure values is relatively small, the choice of $\epsilon$ should be guided by the properties of the test statistic. For instance, in the case of sample variance-based statistics, the minimal recommended value is $\epsilon = 2$. If the population is large enough to yield a substantial number of super-focal units, it may be advisable to select $\epsilon \geq 30$, following established rules of thumb for reliable variance estimation; see, e.g., \cite{van2011statistical}.

 Using the Beta(0.5, 2) DGP described in Section 5.1 of the main text,
 we compute the empirical size $(\sigma_\tau=0)$ and power $(\sigma_\tau=1)$ of Procedure 1 (in the main text)  for different values of $\epsilon$ using the intervals, $\mathcal{I}_{t,\epsilon},$ in \eqref{interval}.
Table S.1 reports the corresponding rejection probabilities and the average computation time in seconds (in parentheses) required to obtain each p-value across iterations.\footnote{We implement our program on a  \texttt{Windows $11$} computer with $16$GB RAM and $11$th Gen Intel(R) Core(TM) i$7-1165$G7 @ $2.80$GHz   $2.80$ GHz processor using  \texttt{R}, version $4.4.2.$ We use the \texttt{doparallel} package  in  \texttt{R} with 7 cores.} 

\begin{table}[ht]\label{rejprobs}
\caption{Sensitivity of  Procedure 1 to the choice of  $\epsilon$ using $\mathcal{I}_{t,\epsilon}$ in \eqref{interval}}
\centering
 \renewcommand{\arraystretch}{1}
\begin{adjustbox}{width=12cm}
  \begin{tabular}{ p{2cm} p{2cm} p{2cm} p{2cm} p{2cm} p{2cm}}
    \hline
     {} &{$\epsilon=2$}  & {$\epsilon=10$} & {$\epsilon=15$} & {$\epsilon=25$}   \\
      \hline
  $\sigma_\tau=0$ & $\underset{(3.26178)}{0.007}$ & $\underset{(3.243553)}{0.006}$  &  $\underset{(3.26292)}{0.004}$ & $\underset{(4.546059)}{0.001}$\\
                 &                                &        &        &\\
  $\sigma_\tau=1$ & $\underset{(3.265274)}{0.995} $& $\underset{(3.245383)}{0.993}$  &  $\underset{(3.523879)}{0.993}$ & $\underset{(4.48634)}{0.996}$\\
    \hline
  \end{tabular}
  \end{adjustbox}
\end{table}
The results indicate that the performance of Procedure 1 is relatively robust to the choice of intervals \( \mathcal{I}_{0,1} \) and \( \mathcal{I}_{1,1} \). In particular, varying \( \epsilon \) between 2 and 25 induces only modest changes in empirical size and power. However, the increase in \( \epsilon \) from 2 to 25 increases the average run time by approximately 27\%.

\subsection{Choice of the Random Focal Assignment Size $B.$}
As discussed in Section~4 of the main text, computing exact $p$-values based on either the uniform or inverse probability weighting scheme becomes computationally burdensome when the number of focal assignments is large. To address this issue, we recommend the use of Monte Carlo methods to approximate the $p$-values. Specifically, we propose drawing an independent and identically distributed (i.i.d.) sample of size $B$ from the relevant focal assignment set. In what follows, we provide guidance on selecting $B$.

The choice of \( B \) depends on the size of the focal assignment set, which, unlike in standard randomization procedures, is not fixed in advance. In our framework, this set is defined relative to the observed treatment vector \( \mathbf{t}^{\mathrm{obs}} \) and varies across realizations. Specifically,
\[
\mathcal{T}_\epsilon(\mathbf{t}^{\mathrm{obs}}) := \left\{ \mathbf{t}' \in \mathcal{T}_0 : \hat{R}(0, \mathbf{t}';  \mathcal{S}(\mathbf{t}^{\mathrm{obs}})) \in \mathcal{I}_{0,\epsilon} \text{ and } \hat{R}(1, \mathbf{t}'; \mathcal{S}(\mathbf{t}^{\mathrm{obs}})) \in \mathcal{I}_{1,\epsilon} \right\},
\]
where
\[
\hat{R}(t, \mathbf{t}'; \mathcal{S}(\mathbf{t}^{\mathrm{obs}}) )\coloneqq \sum_{i \in \mathcal{S}(\mathbf{t}^{\mathrm{obs}})} \mathbbm{I}\{ t'_i = t,\, \pi_i(\mathbf{t}') = \pi_k \}, \quad \text{for } t \in \{0,1\}.
\]
The random nature of \( \mathcal{T}_\epsilon(\mathbf{t}^{\mathrm{obs}}) \) implies that the number of feasible draws for the Monte Carlo procedure is data-dependent.

Given \( \mathcal{T}_0 \), enumerating the assignment vectors in \( \mathcal{T}_\epsilon(\mathbf{t}^{\mathrm{obs}}) \)---which guides the choice of $B$--- can be formulated as a constrained graph coloring (CGC) problem. While this problem is tractable when \( \mathcal{T}_0 \) is small, it becomes NP-hard for moderate to large instances.

To address this challenge, we propose a simple adaptive procedure for selecting \( B \). First, set \( B \) as large as computationally feasible (e.g., 599). Second, implement the sampling procedure under a time threshold and terminate execution if the threshold is exceeded, thereby avoiding potential infinite loops caused by rare constraint-satisfying assignments. Third, repeat the process with a smaller value of \( B \) until  \( p \)-values are obtained.

Following the guidance of \citet{davidson2000bootstrap} in the context of bootstrap sample sizes, we consider values of \( B \) such as 99, 199, 299, and so on.\footnote{We thank Russell Davidson for suggesting the applicability of the recommendation in \citet{davidson2000bootstrap} to randomization-based inference.} A formal justification for this rule in the current setting is left for future research.

\section{Pitman--Morgan Test }\label{PT discussion}

The Pitman--Morgan test \citep{pitman1939note, morgan1939test} provides a classical method for testing the equality of variances in paired samples. Let $\{(X_i, Y_i)\}_{i=1}^n$ denote a random sample of paired observations, where $X_i$ and $Y_i$ are real-valued measurements on the same statistical unit. The null hypothesis of interest is:
\begin{equation}
H_0: \operatorname{Var}(X_i) = \operatorname{Var}(Y_i),
\end{equation}
under the maintained assumption that $(X_i, Y_i)$ are jointly normally distributed and identically distributed across $i$.

The key insight of the test is that under joint normality, equality of variances implies zero correlation between the sum and difference of the paired measurements. Define $D_i = Y_i - X_i$ and $S_i = Y_i + X_i$. Then, under $H_0$,
\begin{equation}
\operatorname{Cov}(D_i, S_i) = \operatorname{Var}(Y_i) - \operatorname{Var}(X_i) = 0,
\end{equation}
so the test reduces to examining whether the sample correlation between $D_i$ and $S_i$ is significantly different from zero. Let $r_{DS}$ denote the Pearson sample correlation between $\{D_i\}$ and $\{S_i\}$. The test statistic is given by:
\begin{equation}\label{PT}
t = \frac{r_{DS} \sqrt{n - 2}}{\sqrt{1 - r_{DS}^2}},
\end{equation}
which follows a Student’s $t$-distribution with $n - 2$ degrees of freedom under the null. 

In the main text, we adapt the test to fit the framework of the paper. Given that we observe only one potential outcome, the outcomes are not paired, as in the classical setting for which the Pitman--Morgan test was developed. Additionally, the joint distribution of potential outcomes may not be bivariate normal for any given exposure value. However, since reference distributions are derived via randomization in this paper, we can employ the test statistic in \eqref{PT} under settings that deviate from the classical setting where the null distribution was derived.

We adapt the Pitman--Morgan statistic by constructing matched pairs of treated and control units within each exposure level. Specifically, for a given exposure category, treated units are randomly matched to control units.\footnote{Matching can alternatively be performed using a similarity metric such as the propensity score.} When the number of treated and control units is unequal, the larger group is trimmed by randomly discarding units to ensure a one-to-one match. Once matched pairs are formed, the test statistic is computed following the classical Pitman--Morgan procedure.

A principal limitation of the proposed matched  Pitman--Morgan statistic is the potential loss of observations when the number of treated and control units within an exposure level is substantially imbalanced. To enforce one-to-one matching, the procedure requires discarding units from the larger group, thereby reducing the effective sample size. The resulting decrease in sample size may lead to diminished statistical power and increased variability in the test statistic, potentially compromising the precision and reliability of the inference.

\section{Additional Simulation Results}
\subsection{Simulation Evidence of the Pairwise  Dominance Condition} \label{sto dom sim}
Using the simulation design described in Section~5.1, we compare the empirical cumulative distribution functions (ECDFs) of the observed test statistics (in blue) and the mixture distribution of imputed test statistics under the null (in red). Four test statistics are considered.

Figure~\ref{fig:ECDF-1-ln} reveals that the absolute ratio of variance (ratio), maximum ratio, and shifted Kolmogorov--Smirnov statistics satisfy the pairwise dominance condition introduced in Section~3.1 of the paper. Using the uniform mixture of the randomized distributions---constructed from 1000 randomly drawn observed assignments---as the reference distribution, we find that the mixture distribution of imputed test statistics first-order stochastically dominates the distribution of observed test statistics. As a result, the pairwise dominance condition is satisfied in these cases.

In contrast, the Pitman--Morgan statistic does not satisfy the pairwise dominance condition. Nevertheless, it meets the sufficient conditions of Corollary~1. In particular, the ECDF of the mixture distribution crosses that of the observed test statistic, indicating that the distributions are not uniformly ordered but still satisfy the relaxed asymptotic condition. 
\begin{figure}[h]
    \centering
    \includegraphics[width=1\linewidth]{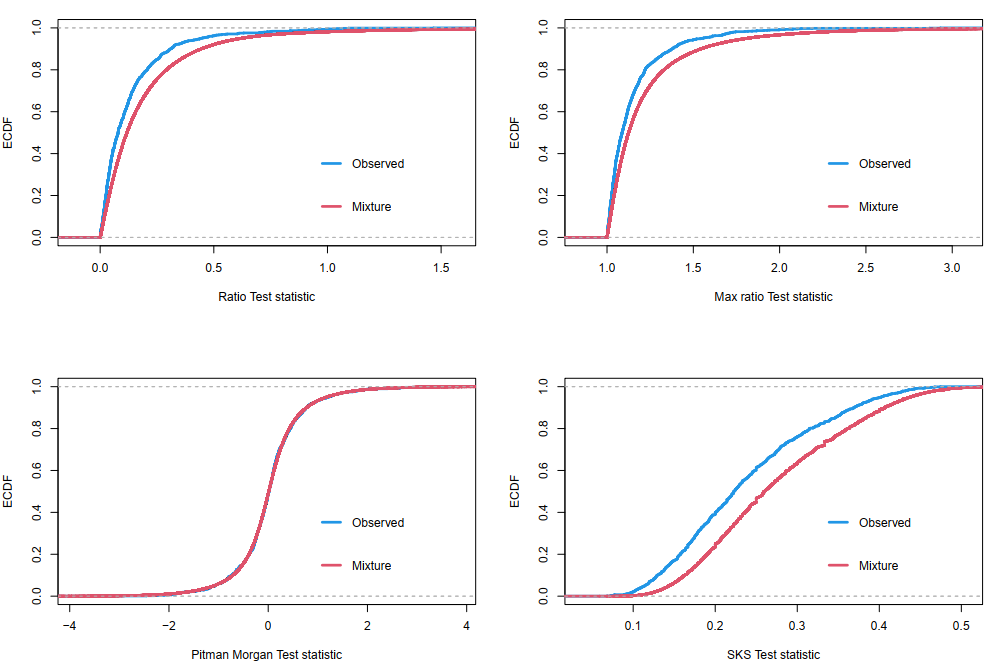}
    \caption{ECDFs of the observed test statistics, and the mixture distribution of imputed test statistics under the null, based on  Procedure 1. The data are generated under a design with a Beta(0.5, 2) distribution. See Section 5.1 of the main text for the detailed simulation design.}
    \label{fig:ECDF-1-ln}
\end{figure}

To empirically justify the use of the mixture distribution as a reference for characterizing asymptotic size distortion (see Section~3.1.1 of the main text), we report the empirical size corresponding to each panel in Figure~\ref{fig:ECDF-1-ln}, computed using Procedure 1. For the ratio statistic, the empirical size is 0.007; for the maximum ratio statistic, it is 0.003; the Pitman--Morgan statistic yields a size of 0.035; and the shifted Kolmogorov--Smirnov statistic yields 0.003. These results illustrate that greater divergence between the mixture distribution of imputed test statistics and the distribution of observed statistics is associated with more pronounced size distortion.

Finally, in Figure \ref{fig:ECDF-1-n}, we plot the distribution of p-values for each of the test statistics. 
\begin{figure}[H]
    \centering
    \includegraphics[width=1\linewidth]{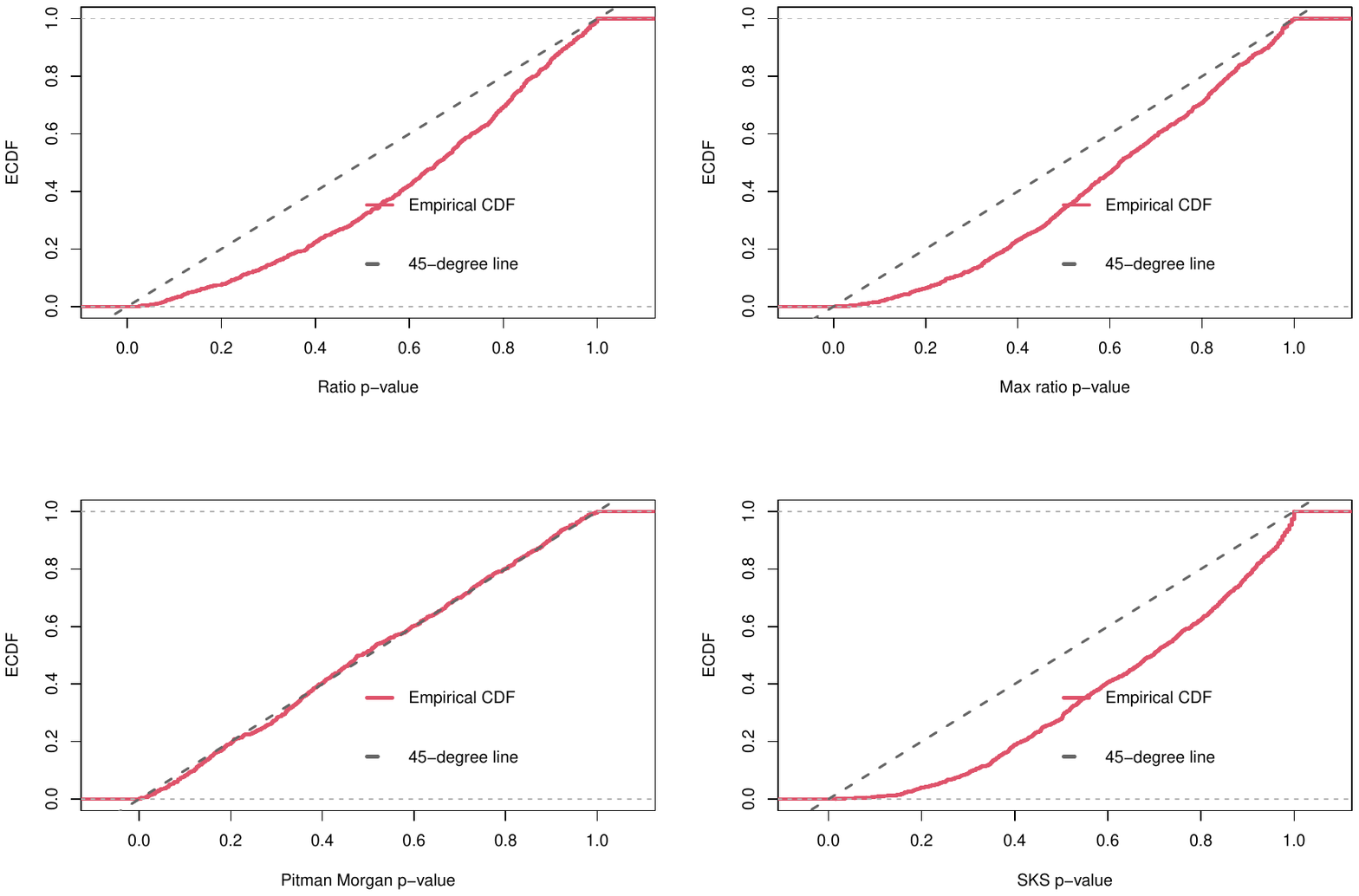}
    \caption{ECDFs of p-values and standard uniform distribution.}
    \label{fig:ECDF-1-n}
\end{figure}
Corroborating the theoretical results in the paper, the empirical distribution functions of the p-values associated with the ratio, maximum ratio, and shifted Kolmogorov--Smirnov statistics first-order stochastically dominate the standard uniform distribution. In contrast, the distribution of p-values derived from the Pitman--Morgan statistic crosses the standard uniform distribution.
\subsection{Finite Sample Properties for Multi-valued Exposure Mapping}
 The statistical power of the proposed procedures in this paper relies on having a large set of focal assignments for imputing test statistics. However, the more possible exposure mapping realizations there are (i.e., the larger the number of exposure values), the smaller the focal assignment set for a fixed population size.  Consequently, to achieve desirable finite-sample power, the population size must grow proportionally with the number of exposure values.

In this subsection, we examine the performance of the proposed procedures under a multi-valued exposure mapping. Specifically, we replicate the Monte Carlo designs described in Section 5.1 of the main text using the exposure mapping 
 \begin{equation*}
     \pi_i(\mathbf{T}) := \mathbbm{I}\Bigg(\sum_{j=1}^NT_jA_{ij}\Bigg)\in \{0,1, 2,3,4, 5\}.
 \end{equation*}
We compute the rejection probabilities for $H^{\tau}_{0},$ for exposure level  $\pi_k=2,$ with population size fixed at 800. All other parameters are held constant, as in Section 5.1.

Figure \ref{multiplepowercurves} presents the rejection probability curves for the different test statistics in  Procedure 1.  Here, the difference between the H\'ajek statistics and the uniformly weighted statistics is significant due to the large variation in inclusion probabilities induced by the exposure mapping. However, the size and power are consistent with those reported in Section 5.1, corroborating our theoretical findings.
\begin{figure}[H]
\setcounter{subfigure}{0}
\centering
\caption{Empirical rejection probabilities at the $5\%$ significance level.}
\subfigure[Power curve for variance-type test statistics]{
    \includegraphics[width=0.47\columnwidth, keepaspectratio]{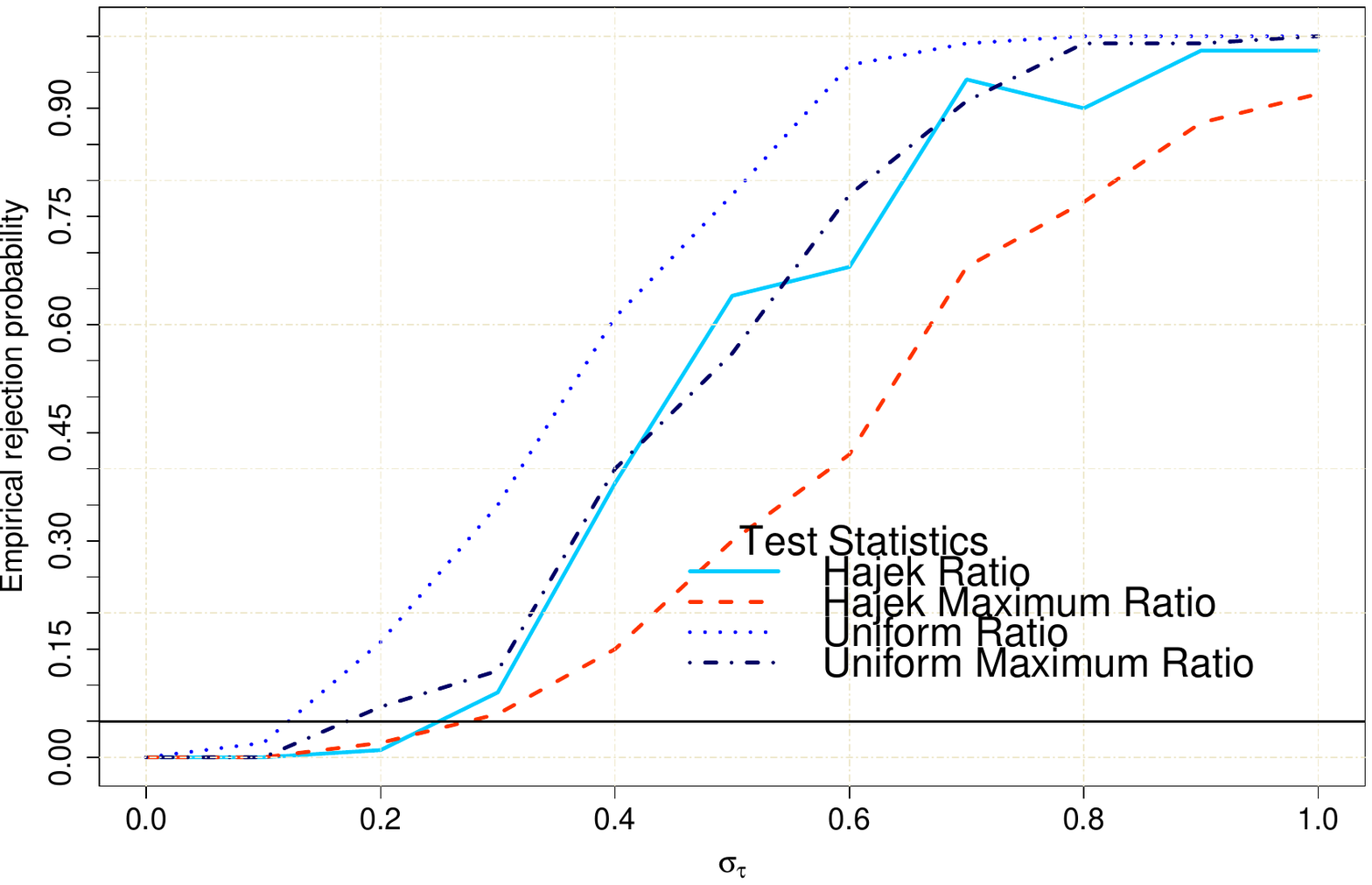}
}
\subfigure[Power curve for uniformly weighted test statistics]{
    \includegraphics[width=0.47\columnwidth, keepaspectratio]{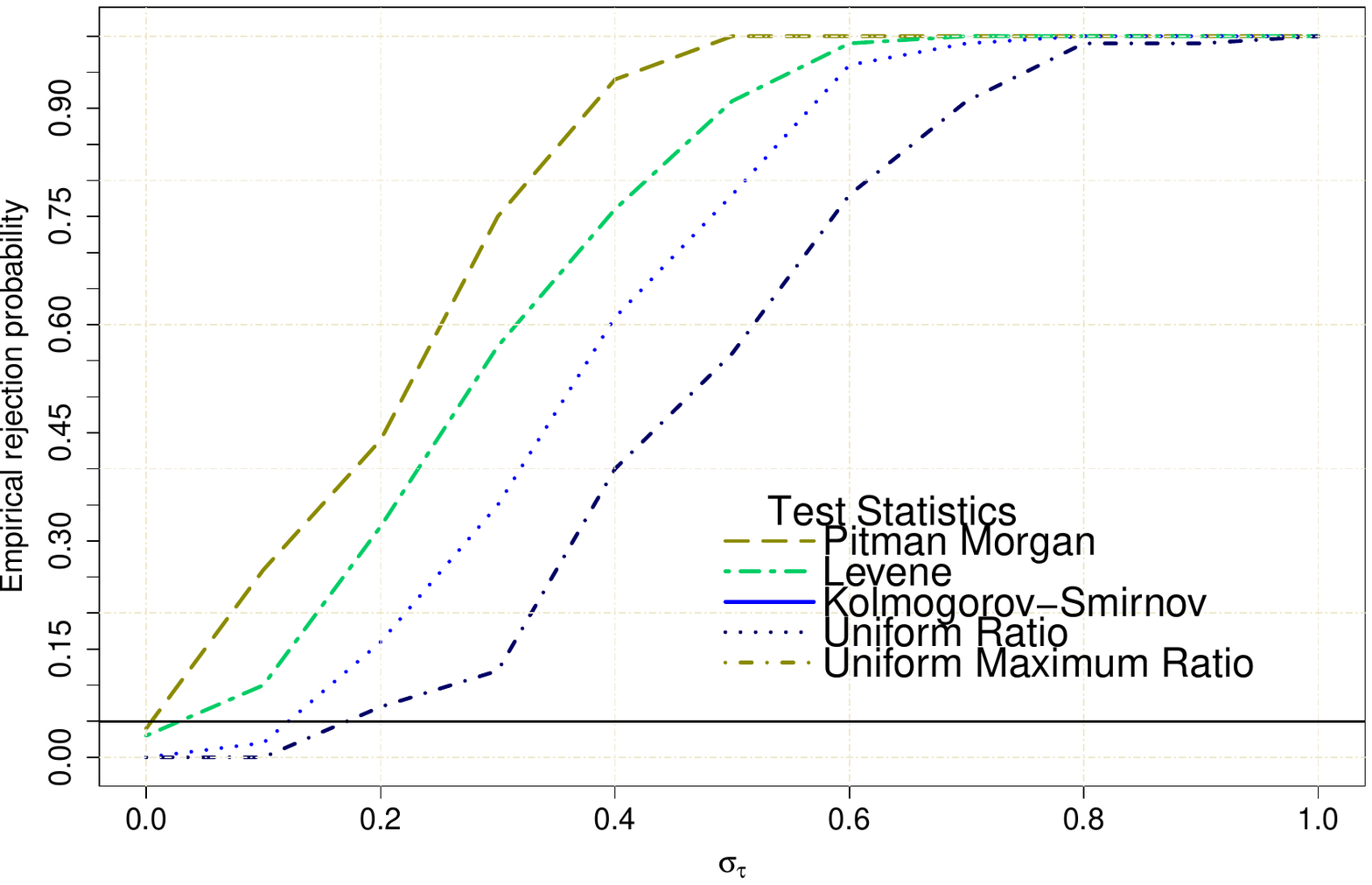}       
}
\label{multiplepowercurves}
\end{figure}

\section{Misspecification of Exposure Mapping}
In the presence of network interference, researchers often rely on exposure mappings to summarize each unit’s treatment-relevant environment \citep{aronow2017estimating}. These mappings play a central role in defining the hypotheses and constructing the test statistics employed in this paper. Recall Assumption 2 in the paper, which states that the exposure mapping is correctly specified.  However, exposure mappings are typically based on simplifications about spillover transmission mechanisms that may not hold in practice. This section examines the consequences of misspecifying the exposure mapping. In particular, according to \cite{aronow2017estimating}, the exposure mapping is misspecified if 
there exists some $i\in [N],$ $t\in\{0,1\}$ but $\mathbf{t}, \mathbf{t}' \in \mathcal{T}_0$ such that 
$\pi_i(\mathbf{t}) = \pi_i(\mathbf{t}')$ and $Y_{i}(t,\pi_i(\mathbf{t})) \neq Y_{i}(t,\pi_i(\mathbf{t}'))$. In that case
$
Y_i = \sum_{\mathbf{t} \in \mathcal{T}_0} \mathbbm{I}(\mathbf{T} = \mathbf{t} ) \cdot Y_{i}(\mathbf{t}), \quad \forall i \in [N].
$

As demonstrated in Section 8 of \citet{aronow2017estimating}, the Horvitz--Thompson estimator \citep{horvitz1952generalization} of the difference in mean potential outcomes across exposure values yields an unbiased estimate of the difference in population mean of individual average potential outcomes given different restrictions on the set of treatments implied in exposure values proposed in \cite{hudgens2008toward}. Accordingly, when the exposure mapping is misspecified, employing the Horvitz--Thompson difference-in-means estimator to test the hypothesis
\[
H_0^0: Y_i(1,\pi_k) - Y_i(0,\pi_k) = 0 \quad \text{for all } i \in [N],
\]
in effect, corresponds to testing the hypothesis
\[
\bar{H}_0^0: \bar{Y}_{i}(1; \pi_k) -\bar{Y}_{i}(0; \pi_k)=0 ), \quad \forall i \in [N],
\]
where
\[
\bar{Y}_{i}(t; \pi_k) = \sum_{\mathbf{t} \in \mathcal{T}_0 : t_i = t, \, \pi_i(\mathbf{t}) = \pi_k} Y_i(\mathbf{t}) \cdot \Pr(\mathbf{T} = \mathbf{t} \mid T_i = t, \pi_i(\mathbf{T}) = \pi_k).
\]
This expression represents the average potential outcome for unit $i$ under treatment status $t$ and exposure level $\pi_k$, averaged over the distribution of assignment vectors consistent with the conditioning event. Thus, in the presence of exposure mapping misspecification, the proposed procedures test a related null hypothesis that may nest the original null hypothesis.  

We conjecture that this correspondence extends to all null hypotheses encompassed by \(H_0^G\). That is, for each such null hypothesis, there exists a related null for which the proposed testing procedure yields valid inference under misspecification of the exposure mapping. A formal analytical investigation of this conjecture is left for future research.

\section{Constant Treatment Effects under  Interference}\label{int vs noint}
 To underscore the novelty of the testing problem addressed in the paper, we compare the hypothesis of constant treatment effects (CTEs) under the assumption of no interference---as in \citet{chung2021permutation} and \citet{ding2016randomization}---with the corresponding hypotheses under network interference considered here.

 Under the standard Stable Unit Treatment Value Assumption (SUTVA), the null hypothesis of interest in the aforementioned literature takes the form:
$
H^{\text{SUTVA}}_{0}:
   Y_i(1, \pi) - Y_i(0, \pi')= \tau \,\,\text{for some} \,\, \tau, \,\,\text{and}\,\, \pi, \pi'\in \{a\},\,\, \nonumber \,\,\forall\,\, i\in [N].
$
What, then, is the implication of rejecting or failing to reject $H^{\text{SUTVA}}_{0}$ for hypotheses of constant treatment effects under network interference---such as $H^{\tau}_0$? 
$H^{\text{SUTVA}}_{0}$ may equivalently be expressed as $H^{\text{SUTVA}}_0: F_{1}(y)= F_{0}(y -\tau), \forall y \in \mathbbm{Y}, \text{some}\,\, \tau,$  where $F_{1}$ and $F_{0}$ denote the empirical distributions of the classical potential outcomes $Y(1)$ and $Y(0)$, respectively \citep{ding2016randomization}. 

Interference introduces fundamental differences. First, the vectors $(Y(1), Y(0))$ and\\ $(Y(1,\pi), Y(0,\pi))$ for $\pi \in \boldsymbol{\Pi}$ generally have different empirical distributions. Moreover, the classical potential outcomes $(Y(0), Y(1))$ are undefined in settings with interference, as outcomes may depend on others' treatment assignments. This distinction underscores that rejecting (or failing to reject) $H^{\text{SUTVA}}_{0}$ has no direct implications for tests of CTEs in the presence of network interference.

\section{Classes of Null Hypotheses under $H_0^{G}$} \label{App 3}
Recall that the general null hypothesis of interest is 
\begin{align} 
H^{G}_{0}:
   &Y_i(t, \pi) - Y_i(t', \pi')= \tau(t, t', \pi, \pi', X_i) \,\,\text{for some function} \,\, \tau(\cdot), \\
   & \text{where}\,\, t,t'\in \{0,1\},\,\, \pi, \pi' \in \mathbf{\Pi},\,\,X_i \in \mathbbm{X} \,\, \text{and} \,\,\,\, i\in [N] \nonumber.
\end{align}
In this section, we discuss three classes of null hypotheses that $H_0^{G}$ represents.

\paragraph*{(i) No treatment effects under network interference}

The broadest null hypothesis, asserting no difference between any pair of potential outcomes for any unit, arises by setting $\tau(t, t', \pi, \pi', X_i)=0$ in $H_0^G$.  Formally, this corresponds to
\begin{equation} \label{no effect}
H^{NE}_{0}:
   Y_i(t, \pi) - Y_i(t', \pi')=0 \,\ \forall\,\, t,t' \in \{0,1\},\,\,  \forall \,\, \pi, \pi' \in \mathbf{\Pi}, \text{and} \,\,\forall\,\, i\in [N].
\end{equation}
This is a \textit{sharp} null hypothesis and can be tested using Fisher’s randomization test. It rules out direct and spillover effects, making it strong and often uninformative in applied settings. Consequently, empirical researchers typically focus on testing for the presence of either direct or spillover effects.

Modifying $H^{NE}_{0}$ slightly gives us the following null hypotheses of \textit{no direct effects} and \textit{no spillover effects} in  \eqref{no direct effect} and \eqref{no spillover effect} respectively.

\begin{equation} \label{no direct effect}
H^{NDE}_{0}:
   Y_i(1, \pi) - Y_i(0, \pi)=0 \,\ \forall\,\,  \pi  \in \mathbf{\Pi}, \text{and} \,\,\forall\,\, i\in [N].
\end{equation}

\begin{equation} \label{no spillover effect}
H^{NSE}_{0}:
   Y_i(t, \pi) - Y_i(t, \pi')=0 \,\ \forall\,\,  \pi\neq \pi'  \in \mathbf{\Pi}, \forall\,\, t\in\{0,1\}, \text{and} \,\,\forall\,\, i\in [N].
\end{equation}
These hypotheses are of particular empirical interest, as they enable independent assessment of direct and indirect treatment effects. However, in contrast to $H^{NE}_{0},$ they are \textit{not sharp} due to the multiplicity of potential outcomes.

\paragraph*{(ii) Constant direct treatment effects under network interference}

 Based on $H_0^G,$ we can generate several hypotheses that imply different notions of constant direct treatment effects.  We focus on three leading examples, beginning with the strongest form:
\begin{equation} \label{null 1}
H^{CDE}_{0}:
   Y_i(1, \pi) - Y_i(0, \pi)= \tau  \,\,\text{for some}\,\, \tau\in \mathbbm{R},\,\,  \forall\,\,  \pi \in \mathbf{\Pi},\,\, \text{and}\,\,\forall \,\, i\in [N].
\end{equation}
This hypothesis asserts that the direct treatment effect is constant across the entire population, regardless of network exposure.  Testing $H^{CDE}_{0}$ is crucial for designing treatment assignment rules aimed at maximizing direct welfare. For instance, failing to reject this null suggests that an optimal direct welfare-maximizing policy would treat all units if $\tau>0$ and none if $\tau\leq 0.$ See \cite{han2022statistical} and \cite{viviano2024policy}  for discussions on targeting rules under interference.

A special case of $H^{CDE}_{0}$ is considered in \citet{ding2016randomization} and \citet{chung2021permutation}, where the authors assume the absence of interference---that is, $\boldsymbol{\Pi}$ is a singleton---and set $\tau$ equal to the average treatment effect (ATE).  In this setting, the null hypothesis is \textit{not sharp}, as it depends on the unknown ATE, a nuisance parameter. 
In contrast, even if we set $\tau=ATE$ in \eqref{null 1},  the resulting hypothesis remains \textit{non-sharp}, not only due to the nuisance parameter $\tau,$  but also because of the multiplicity of potential outcomes under interference.  

The second hypothesis of interest within the class (ii) asserts that \textit{direct treatment effects only vary systematically across treatment exposure values}, formally stated as:
\begin{equation}\label{null pi}
 H^{\Pi}_{0}:
    Y_i(1, \pi) - Y_i(0, \pi)= \tau (\pi)   \,\,\text{for some function}\,\, \tau(\cdot), \,\, \forall\,\,  \pi \in \mathbf{\Pi},\,\, \text{and}\,\,\forall \,\,i\in [N].
\end{equation}
Failure to reject $H_0^\Pi$ may indicate that a policymaker interested in direct effects could design treatment allocation strategies based on the exposure variable. If  $\tau (\cdot)$  is the average treatment effect as a function of $\pi$, then  $H^{\Pi}_{0}$ will be non-sharp because of the nuisance parameters $\{\tau(\pi): \pi \in \mathbf{\Pi}\},$ and the multiplicity of potential outcomes.

The third hypothesis within this class asserts that \textit{direct treatment effects only vary systematically across pretreatment and exposure values}:

\begin{equation}\label{null x pi}
 H^{X,\Pi}_{0}:
    Y_{i}(1, \pi) - Y_{i}(0, \pi)= \tau(\pi, X_i)  \, \text{for some }\, \tau(\cdot, \cdot),\,\forall\pi \in \mathbf{\Pi},\,\forall\, X_i \in \mathbbm{X}\,\text{and} \,\forall\,i\in [N].  
\end{equation}
Testing $H_{0}^{X,\Pi}$  is particularly relevant in designing covariate-dependent eligibility rules, enabling the efficient allocation of scarce resources to maximize direct welfare within social networks.

\paragraph*{(iii) Constant indirect (spillover) effects under network interference}
Several null hypotheses represent distinct notions of constant spillover (or indirect) effects and can be derived as special cases of $H_0^G$. We discuss two such examples.

The first example is the null hypothesis:
\begin{align}\label{no hse treat}
 H^{^{SE(T)}}_{0}:
    &Y_{i}(t, \pi) - Y_{i}(t, \pi')= \tau(t) \,\, \text{for some}\,\, \tau(\cdot),\,\, \text{for}\,\, t \in \{0,1\}  \,\, \forall\,\,  \pi\neq \pi' \in \mathbf{\Pi},\,\, \text{for} \,\,i\in [N]. 
\end{align}

This hypothesis asserts that indirect effects exist but are constant within each treatment arm. It resembles the null hypothesis tested in \cite{aronow2012general} and  \textit{Hypothesis 2} in \cite{athey2018exact}. However, there are important conceptual distinctions. In the cited references, failing to reject the null may reflect the absence of interference, the presence of constant direct effects, or both. In contrast, under \eqref{no hse treat}, failing to reject directly implies the presence of interference with homogeneous spillover effects within each treatment level. Moreover, our framework permits testing such hypotheses for subsets of exposure values, a flexibility not accommodated in the referenced procedures. 

The second example is the null hypothesis:
\begin{align}\label{no hse exposure}
 H^{^{SE(\Pi)}}_{0}:
    &Y_{i}(t, \pi) - Y_{i}(t, \pi')= \tau(\pi,\pi') \,\, \text{for some}\,\, \tau (\cdot, \cdot),\,\, \text{for}\,\, t \in \{0,1\}  \,\, \forall\,\,  \pi\neq \pi' \in \mathbf{\Pi}, 
    \text{for} \,\,i\in [N]. 
\end{align}
This hypothesis allows spillover effects to vary across exposure levels but assumes they are invariant to the individual's treatment status. In other words, spillover effects depend only on the exposure contrast $(\pi, \pi')$, not on whether a unit is treated.

The non-sharpness of both hypotheses in \eqref{no hse treat} and \eqref{no hse exposure} arises from the same structural features that make the hypotheses under category (ii) above non-sharp. 

\section{Conditional Randomization Inference: A Review} \label{CRI}
This section reviews the general framework proposed by \citet{basse2019randomization} for constructing conditional randomization tests under interference in the presence of non-sharp null hypotheses.
Building on the foundational ideas of \citet{aronow2012general} and \citet{athey2018exact}, this framework formalizes a conditioning approach that enables valid inference despite the challenges posed by interference. Conditional randomization inference (CRI) methods typically involve selecting a subset of units --- referred to as focal units --- and a subset of treatment assignment vectors—referred to as focal assignments—to define a conditional reference distribution under the null. 

While this approach provides a powerful strategy for addressing non-sharpness, identifying appropriate conditioning sets remains nontrivial and is the subject of ongoing research.

\citet{athey2018exact} propose a conditioning procedure that discourages the use of functions of the observed treatment assignment as conditioning variables. In contrast, \citet{basse2019randomization} demonstrates that conditioning on specific functions of the treatment assignment can enhance statistical power while preserving validity under appropriate conditions. This divergence highlights an important trade-off in the design of conditional randomization tests: the tension between strict randomization orthodoxy and practical considerations related to power.

The formal concept presented by \cite{basse2019randomization} involves selecting a space of conditioning events defined as 
$
\mathbbm{C} = \{(\mathcal{U}, \mathcal{T}): \mathcal{U} \in \mathbbm{U}, \mathcal{T} \in \mathbbm{T}\},
$ 
where \(\mathbbm{U}\) denotes the power set of units and \(\mathbbm{T}\) represents the power set of assignments that adhere to the design. Subsequently, a conditional distribution 
$
m(\mathcal{C}|\mathbf{T})
$ 
is chosen on this event space, ensuring that, conditional on \(\mathcal{C} \in \mathbbm{C}\), the test statistic 
$
z(\mathbf{Y}(\mathbf{T}), \mathbf{T}, \boldsymbol{\pi}(\mathbf{T}), \mathcal{C})
$ 
is imputable under the null hypothesis. The authors refer to \(m(\mathcal{C}|\mathbf{T})\) as the \textit{conditioning mechanism,} which, together with the treatment assignment mechanism 
$
p(\cdot),
$ 
induces the joint distribution 
$
\Pr(\mathbf{T}, \mathcal{C}) = m(\mathcal{C}|\mathbf{T}) \cdot p(\mathbf{T}).
$

The conditional randomization testing procedure of \cite{basse2019randomization} for any generic null hypothesis, conditioning mechanism $m(\mathcal{C}|\mathbf{T}),$  and test statistic $z,$ is summarized as: 
\begin{enumerate}
    \item Draw  $\mathbf{t}^{\text{obs}} \sim p(\mathbf{T}^{\text{obs}}) $ and observe $\mathbf{Y}^{\text{obs}}$.
 
    \item Draw  $\mathcal{C} \sim m(\mathcal{C}|\mathbf{T}^{\text{obs}})$ and compute the observed test statistic $z^{obs}:=z(\mathbf{Y}^{obs}, \mathbf{t}^{\text{obs}},\boldsymbol{\pi}(\mathbf{t}^{\text{obs}}); \mathcal{C} )$.
  
    \item Randomize treatment under the null and compute the test statistic $z(\mathbf{Y}(\mathbf{T}), \mathbf{T},\boldsymbol{\pi}(\mathbf{T}); \mathcal{C})$ under the imputed outcomes.

    \item Compute the p-value as $ \mathbbm{E}[z(\mathbf{Y}(\mathbf{T}), \mathbf{T},\boldsymbol{\pi}(\mathbf{T}); \mathcal{C}) > z^{obs}\}| \mathcal{C} ] $, with the expectation taken with respect to the randomization distribution  $P(\mathbf{T}|\mathcal{C}) = m(\mathcal{C}|\mathbf{T})\cdot p(\mathbf{T})/P(\mathcal{C}).$
\end{enumerate}

The authors proved that the above procedure is valid under any conditioning event $\mathcal{C}.$ However, the choice of $\mathcal{C}$ has implications on the power of the test. 
While the ideas in \cite{basse2019randomization} are innovative, they did not provide general guidance on how to pick the conditioning event space and the conditioning mechanism. Instead, they focused on two-stage randomized trials and proposed conditioning mechanisms that decompose as  
\begin{align} \label{basse cm-r}
    m(\mathcal{C}|\mathbf{T}^{obs}) = f(\mathcal{U}|\mathbf{T}^{obs}) \cdot g(\mathcal{T}|\mathcal{U},  \mathbf{T}^{obs}  ),
\end{align}
where $f$ and $g$ are distributions over $\mathbbm{U}$ and $\mathbbm{T}$ respectively.
Intuitively, \eqref{basse cm-r} suggests a two-step conditioning procedure: focal units are chosen based on the observed treatment assignment, then focal assignments are selected based on the focal units and observed treatment assignment.
Although this two-step conditioning procedure is reasonable in most settings, there is no guidance on each step.

Recognizing the absence of clear guidance on the choice of conditioning event, \cite{puelz2022graph} proposed a general approach to selecting a conditioning event space and a mechanism motivated by graph theory. Their innovation builds on the insights of \cite{basse2019randomization}. For any null hypothesis, \cite{puelz2022graph}  proposed a conditioning procedure that is summarized as follows:
\begin{enumerate}
    \item Map units and treatment assignment vectors under the design based on the restrictions imposed by the null hypothesis to form a bipartite graph referred to as the \textit{null exposure graph}.

    \item  Earmark all the subgraphs in the null exposure graph where all units are connected to all assignments. These subgraphs are called \textit{bicliques}. Within each biclique, the null is sharp.

    \item Select a subset of the bicliques that partition the set of all assignments connected to at least one unit in the null exposure graph to form the event space $\mathbbm{C}.$ This subset of bicliques is referred to as the \textit{biclique decomposition} by the authors.

    \item The event space $\mathbbm{C}$ of the biclique decomposition of the null exposure graph induces a degenerate conditioning mechanism $m(\mathcal{C}|\mathbf{T})=\mathbbm{I}\{\mathbf{T} \in \mathcal{T}(\mathcal{C})\},$  where $\mathcal{T}(\mathcal{C})$ denotes the set of treatment assignments for a given $\mathcal{C} \in \mathbbm{C}.$
\end{enumerate} 
This conditioning procedure, as shown by \cite{puelz2022graph}, leads to valid conditional randomization tests that demonstrate improved power compared to those reported in the literature.

\subsection{Proposed Procedure  Versus  Existing Procedures }\label{comparison}
The practical implementation of the test for $H^{\tau}_{0}$ using  Procedure 1, combined with the accompanying validity results in the Theorems of the main text, facilitates a comparison between our proposed conditioning procedure and others in the literature.

First, while our conditioning approach adheres to the sequence outlined by the decomposition in \cite{basse2019randomization}, the execution of each step is inherently tailored to the specific null hypothesis. 
Intuitively, any test applying our proposed conditioning procedure will achieve at least the same power as the procedure in  \cite{basse2019randomization}. This improved power is due to the procedure ensuring a larger set of focal assignments for any observed assignment. 

Compared to the biclique-based approach of \cite{puelz2022graph}, our conditioning procedure sacrifices generality for increased power. While constructing a null exposure graph between the super-focal units and treatment assignments, as proposed in \cite{puelz2022graph}, is a feasible approach, it would result in tests with reduced statistical power since there may be no bicliques or bicliques of small sizes, as shown in our simulation exercise. 
See Theorem \ref{power theorem} below for a formal basis for this power comparison.

Second, our selection mechanism of focal assignments and focal units is intertwined through the super-focal units. Our proposed procedure deliberately chooses focal assignments to maximize the number of focal units while optimizing the number of focal assignments. In contrast, the approach in \cite{basse2019randomization} advocates for a prior selection of focal units---whether random or non-random---based solely on the observed treatment assignment without considering the treatment assignment vectors. This inter-dependency in our procedure ensures a larger subset of treatment assignments, leading to higher statistical power and reduced size distortion.

Third, a notable feature of our conditioning procedure is its allowance for focal units to vary across focal assignments. Consequently, the number of focal units may differ across these assignments. This flexibility increases the cardinality of the subset of treatment assignment vectors, thereby improving statistical power and size.
For example, consider a population of four units, where two are treated and paired into two dyads. When testing the null hypothesis of \textit{no spillovers}, the methods of \cite{athey2018exact} and \cite{basse2019randomization} yield a subset of treatment assignment vectors of size two. In contrast, permitting focal units to vary across treatment assignments results in a subset of size four. Depending on the test statistic, all four treatment assignments may contribute to approximating the null distribution of the test statistic, thereby improving statistical power.

The power of the randomization testing procedure proposed in this paper is influenced by several factors, including sample size, the interference structure, exposure mapping, and the test statistic. However, the preceding discussion suggests that our testing procedure achieves higher power, conditional on these factors.

To facilitate a formal comparison of statistical power with the biclique-based procedure proposed in \cite{puelz2022graph}, we focus on the null hypothesis $$H_0^0: Y_i(1,\pi_k)-Y_i(0,\pi_k)=0,\,\, \text{for some}\,\, \pi_k \in \mathbf{\Pi}, \,\, \text{for all}\,\, i\in [N].$$ 
In Theorem \ref{power theorem}, we analyze the statistical power of this null hypothesis using our proposed testing procedure (Procedure 1 in the main paper) within a simplified version of our model, similar to that employed in Theorem 3 in \cite{puelz2022graph}.
\begin{theorem}{(Power of the CRT Procedure for a Zero Treatment Effects Null).} \label{power theorem}\\
     For any arbitrary observed  assignment, $\mathbf{t}^{obs},$ let $\mathcal{C}(\mathbf{t}^{obs})=(\mathcal{S}(\mathbf{t}^{obs}),\mathcal{U}(\mathbf{t}^{obs}, \cdot),\mathcal{T}_{\epsilon}(\mathbf{t}^{obs}) )$ be the  conditioning event described in  Procedure 1.  Let  $\mathcal{T}_{\epsilon,n}(\mathbf{t}^{obs}):=\{\mathbf{t} \in \mathcal{T}_{\epsilon}(\mathbf{t}^{obs}): |\mathcal{U}(\mathbf{t}^{obs},\mathbf{t})|=n \}.$  Also, set $|\mathcal{T}_{\epsilon,n}(\mathbf{t}^{obs})|=m(n).$ Consider the null hypothesis $H^{0}_{0}$ and suppose that $z(\cdot)$ is a test statistic that is a linear function of outcomes which satisfies the sufficient conditions in Theorem 1 and Corollary 1 of the paper. Furthermore, let the randomization distribution and the null distribution conditional on $\mathcal{T}_{\epsilon,n}(\mathbf{t}^{obs})$ be denoted by $\hat F_{1,n,m(n)}$ and  $\hat F_{0,n,m(n)}$ respectively.  That is,  
    \begin{equation}\label{power eq.}
        z( \mathbf{Y}(\mathbf{T}), \mathbf{T} , \mathbf{A};|\mathcal{U}(\mathbf{t}^{obs},\mathbf{T}) |=n)\sim \hat F_{1,n,m(n)}\,\, \text{and}\,\,\, z( \mathbf{Y}^{obs}, \mathbf{T}^{obs}  , \mathbf{A}; |\mathcal{U}(\mathbf{t}^{obs},\mathbf{T}) |=n)\sim \hat F_{0,n,m(n)},\,\,
    \end{equation} 
where $\mathbf{T}\sim \Pr_{\mathbf{T}|\mathcal{T}_{\epsilon,n}(\mathbf{t}^{obs})}(\mathbf{T}).$

 In addition, assume that for any fixed $n>0:$
 \begin{itemize}
     \item[(A.1)] There exist continuous CDFs $F_{1,n}$ and $F_{0,n}$ such that $\hat F_{1,n,m(n)}$ and $ \hat F_{0,n,m(n)}$ in \eqref{power eq.} are the empirical distribution functions over $m(n)$ independent samples from $F_{1,n}$ and $F_{0,n}$ respectively.
     \item[(A.2)] There exist a continuous CDF $F$ such that $F_{0,n}$  is a scaled $F,$ i.e.,  $$\exists \,\, \sigma_n>0: F_{0,n}(z)=F(z/\sigma_n) \,\, \forall z \in \mathbbm{R}.$$
     \item[(A.3)] Treatment effects are additive and constant   i.e., $$\exists \,\,\text{a constant}\,\, \tau \in \mathbbm{R} : F_{1,n}(z )=F_{0,n}(z-\tau)\,\, \forall\,\,  z\in \mathbbm{R}.$$
 \end{itemize}
 
Let $\beta_{n,m(n)}(\tau)$ denote the power function of Procedure 1 conditional on focal units and assignments of size $(n,m(n))$ under a fixed alternative of positive treatment effect $\tau,$ i.e., for any $\alpha \in (0,1),$
  \begin{equation} \label{power 1}
   \beta_{n,m(n)}(\tau)= \Pr{}_{\mathbf{T}^{obs}}\left(pval_{_k}(\mathbf{Y}^{obs}, \mathbf{T}^{obs} ;\mathcal{C}(\mathbf{T}^{obs}))\leq \alpha\,\,\Bigg|\,\,\mathcal{T}= \mathcal{T}_{\epsilon,n}\right),
\end{equation} 
where $\mathcal{T}_{\epsilon,n}:=\bigcup_{\mathbf{t}^{obs} \in \mathcal{T}_0}\mathcal{T}_{\epsilon,n}(\mathbf{t}^{obs}).$
 Furthermore, let $\beta(\tau)$ represent the unconditional power function of our proposed testing procedure under a fixed alternative of positive treatment effect $\tau,$ i.e., for any $\alpha \in (0,1)$
  \begin{equation} \label{power 1}
   \beta(\tau)= \sum_{j=1}^{|\bar{\mathcal{U}}|}\beta_{j,m(j)}(\tau)\cdot \Pr\left( \mathcal{T}= \mathcal{T}_{\epsilon,j}\right),
\end{equation} 
where $|\bar{\mathcal{U}}|=\max_{\mathbf{t}^{obs} \in \mathcal{T}_0 }(\max_{\mathbf{t}\in \mathcal{T}_{\epsilon}(\mathbf{t}^{obs}) } |\mathcal{U}(\mathbf{t}^{obs}, \mathbf{t})|).$ 
Fix a small $\delta>0.$ Then, for large enough $m(j)'s,\forall\,\, 1\leq j\leq |\bar{\mathcal{U}}|$ and $\bar m=\max_{1\leq j\leq |\bar{\mathcal{U}}|}m(j)\cdot\mathbbm{I}\{\Pr\left( \mathcal{T}= \mathcal{T}_{\epsilon,j}\right)\neq 0\}$
\begin{align}\label{power eq 1}
     \beta(\tau)\geq 1-\sum_{j=1}^{|\bar{\mathcal{U}}|}\left\{F\left(F^{-1}(1-\alpha)-\frac{\tau}{\sigma_j}\right)\right\}\cdot\Pr\left( \mathcal{T}= \mathcal{T}_{\epsilon,j}\right) -\mathcal{O}({\bar m}^{-0.5+\delta}).
\end{align}
Finally, if $F(v)$ is almost-sigmoidal, i.e., $\sup_{v \in \mathbbm{R}}\left|F(v)-\frac{1}{1+e^{-bv}}\right|\leq \varepsilon$ for fixed $b, \varepsilon>0$
and $\sigma_j=\mathcal{O}(j^{-1/2})\,\,\, \forall\,\, 1\leq j\leq |\bar{\mathcal{U}}|,$ then for some fixed $A, a>0$
\begin{align}\label{power eq 2}
 \beta(\tau)\geq\sum_{j=1}^{|\bar{\mathcal{U}}|}\frac{\Pr\left( \mathcal{T}= \mathcal{T}_{\epsilon,j}\right) }{1+Ae^{-\alpha\tau\sqrt{j}}}-\mathcal{O}(\bar{m}^{-0.5+\delta})-\varepsilon.
\end{align}
\end{theorem}
Condition (A.1) links the empirical distributions of the test statistics to the marginal continuous distributions $F_{0,n}$ and $F_{1,n}$ for any fixed $n.$  Condition (A.2) asserts that  $n$ affects the null distribution only as a scale
parameter. Finally, condition (A.3) provides the relationship between the null distribution of the test statistic and its distribution under the alternative. As in Theorem 3 of \cite{puelz2022graph}, these conditions describe a simplified model that is a useful approximation for large focal assignment sets and facilitates the derivation of power heuristics.

Under conditions (A.1)--(A.3), \eqref{power eq 1} and \eqref{power eq 2}  demonstrates that the statistical power of our procedure---for the zero treatment effect null hypothesis in the presence of interference---increases with the size of the focal units and assignments.  Specifically, the smaller the standard deviations, $\sigma_{j}'s,$ the quicker the test achieves maximum power. Compared to the biclique-based CRT---which has only one term in the summation in \eqref{power eq 1}---our proposed CRT reaches maximum power at a slower rate.  This is intuitive: some focal assignments may correspond to fewer focal units (than those in the biclique-based CRT), reducing the precision of the imputed test statistics. Consequently, the wider the intervals $\mathcal{I}_{0,\epsilon}$ and $\mathcal{I}_{1,\epsilon}$---that regulate the number of focal units---the slower the test converges to maximum power.

The number of focal assignments affects the maximum power of the test. 
Our proposed CRT achieves higher maximum power than the biclique-based CRT as long as $\bar{m}$ exceeds the size of the chosen biclique. The tuning parameter $\epsilon$ influences the maximum power. A smaller $\epsilon$, which results in a larger set of focal assignments, increases $\bar{m}$ in expectation and raises maximum power.  Hence, the choice of $\mathcal{I}_{t,\epsilon}$ affects estimation precision (which controls the rate at which the test achieves maximum power) and the maximum statistical power.

\section{Lemmas and Main Proofs}\label{proofs}
This section contains all the proofs of our theoretical results. To prove the main results, we require the following Lemmas.
\begin{lemma}{(The probability integral transform theorem)} \label{lemma 1}\\
Assume a random variable $A$ has a continuous distribution for which the cumulative distribution function (CDF) is $F_A$. Then, the random variable $B$, which is defined as $B = F_A(A)$, has a standard uniform distribution.
\end{lemma}

\begin{lemma}\label{stein}
Let \( X_1, \ldots, X_n \) be random variables such that \( \mathbb{E}[X_i^4] < \infty \),  
\( \mathbb{E}[X_i] = 0 \), \( \sigma^2 = \operatorname{Var}\left(\sum_{i=1}^n X_i\right) \),  
and define \( W = \sum_{i=1}^n X_i / \sigma \).  

Let the collection \( (X_1, \ldots, X_n) \) have dependency neighborhoods \( N_i \),  
for \( i = 1, \ldots, n \), and define  
\[
D := \max_{1 \leq i \leq n} |N_i|.
\]  
Then, for \( Z \sim \mathcal{N}(0, 1) \),  
\[
d_W(W, Z) \leq \frac{D^2}{\sigma^3} \sum_{i=1}^n \mathbbm{E}|X_i|^3 
+ \frac{\sqrt{28} D^{3/2}}{\sqrt{\pi} \sigma^2} \left( \sum_{i=1}^n \mathbbm{E}[X_i^4] \right)^{1/2}.
\]    
\end{lemma}
\begin{proof}
    See Proof of Theorem 3.6 in \cite{ross2011fundamentals}.
\end{proof}

 \begin{lemma}{(Theorem 3 \cite{puelz2022graph})}\label{PL Th}\\
Consider a conditioning procedure where the focal units are fixed across treatment assignments and let $\mathcal{C} = (\mathcal{U},\mathcal{T})$ be a given conditioning event based on it. Let $(|\mathcal{U}|, |\mathcal{T}|)= (n, m)$ denote the size of the focal units and focal treatment assignments. Let
the randomization distribution and the null distribution be denoted, respectively, by 
\begin{align} \label{t1 puelz}
    z( \mathbf{Y}(\mathbf{T}), \mathbf{T} , \mathbf{A};|\mathcal{U}|=n)\sim \hat F_{1,n,m},\,\, \text{and}\,\, z( \mathbf{Y}^{obs}, \mathbf{T}  , \mathbf{A}; |\mathcal{U} |=n) \sim \hat F_{0,n,m},\,\, \text{where}\,\, \mathbf{T}\sim P(\mathbf{T}|\mathcal{C}), 
\end{align}
 where the null hypothesis is  
  \begin{align} \label{power null}
H^{0}_{0}:
   &Y_i(1, \pi_k) - Y_i(0, \pi_k)= 0 \,\,\,  \text{for some} \,\,  \pi_k \in \mathbf{\Pi},\,\, \forall \,\,i\in [N].
\end{align} 
Suppose that for any fixed $n > 0$:\\
\indent (B1) There exist continuous CDFs $F_{1,n}$ and $F_{0,n}$ such that $\hat F_{1,n,m}$ and $\hat F_{0,n,m}$ in \eqref{t1 puelz} are the empirical
distribution functions over $m$ independent samples from $F_{1,n}$ and $F_{0,n}$, respectively.\\
 \indent (B2) There exist a continuous CDF $F$ such that $F_{0,n}$  is a scaled $F,$ i.e.,  $$\exists \,\, \sigma_n>0: F_{0,n}(z)=F(z/\sigma_n) \,\, \forall z \in \mathbbm{R}.$$
  \indent (B3) Treatment effects are additive and constant  i.e., $$\exists \,\,\text{a constant}\,\, \tau \in \mathbbm{R} : F_{1,n}(z )=F_{0,n}(z-\tau)\,\, \forall\,\,  z\in \mathbbm{R}$$
  Let $\beta_{n,m}(\tau)$ represent the power of a testing procedure on fixed focal units and assignments of size $(n,m)$ and under a fixed alternative of positive treatment effect $\tau,$ i.e., for any $\alpha \in (0,1)$
  \begin{equation} \label{power 1}
   \beta_{n,m}(\tau)= \Pr{}_{\mathbf{T}^{obs}}\left(pval_{_k}(\mathbf{T}^{obs}, \mathbf{Y}^{obs};\mathcal{C})\leq \alpha\,\,\Big|\,\,\mathcal{C}\right).
\end{equation} 
Fix any small $\delta > 0$. Then, for large enough $m$,
\begin{align}  \label{t2 puelz}
    \beta_{n,m}(\tau)\geq 1 - F\left(F^{-1}(1 - \alpha) - \frac{\tau}{\sigma_n}\right) - \mathcal{O}(m^{-0.5+\delta}).
\end{align}

If, in addition, $\sup_{v \in \mathbbm{R}} |F(v) - \frac{1}{1 + e^{-bv}}| \leq \epsilon$ for fixed $b$, $\epsilon > 0$, and $\sigma_n = \mathcal{O}\left(\frac{1}{\sqrt{n}}\right)$, then for some
fixed $A$, $a > 0$:
\begin{align}\label{t3 puelz}
   \beta_{n,m}(\tau) \geq \frac{1}{1 + A e^{-a \tau \sqrt{n}}} - \mathcal{O}(m^{-0.5 + \delta}) - \epsilon 
\end{align}
    \end{lemma}
\begin{proof}
    See Proof of Theorem 3 in Supplementary Material E in \cite{puelz2022graph}
\end{proof}

\subsection*{Proof of Proposition \ref{Validity of the individual unadjusted tests}}
\noindent \textbf{Proposition \ref{Validity of the individual unadjusted tests} (Finite Sample Validity of the Randomization Test for  $H^{\tau}_{0}$)}\\
\textit{Suppose Assumptions \ref{direct neighbors}-\ref{Overlap} hold. Assume $z(\cdot)$ is an unbiased test statistic, with the condition  that 
\vspace{-0.5cm}
\small
\begin{align}\label{sto dom}
   \Pr{}&_{\mathbf{T}|\mathcal{T}_\epsilon(\mathbf{t}^{obs})}\left( z( \mathbf{Y}(\mathbf{T}), \mathbf{T} , \boldsymbol{\pi}(\mathbf{T});\mathcal{U}(\mathbf{t}^{obs},\mathbf{T}) )\leq z^{obs}\Big|\mathcal{C}(\mathbf{t}^{obs}), H^{\tau}_{0} \right)\leq \nonumber\\
  &\Pr{}_{\mathbf{T}^{obs}}\left( z(\mathbf{Y}^{obs}, \mathbf{T}^{obs} , \boldsymbol{\pi}(\mathbf{T}^{obs});\mathcal{S}(\mathbf{T}^{obs}) )\leq z^{obs}\Big|
  H^{\tau}_{0} \right)
\end{align}
\normalsize
for all $\mathbf{t}^{obs}\in \mathcal{T}_0,$ where $z^{\text{obs}}:=z( \mathbf{Y}^{obs}, \mathbf{t}^{obs}  , \boldsymbol{\pi}(\mathbf{t}^{obs}); \mathcal{S}(\mathbf{t}^{obs})).$
Then,  the randomization testing procedure in Procedure \ref{alg:unadjusted CDE} based on unbiased p-values is unconditionally valid  at any significant level $\alpha,$ i.e.,     
\vspace{-0.3cm}
\begin{equation} \label{size 1}
    \Pr{}_{\mathbf{T}^{obs}}(pval_{_k}(\mathbf{Y}^{obs}, \mathbf{T}^{obs}, \boldsymbol{\pi}(\mathbf{T}^{obs});\mathcal{C}(\mathbf{T}^{obs}))\leq \alpha|H^{\tau}_{0})\leq \alpha.
\end{equation} }

\begin{proof}
To prove Proposition \ref{Validity of the individual unadjusted tests}, we have to show that: (i) the test statistic is computable under  $H^{\tau}_{0}$ at any focal assignment   $\mathbf{t}\in \mathcal{T}_{\epsilon}(\mathbf{t}^{obs})$ using focal units $\mathcal{U}(\mathbf{t}^{obs}, \mathbf{t})$ for every $\mathbf{t}^{obs}\in \mathcal{T}_0$  
 and (ii) the standard uniform distribution does not stochastically dominate the distribution of p-values under the null.\\

\noindent  (i) \underline{Computability of test statistic under the Null:}\\ 
 Under Assumptions \ref{direct neighbors} and \ref{exp map}, the test statistic depends on observed data $(\mathbf{Y}^{obs}, \mathbf{T}^{obs}, \boldsymbol{\pi}(\mathbf{T}^{obs})).$  If $H^{\tau}_{0}$ is true, then the observed outcome is related to the potential outcome at focal assignments for the focal units. Under Assumption \ref{Overlap}, $|\mathcal{S}(t^{obs})|>0$ and the observed test statistics are computable.
 
 In addition, for a well-defined interval, $I_{t,\epsilon},$ in Procedure \ref{alg:unadjusted CDE}, the focal units that correspond to each focal assignment are sufficient to compute the test statistic, thus 
 $z(\mathbf{Y}(\mathbf{t}),  \mathbf{t},\boldsymbol{\pi}(\mathbf{t});\mathcal{U}(\mathbf{t}^{obs}, \mathbf{t})) $ is well-defined for all $\mathbf{t}\in \mathcal{T}_{\epsilon}(\mathbf{t}^{obs})$ and for any $\mathbf{t}^{obs}\in \mathcal{T}_0.$\\

\noindent (ii) \underline{Distribution of p-values under the null.}\\
By definition, the p-value at observed assignment $\mathbf{t}^{obs} \in \mathcal{T}_0$ is \begin{align*}
    pval_{_k}(\mathbf{Y}^{obs},\mathbf{t}^{obs},& \boldsymbol{\pi}(\mathbf{t}^{obs});\mathcal{C}(\mathbf{t}^{obs})):=
    \Pr{}_{\mathbf{T}|\mathcal{T}_\epsilon(\mathbf{t}^{obs})}( z( \mathbf{Y}(\mathbf{T}), \mathbf{T} , \boldsymbol{\pi}(\mathbf{T});\mathcal{U}(\mathbf{t}^{obs},\mathbf{T}) )\geq z^{\text{obs}} )|\mathcal{C}(\mathbf{t}^{obs}),H^{\tau}_{0} ),
\end{align*}
where $z^{\text{obs}}:=z( \mathbf{Y}^{obs}, \mathbf{t}^{obs}  , \boldsymbol{\pi}(\mathbf{t}^{obs}); \mathcal{S}(\mathbf{t}^{obs})).$ 
\small
\begin{align}
    &\Pr{}_{\mathbf{T}^{obs}}(pval_{_k}( \mathbf{Y}^{obs},\mathbf{T}^{obs},\boldsymbol{\pi}(\mathbf{T}^{obs});\mathcal{C}(\mathbf{T}^{obs}))\leq \alpha|H^{\tau}_{0} )=\nonumber\\
     &\Pr{}_{\mathbf{T}^{obs}}\left(\Pr{}_{\mathbf{T}|\mathcal{T}_\epsilon(\mathbf{T}^{obs})}( z( \mathbf{Y}(\mathbf{T}), \mathbf{T} , \boldsymbol{\pi}(\mathbf{T});\mathcal{U}(\mathbf{T}^{obs},\mathbf{T}) )\geq z( \mathbf{Y}^{obs}, \mathbf{T}^{obs}  , \boldsymbol{\pi}(\mathbf{T}
     ^{obs}); \mathcal{S}(\mathbf{T}^{obs})) |\mathcal{C}(\mathbf{T}^{obs}),  H^{\tau}_{0} )\leq \alpha\right)\nonumber\\
     &\leq \Pr{}_{\mathbf{T}^{obs}}\left(\Pr{}_{\mathbf{T}^{obs}}( \tilde{z}( \mathbf{Y}^{obs}, \mathbf{T}^{obs}  ,\boldsymbol{\pi}(\mathbf{T}^{obs}); \mathcal{S}(\mathbf{T}^{obs}))\geq z( \mathbf{Y}^{obs}, \mathbf{T}^{obs}  , \boldsymbol{\pi}(\mathbf{T}^{obs}); \mathcal{S}(\mathbf{T}^{obs}))| H^{\tau}_{0} )\leq \alpha\right)\nonumber\\
     &\leq \alpha,\nonumber
\end{align}
\normalsize
where $\tilde{z}( \mathbf{Y}^{obs}, \mathbf{T}^{obs}  ,\boldsymbol{\pi}(\mathbf{T}^{obs}); \mathcal{S}(\mathbf{T}^{obs}))$ is an i.i.d copy of ${z}( \mathbf{Y}^{obs}, \mathbf{T}^{obs}  ,\boldsymbol{\pi}(\mathbf{T}^{obs}); \mathcal{S}(\mathbf{T}^{obs})).$
The first inequality is due to the pairwise pointwise dominance assumption under the null hypothesis in \eqref{sto dom}.  The second inequality is via an application of Lemma \ref{lemma 1} to discrete random variables.
Therefore, the procedure is unconditionally valid given \eqref{sto dom}.

\end{proof}

\subsection*{Proof of Theorem \ref{Asymptotic Validity}}
\noindent \textbf{Theorem \ref{Asymptotic Validity} (Asymptotic Validity of the Randomization Test for  $H^{\tau}_{0}$)}\\
  \textit{Suppose Assumptions \ref{direct neighbors}-\ref{Overlap} hold. Let $z(\cdot)$ denote a consistent non-negative test statistic, with support $[\underline{z},\infty).$ If $z(\mathbf{Y}^{obs}, \mathbf{T}^{obs} , \boldsymbol{\pi}(\mathbf{T}^{obs});\mathcal{S}(\mathbf{T}^{obs}) )=\underline{z}+o_{p}(1)$ under $H^{\tau}_{0},$ then 
    \begin{align}\label{asysto dom}
   \Pr{}&_{\mathbf{T}|\mathcal{T}_\epsilon(\mathbf{t}^{obs})}\left( z( \mathbf{Y}(\mathbf{T}), \mathbf{T} , \boldsymbol{\pi}(\mathbf{T});\mathcal{U}(\mathbf{t}^{obs},\mathbf{T}) )\leq z^{obs}\Big|\mathcal{C}(\mathbf{t}^{obs}), H^{\tau}_{0} \right)\leq \nonumber\\
  &\Pr{}_{\mathbf{T}^{obs}}\left( z(\mathbf{Y}^{obs}, \mathbf{T}^{obs} , \boldsymbol{\pi}(\mathbf{T}^{obs});\mathcal{S}(\mathbf{T}^{obs}) )\leq z^{obs}\Big|
  H^{\tau}_{0} \right)\,\,\, \text{as}\,\,\, N\to \infty.
\end{align}
Consequently, Procedure  \ref{alg:unadjusted CDE} is unconditionally valid at any significant level $\alpha\in(0,1)$ as $N\to\infty.$}
\begin{proof}
    Under Assumptions \ref{direct neighbors} and \ref{exp map}, the test statistic depends on observed data $(\mathbf{Y}^{obs}, \mathbf{T}^{obs}, \boldsymbol{\pi}(\mathbf{T}^{obs})).$  If $H^{\tau}_{0}$ is true, then the observed outcome is related to the potential outcome at focal assignments for the focal units. For a well-defined interval, $I_{t,\epsilon},$ in Procedure \ref{alg:unadjusted CDE}, the focal units that correspond to each focal assignment are sufficient to compute the test statistic, thus 
  $z(\mathbf{Y}(\mathbf{t}),  \mathbf{t},\boldsymbol{\pi}(\mathbf{t});\mathcal{U}(\mathbf{t}^{obs}, \mathbf{t})) $ is well-defined for all $\mathbf{t}\in \mathcal{T}_{\epsilon}(\mathbf{t}^{obs})$ and for any $\mathbf{t}^{obs}\in \mathcal{T}_0.$

 Next, under Assumption \ref{Overlap}, for any $\mathbf{t}^{obs}\in\mathcal{T}_0,$  $\eta\cdot N <\sum_{i=1}^N \mathbbm{I}(t_i=1, \pi_i(\mathbf{t}^{obs})=\pi)< (\zeta-\eta)\cdot N.$ Let $\mathcal{S}_t(\mathbf{t}^{obs}):=\sum_{i=1}^N \mathbbm{I}(t_i=1, \pi_i(\mathbf{t}^{obs})=\pi),$ then for any $t\in\{0,1\}$  $|\mathcal{S}_t(\mathbf{t}^{obs})|\to \infty$ as $N\to \infty.$ 
 
Next, if $z(\mathbf{Y}^{obs}, \mathbf{T}^{obs} , \boldsymbol{\pi}(\mathbf{T}^{obs});\mathcal{S}(\mathbf{T}^{obs}) )=\underline{z}+o_{p}(1)$ under the null, then  by definition,\\ for all $\vartheta>0$ as $N\to \infty$ 
 \vspace{-0.5cm}
\begin{align*}
    \Pr{}_{\mathbf{T}^{obs}}(|z(\mathbf{Y}^{obs}, \mathbf{T}^{obs} , \boldsymbol{\pi}(\mathbf{T}^{obs});\mathcal{S}(\mathbf{T}^{obs}))&-\underline{z}|\geq \vartheta|
  H^{\tau}_{0})\\
  &= \Pr{}_{\mathbf{T}^{obs}}(z(\mathbf{Y}^{obs}, \mathbf{T}^{obs} , \boldsymbol{\pi}(\mathbf{T}^{obs});\mathcal{S}(\mathbf{T}^{obs}))-\underline{z}\geq \vartheta|
  H^{\tau}_{0})\\
    &=\Pr{}_{\mathbf{T}^{obs}}(z(\mathbf{Y}^{obs}, \mathbf{T}^{obs} , \boldsymbol{\pi}(\mathbf{T}^{obs});\mathcal{S}(\mathbf{T}^{obs}))\geq \underline{z}+\vartheta|
  H^{\tau}_{0})\\
    &=0,
\end{align*}
where the first equality is due to the fact that $z(\cdot)$ is non-negative, and $\underline{z}$ is the minimum value in its support.

Next, note that without additional restrictions, as $N\to \infty$ it is possible that  $|\mathcal{U}(\mathbf{t}^{obs},\mathbf{t})|<\infty$ for all $\mathbf{t}\in \mathcal{T}_\epsilon(\mathbf{t}^{obs}),$ thus, by the non-negativity property of a probability measure, as $N\to \infty$
\begin{align*}
    \Pr{}_{\mathbf{T}|\mathcal{T}_\epsilon(\mathbf{t}^{obs})}( z( \mathbf{Y}(\mathbf{T}), \mathbf{T} , \boldsymbol{\pi}(\mathbf{T});\mathcal{U}(\mathbf{t}^{obs},\mathbf{T}))& \geq \underline{z}+\vartheta|\mathcal{C}(\mathbf{t}^{obs}), H^{\tau}_{0})\geq 0\\
    &=\Pr{}_{\mathbf{T}^{obs}}(z(\mathbf{Y}^{obs}, \mathbf{T}^{obs} , \boldsymbol{\pi}(\mathbf{T}^{obs});\mathcal{S}(\mathbf{T}^{obs}))\geq \underline{z}+\vartheta| H^{\tau}_{0})
\end{align*} 
 for all $\mathbf{t}^{obs}\in \mathcal{T}_0$ and $\vartheta>0.$ Since $\underline{z}+\vartheta,$ for $\vartheta>0$ encompasses the support of $z(\cdot),$ we have shown that $z( \mathbf{Y}(\mathbf{T}), \mathbf{T} , \boldsymbol{\pi}(\mathbf{T});\mathcal{U}(\mathbf{t}^{obs},\mathbf{T})$ first-order stochastically dominates\\ $z(\mathbf{Y}^{obs}, \mathbf{T}^{obs} , \boldsymbol{\pi}(\mathbf{T}^{obs});\mathcal{S}(\mathbf{T}^{obs}))$ which implies condition \eqref{asysto dom}. 
\end{proof}

\subsection*{Proof of Theorem \ref{test statistics}}
\noindent \textbf{Theorem \ref{test statistics}}
  \textit{If Assumptions \ref{direct neighbors}--\ref{dependency graph} hold, then the statistics $z_{_{AR}}(\cdot),$ $z_{_{MR}}(\cdot),$ and $z_{_{SK}}(\cdot)$ satisfy the asymptotic pairwise condition in \eqref{asysto dom} under $H^{\tau}_{0}.$  Hence, using these statistics,  $$\lim_{N\to \infty}\Pr{}_{\mathbf{T}^{obs}}(pval_{_k}(\mathbf{Y}^{obs}, \mathbf{T}^{obs}, \boldsymbol{\pi}(\mathbf{T}^{obs});\mathcal{C}(\mathbf{T}^{obs}))\leq \alpha|H^{\tau}_{0})\leq \alpha$$ for all $\alpha\in (0,1).$ }

\begin{proof}
    For brevity, we focus on the proof that $z_{_{MR}}(\cdot)$ satisfies condition \eqref{asysto dom}. The proof of $z_{_{AR}}(\cdot)$ and $z_{_{SK}}(\cdot)$ follow similar arguments.

 Assumptions \ref{direct neighbors}- \ref{Overlap} guarantees the same conditions as in the proof of Theorem \ref{Asymptotic Validity}.
    
Given that $\hat{\sigma}^2_{t}(\pi_k)$ is the \cite{yates1953selection} estimator of the population variance of the values $Y_1(t,\pi_k)\cdots Y_N(t,\pi_k)$ --- denoted as $${\sigma}^2_{t}(\pi_k)= (N-1)^{-1}\sum_{i=1}^N\left(Y_i(t,\pi_k)-N^{-1}\sum_{i=1}^NY_i(t,\pi_k)\right)^2.$$ 
    First, we show that $\hat{\sigma}^2_{t}(\pi_k)$ is an unbiased estimator of ${\sigma}^2_{t}(\pi_k)$
    \begin{align*}
        \mathbbm{E}_{\mathbf{T}}[\hat{\sigma}^2_{t}(\pi_k)]=& \mathbbm{E}_{\mathbf{T}}\left[\frac{1}{(N(N-1))}\sum_{i=1}^N\sum_{j>i}\frac{D_iD_j (Y_i-Y_j)^2}{\Pr_{\mathbf{T}}(D_i=1, D_j=1)}\right]\\
        =&\mathbbm{E}_{\mathbf{T}}\left[\frac{1}{(N(N-1))}\sum_{i=1}^N\sum_{j>i}\frac{D_iD_j (Y_i(t,\pi_k)-Y_j(t, \pi_k))^2}{\Pr_{\mathbf{T}}(D_i=1, D_j=1)}\right]\\
        =&\frac{1}{(N(N-1))}\sum_{i=1}^N\sum_{j>i}\frac{\Pr_{\mathbf{T}}(D_i=1, D_j=1) (Y_i(t,\pi_k)-Y_j(t, \pi_k))^2}{\Pr_{\mathbf{T}}(D_i=1, D_j=1)}\\
        =&\frac{1}{(N(N-1))}\sum_{i=1}^N\sum_{j>i}(Y_i(t,\pi_k)-Y_j(t, \pi_k))^2\\
        =&{\sigma}^2_{t}(\pi_k)
    \end{align*}
    Note that the last equality holds because the expression in the fourth equality is the U-statistic form of the population variance \cite[p.~573]{lehmann2022general}.

    Second, we derive the upper bound of the variance of $\hat{\sigma}^2_{t}(\pi_k):$
    \small
      \begin{align*}
        Var{}_{\mathbf{T}}[\hat{\sigma}^2_{t}(\pi_k)]= & \frac{1}{(N(N-1))^2}\sum_{i_1=1}^N\sum_{i_2>i_1}\sum_{i_3=1}^N\sum_{i_4>i_3}\frac{(y_{i_1}(t,\pi_k)-y_{i_2}(t, \pi_k))^2(y_{i_3}(t,\pi_k)-y_{i_4}(t, \pi_k))^2}{\Pr_{\mathbf{T}}(D_{i_1}=1, D_{i_2}=1)\Pr_{\mathbf{T}}(D_{i_3}=1, D_{i_4}=1)}\cdot\\
         &\quad \quad\quad\quad\quad\quad\quad\quad\quad \quad\quad\quad\quad\quad\quad\quad Cov{}_{_\mathbf{T}}(D_{i_1}D_{i_2}, D_{i_3}D_{i_4}  )\\
         =& \frac{1}{(N(N-1))^2}\sum_{i_1=1}^N\sum_{i_2>i_1}\sum_{i_3=1}^N\sum_{i_4>i_3}\frac{(y_{i_1}(t,\pi_k)-y_{i_2}(t, \pi_k))^2(y_{i_3}(t,\pi_k)-y_{i_4}(t, \pi_k))^2}{\Pr_{\mathbf{T}}(D_{i_1}=1, D_{i_2}=1)\Pr_{\mathbf{T}}(D_{i_3}=1, D_{i_4}=1)}\cdot\\
         &\left(\Pr{}_{_\mathbf{T}}(D_{i_1}=1, D_{i_2}=1 D_{i_3}=1, D_{i_4}=1 ) - \Pr{}_{_\mathbf{T}}(D_{i_1}=1, D_{i_2}=1)\cdot \Pr_{_\mathbf{T}}( D_{i_3}=1, D_{i_4}=1 )\right)\\
          \leq& \frac{1}{(N(N-1))^2}\sum_{i_1=1}^N\sum_{i_2>i_1}\sum_{i_3=1}^N\sum_{i_4>i_3}\frac{(y_{i_1}(t,\pi_k)-y_{i_2}(t, \pi_k))^2(y_{i_3}(t,\pi_k)-y_{i_4}(t, \pi_k))^2}{\Pr_{\mathbf{T}}(D_{i_1}=1, D_{i_2}=1)\Pr_{\mathbf{T}}(D_{i_3}=1, D_{i_4}=1)}\cdot\\
         &\quad \quad\quad\quad\quad\quad\quad\quad\quad \quad\quad\quad\quad\quad\quad\quad\Pr{}_{_\mathbf{T}}(D_{i_1}=1, D_{i_2}=1 D_{i_3}=1, D_{i_4}=1 )\\
          \leq& \frac{1}{(N(N-1))^2}\sum_{i_1=1}^N\sum_{i_2>i_1}\sum_{i_3=1}^N\sum_{i_4>i_3}c_4\cdot\Pr{}_{_\mathbf{T}}(D_{i_1}=1, D_{i_2}=1 D_{i_3}=1, D_{i_4}=1 )\\
           \leq& \frac{c_4}{(2N(N-1))^2}\sum_{i_1=1}^N\sum_{i_2=1}^N\sum_{i_3=1}^N\sum_{i_4=1}^N h_{i_1i_2i_3i_4}
    \end{align*}
    \normalsize
    The first inequality holds since $0\leq\sum_{i_1=1}^N\sum_{i_2>i_1}\sum_{i_3=1}^N\sum_{i_4>i_3}(y_{i_1}(t,\pi_k)-y_{i_2}(t, \pi_k))^2(y_{i_3}(t,\pi_k)-y_{i_4}(t, \pi_k))^2<\infty$ under Assumption \ref{bounded} (i). Assumption \ref{bounded} (i) and(ii) also implies that $|y_i(t, \pi)|/\Pr_{\mathbf{T}}(D_{i_1}=1, D_{i_2}=1)\Pr_{\mathbf{T}}(D_{i_3}=1, D_{i_4}=1)\leq c_4<\infty.$

    Using the chebyshev inequality, $\hat{\sigma}^2_{t}(\pi_k)$ is a consistent estimator of ${\sigma}^2_{t}(\pi_k)$ as long as Assumption \ref{dependency graph} holds. Thus so far we have established that 
     $$\hat{\sigma}^2_{t}(\pi_k)-{\sigma}^2_{t}(\pi_k)=o_p(1)\,\,\,\,\text{for}\,\,\,\, t\in \{0,1\}.$$

    Since  the function $f(x,y)=x/y$ is continuous for $y>0$, by the continuous mapping theorem (CMT) and  Assumption \ref{bounded} (iii):
    $$\hat{\sigma}^2_{t}(\pi_k)/\hat{\sigma}^2_{t'}(\pi_k) -{\sigma}^2_{t}(\pi_k)/ {\sigma}^2_{t'}(\pi_k)=o_p(1)\,\,\,\,\text{for}\,\,\,\, t\neq t'\in \{0,1\}.$$ Note that ${\sigma}^2_{t}(\pi_k)/ {\sigma}^2_{t'}(\pi_k)<\infty \,\,\,\text{for}\,\,\,\, t\neq t'\in \{0,1\}$ by  Assumption \ref{bounded} (iii). Thus, the maximum function is continuous at $({\sigma}^2_{1}(\pi_k)/ {\sigma}^2_{0}(\pi_k), {\sigma}^2_{0}(\pi_k)/ {\sigma}^2_{1}(\pi_k))$ and by the continuous mapping theorem
     $$\max\{\hat{\sigma}^2_{1}(\pi_k)/\hat{\sigma}^2_{0}(\pi_k), \hat{\sigma}^2_{0}(\pi_k)/\hat{\sigma}^2_{1}(\pi_k)\}  -\max\{{\sigma}^2_{1}(\pi_k)/ {\sigma}^2_{0}(\pi_k), {\sigma}^2_{0}(\pi_k)/ {\sigma}^2_{1}(\pi_k)\}=o_p(1).$$

     Under the null, ${\sigma}^2_{1}(\pi_k)/ {\sigma}^2_{0}(\pi_k)= {\sigma}^2_{0}(\pi_k)/ {\sigma}^2_{1}(\pi_k)=\max\{{\sigma}^2_{1}(\pi_k)/ {\sigma}^2_{0}(\pi_k), {\sigma}^2_{0}(\pi_k)/ {\sigma}^2_{1}(\pi_k)\}=1. $ Thus,
 $$z_{_{MR}}(\mathbf{Y}^{obs}, \mathbf{T}^{obs} , \boldsymbol{\pi}(\mathbf{T}^{obs});\mathcal{S}(\mathbf{T}^{obs}) )-1=o_p(1) \,\,\text{under the null.}$$
Finally, since $\max\{\hat{\sigma}^2_{1}(\pi_k)/\hat{\sigma}^2_{0}(\pi_k), \hat{\sigma}^2_{0}(\pi_k)/\hat{\sigma}^2_{1}(\pi_k)\} \in [1, \infty),$ $z_{_{MR}}(\cdot)$ satisfies the sufficient conditions of Theorem \ref{Asymptotic Validity} which implies asymptotic validity as required.
 
\end{proof}

\subsection*{Proof of Corollary \ref{Gen Asymptotic Validity}}
\noindent \textbf{Corollary \ref{Gen Asymptotic Validity} (Restricted Asymptotic Validity of the Randomization Test for  $H^{\tau}_{0}$) }
\textit{Suppose Assumptions \ref{direct neighbors}--\ref{Overlap} hold. Let $z(\cdot)$ denote a consistent test statistic with support $I,$ that may be bounded or unbounded.  Let $z(\mathbf{Y}^{obs}, \mathbf{T}^{obs}, \boldsymbol{\pi}(\mathbf{T}^{obs});\mathcal{S}(\mathbf{T}^{obs}) )=z^{*}+o_{p}(1)$ under $H^{\tau}_{0}$ where $z^{*}<\sup I$ and $z(\mathbf{Y}^{obs}, \mathbf{T}^{obs}, \boldsymbol{\pi}(\mathbf{T}^{obs});\mathcal{S}(\mathbf{T}^{obs}) )\overset{d}{\to}F$  under $H^{\tau}_{0},$ where $F$ is non-degenerate. If $\alpha \in (0, 1-F(z^{*})),$ then 
    \vspace{-0.3cm}
    $$\Pr{}_{\mathbf{T}^{obs}}(pval_{_k}(\mathbf{Y}^{obs}, \mathbf{T}^{obs}, \boldsymbol{\pi}(\mathbf{T}^{obs});\mathcal{C}(\mathbf{T}^{obs}))\leq \alpha|H^{\tau}_{0})\leq \alpha
\,\,\, \text{as}\,\,\, N\to \infty.$$}
\begin{proof}
    Using the same arguments as in the proof of Theorem  \ref{Asymptotic Validity}, we can show that the test statistic depends on observed data $(\mathbf{Y}^{obs}, \mathbf{T}^{obs}, \boldsymbol{\pi}(\mathbf{T}^{obs}))$ if Assumptions \ref{direct neighbors} and \ref{exp map} hold. In addition, Assumption \ref{Overlap} ensures that for any $\mathbf{t}^{obs}\in\mathcal{T}_0,$ and $t\in\{0,1\}$  $|\mathcal{S}_t(\mathbf{t}^{obs})|\to \infty$ as $N\to \infty.$ 
    
   Since $z(\mathbf{Y}^{obs}, \mathbf{T}^{obs} , \boldsymbol{\pi}(\mathbf{T}^{obs});\mathcal{S}(\mathbf{T}^{obs}) )=z^{*}+o_{p}(1)$ under the null,  then for all $\vartheta>0$ as $N\to \infty, $
   $ \Pr{}_{\mathbf{T}^{obs}}(|z(\mathbf{Y}^{obs}, \mathbf{T}^{obs} , \boldsymbol{\pi}(\mathbf{T}^{obs});\mathcal{S}(\mathbf{T}^{obs}))-z^{*}|\geq \vartheta|
  H^{\tau}_{0})=0.$

Also, using the fact that for any arbitrary random variable $E,$ $|E|$ first-order stochastically dominates $E,$ as  $N\to \infty,$ we have 
\vspace{-0.5cm}
\begin{align*}
    \Pr{}_{\mathbf{T}^{obs}}(z(\mathbf{Y}^{obs}, \mathbf{T}^{obs} , \boldsymbol{\pi}(\mathbf{T}^{obs});\mathcal{S}(\mathbf{T}^{obs}))&-z^{*}\geq \vartheta|
  H^{\tau}_{0})\\
  &\leq  \Pr{}_{\mathbf{T}^{obs}}(|z(\mathbf{Y}^{obs}, \mathbf{T}^{obs} , \boldsymbol{\pi}(\mathbf{T}^{obs});\mathcal{S}(\mathbf{T}^{obs}))-z^{*}|\geq \vartheta|
  H^{\tau}_{0})=0\\
\end{align*}

Thus, by the non-negativity property of a probability measure, as $N\to \infty$
$$\Pr{}_{\mathbf{T}^{obs}}(z(\mathbf{Y}^{obs}, \mathbf{T}^{obs} , \boldsymbol{\pi}(\mathbf{T}^{obs});\mathcal{S}(\mathbf{T}^{obs}))-z^{*}\geq \vartheta|
  H^{\tau}_{0})=0$$
Using the same arguments as in the proof of Theorem  \ref{Asymptotic Validity}, note that for all $\mathbf{t}^{obs}\in \mathcal{T}_0,$   
 as $N\to \infty$
 \vspace{-0.5cm}
\begin{align*}
    \Pr{}_{\mathbf{T}|\mathcal{T}_\epsilon(\mathbf{t}^{obs})}( z( \mathbf{Y}(\mathbf{T}), \mathbf{T} , \boldsymbol{\pi}(\mathbf{T});\mathcal{U}(\mathbf{t}^{obs},\mathbf{T}))& \geq z^{*}+\vartheta|\mathcal{C}(\mathbf{t}^{obs}), H^{\tau}_{0})\geq 0 \nonumber \\
    &=\Pr{}_{\mathbf{T}^{obs}}(z(\mathbf{Y}^{obs}, \mathbf{T}^{obs} , \boldsymbol{\pi}(\mathbf{T}^{obs});\mathcal{S}(\mathbf{T}^{obs}))\geq z^{*}+\vartheta| H^{\tau}_{0})
\end{align*} 
 for all $\vartheta>0.$
Thus, for all $\mathbf{t}^{obs}\in \mathcal{T}_0,$   
 as $N\to \infty$
\vspace{-0.5cm}
\begin{align}\label{resticted dominance}
    \Pr{}_{\mathbf{T}|\mathcal{T}_\epsilon(\mathbf{t}^{obs})}( z( &\mathbf{Y}(\mathbf{T}), \mathbf{T} , \boldsymbol{\pi}(\mathbf{T});\mathcal{U}(\mathbf{t}^{obs},\mathbf{T})) \geq z|\mathcal{C}(\mathbf{t}^{obs}), H^{\tau}_{0})\geq 0 \nonumber \\
    &=\Pr{}_{\mathbf{T}^{obs}}(z(\mathbf{Y}^{obs}, \mathbf{T}^{obs} , \boldsymbol{\pi}(\mathbf{T}^{obs});\mathcal{S}(\mathbf{T}^{obs}))\geq z| H^{\tau}_{0}), \,\, \forall z \in (z^{*}, \sup I].
\end{align}

Now, note that by Lemma \ref{lemma 1},
\vspace{-0.5cm}
 as $N\to \infty$
 \small
\begin{align}\label{truncated uniform}
    \Pr{}_{\mathbf{T}^{obs}}\Big(\Pr{}_{\mathbf{T}^{obs}}(& \tilde{z}(\mathbf{Y}^{obs}, \mathbf{T}^{obs} , \boldsymbol{\pi}(\mathbf{T}^{obs});\mathcal{S}(\mathbf{T}^{obs}))\geq z(\mathbf{Y}^{obs}, \mathbf{T}^{obs} , \boldsymbol{\pi}(\mathbf{T}^{obs});\mathcal{S}(\mathbf{T}^{obs}))\mid \nonumber\\
    &\quad \quad \quad   z(\mathbf{Y}^{obs}, \mathbf{T}^{obs} , \boldsymbol{\pi}(\mathbf{T}^{obs});\mathcal{S}(\mathbf{T}^{obs}))>z^*, H^{\tau}_{0})\Big)\sim U[0, 1-F^{obs}(z^*)],
\end{align}
where $\tilde{z}( \mathbf{Y}^{obs}, \mathbf{T}^{obs}  ,\boldsymbol{\pi}(\mathbf{T}^{obs}); \mathcal{S}(\mathbf{T}^{obs}))$ is an i.i.d copy of ${z}( \mathbf{Y}^{obs}, \mathbf{T}^{obs}  ,\boldsymbol{\pi}(\mathbf{T}^{obs}); \mathcal{S}(\mathbf{T}^{obs}))$ and $U[0, 1-F^{obs}(z^*)]$ is the uniform distribution between 0 and $1-F^{obs}(z^*).$
\normalsize
Hence, as $N\to \infty,$
\small
\begin{align}
     &\Pr{}_{\mathbf{T}^{obs}}\Big(\Pr{}_{\mathbf{T}|\mathcal{T}_\epsilon(\mathbf{T}^{obs})}( z( \mathbf{Y}(\mathbf{T}), \mathbf{T} , \boldsymbol{\pi}(\mathbf{T});\mathcal{U}(\mathbf{T}^{obs},\mathbf{T}) )\geq z( \mathbf{Y}^{obs}, \mathbf{T}^{obs}  , \boldsymbol{\pi}(\mathbf{T}^{obs}); \mathcal{S}(\mathbf{T}^{obs})) \mid \nonumber\\
     &\quad \quad \quad \quad \quad \quad \quad \quad \quad\quad \quad \quad \quad \quad   z(\mathbf{Y}^{obs}, \mathbf{T}^{obs} , \boldsymbol{\pi}(\mathbf{T}^{obs});\mathcal{S}(\mathbf{T}^{obs}))>z^*, \mathcal{C}(\mathbf{T}^{obs}),  H^{\tau}_{0} )\leq \alpha\Big)\label{size defn}\\
     &\leq \Pr{}_{\mathbf{T}^{obs}}\Big(\Pr{}_{\mathbf{T}^{obs}}( \tilde{z}( \mathbf{Y}^{obs}, \mathbf{T}^{obs}  ,\boldsymbol{\pi}(\mathbf{T}^{obs}); \mathcal{S}(\mathbf{T}^{obs}))\geq z( \mathbf{Y}^{obs}, \mathbf{T}^{obs}  , \boldsymbol{\pi}(\mathbf{T}^{obs}); \mathcal{S}(\mathbf{T}^{obs}))\mid\nonumber\\ &\quad \quad \quad \quad \quad \quad \quad \quad \quad   z(\mathbf{Y}^{obs}, \mathbf{T}^{obs} , \boldsymbol{\pi}(\mathbf{T}^{obs});\mathcal{S}(\mathbf{T}^{obs}))>z^*, H^{\tau}_{0} )\leq \alpha\Big)\nonumber\,\,\,\,\,[\text{due to inequality \eqref{resticted dominance}}] \nonumber\\
      &\leq \Pr(U\leq \alpha) \,\,\, \,\,\, \,\,\, \quad \quad \quad \quad \quad \quad \quad \quad \quad \quad \quad \quad \quad \quad \quad \quad    [\text{where}\,\, U \sim U[0, 1-F^{obs}(z^*)]\,\, \text{from \eqref{truncated uniform}}]\nonumber\\
     &\leq \alpha \,\,\, \,\,\, \,\,\,   \forall\,\,\, \alpha\in (0, 1-F^{obs}(z^*)]. \nonumber
\end{align}
\normalsize
And since the probability in \eqref{size defn} represents the asymptotic size, the necessary condition of the corollary holds.
\end{proof}

\subsection*{Proof of Theorem \ref{general size bound}}
\noindent \textbf{Theorem \ref{general size bound}}
  \textit{ Suppose the sufficient conditions of Theorem \ref{Asymptotic Validity} hold,  then   
    \vspace{-0.5cm}
    \begin{align}
        \lim_{N\to \infty}|\Pr{}&_{\mathbf{T}^{obs}}(1- F_{\mathrm{mix}}(z(\mathbf{Y}^{obs}, \mathbf{T}^{obs} , \boldsymbol{\pi}(\mathbf{T}^{obs});\mathcal{S}(\mathbf{T}^{obs}))\leq \alpha)-\alpha| \nonumber \\
        \leq& |(1-\alpha)- F_{\mathrm{mix}}(F^{-1}_{mix}(1-\alpha))| + \lim_{N\to \infty}\sqrt{2\kappa\cdot d_W(F_{\mathrm{mix}}, F_{Z^{obs}})}  \label{size d b main},
        \vspace{-1cm}
    \end{align}
    where $\kappa$ is the upper bound of the Lebesgue density of $z(\mathbf{Y}^{obs}, \mathbf{T}^{obs} , \boldsymbol{\pi}(\mathbf{T}^{obs});\mathcal{S}(\mathbf{T}^{obs})).$ 
  }
\begin{proof}
Under the sufficient conditions of Theorem \ref{Asymptotic Validity}, the test statistics are computable, and the asymptotic pairwise dominance condition holds.

Now, note that as $N\to \infty,$ 
\begin{align*}
\Pr{}_{\mathbf{T}^{obs}}(1- F_{\mathrm{mix}}(z(\mathbf{Y}^{obs}, \mathbf{T}^{obs} , \boldsymbol{\pi}(\mathbf{T}^{obs})&;\mathcal{S}(\mathbf{T}^{obs}))\leq \alpha)\\
=&\Pr{}_{\mathbf{T}^{obs}}(z(\mathbf{Y}^{obs}, \mathbf{T}^{obs} , \boldsymbol{\pi}(\mathbf{T}^{obs});\mathcal{S}(\mathbf{T}^{obs}))\geq F_{\mathrm{mix}}^{-1}(1-\alpha))\\
=&1- F_{Z^{obs}}(F_{\mathrm{mix}}^{-1}(1-\alpha)).
\end{align*}

Thus, as $N\to \infty,$ 
\small
\begin{align*}
       |&\Pr{}_{\mathbf{T}^{obs}}(1- F_{\mathrm{mix}}(z(\mathbf{Y}^{obs}, \mathbf{T}^{obs} , \boldsymbol{\pi}(\mathbf{T}^{obs});\mathcal{S}(\mathbf{T}^{obs}))\leq \alpha)-\alpha|\\
       =&|1- F_{Z^{obs}}(F_{\mathrm{mix}}^{-1}(1-\alpha))-(1- F_{\mathrm{mix}}(F_{\mathrm{mix}}^{-1}(1-\alpha)))+(1- F_{\mathrm{mix}}(F_{\mathrm{mix}}^{-1}(1-\alpha)))-\alpha|\\
    \leq& |F_{\mathrm{mix}}(F_{\mathrm{mix}}^{-1}(1-\alpha))-F_{Z^{obs}}(F_{\mathrm{mix}}^{-1}(1-\alpha))|+|(1-\alpha) - F_{\mathrm{mix}}(F_{\mathrm{mix}}^{-1}(1-\alpha))|\\
         \leq& \sup_{z}|F_{\mathrm{mix}}(z)-F_{Z^{obs}}(z)|+|(1-\alpha) - F_{\mathrm{mix}}(F_{\mathrm{mix}}^{-1}(1-\alpha))|\\
           \leq& \sqrt{2\kappa\cdot d_W(F_{Z}, F_{Z^{obs}})} +|(1-\alpha) - F_{\mathrm{mix}}(F_{\mathrm{mix}}^{-1}(1-\alpha))|,
    \end{align*}
      \normalsize where the first inequality is due to the triangle inequality, and the last inequality is based on Proposition 1.2 in \cite{ross2011fundamentals}.
\end{proof}

\subsection*{Proof of Corollary \ref{size bound for normals}}
\noindent \textbf{Corollary \ref{size bound for normals}}
 \textit{ Suppose the sufficient conditions of Corollary \ref{Gen Asymptotic Validity}  hold and $F=N(0,1),$ where $N(0,1)$ denotes the standard normal distribution. Then for all $\alpha \in (0,1)$
    \vspace{-0.5cm}
    \begin{align}
        \lim_{N\to \infty}|\Pr{}_{\mathbf{T}^{obs}}(&1- F_{\mathrm{mix}}(z(\mathbf{Y}^{obs}, \mathbf{T}^{obs}, \boldsymbol{\pi}(\mathbf{T}^{obs});\mathcal{S}(\mathbf{T}^{obs}))\leq \alpha)-\alpha|\nonumber \\
        \leq&  |(1-\alpha)- F_{\mathrm{mix}}(F^{-1}_{mix}(1-\alpha))| + \left(\frac{2}{\pi}\right)^\frac{1}{4}\cdot \sqrt{d_W( F_{\mathrm{mix}}, \Phi)},
        \vspace{-1cm}
    \end{align}
   Moreover, if the test statistic can be expressed as the sum of random variables\\ (say $\sum_{i=1}^N W_i/\sqrt{Var(\sum_{i=1}^N W_i})$ where $W_i =N^{-1}\cdot(Y_i\cdot \mathbbm{I}\{T_i=1, \Pi_i=\pi_k\}/\Pr{}_{_\mathbf{T}}(T_i=1, \Pi_i=\pi_k)-Y_i\cdot\mathbbm{I}\{T_i=0, \Pi_i=\pi_k\}/\Pr{}_{_\mathbf{T}}(T_i=0, \Pi_i=\pi_k)),$ then 
\small
    \begin{align}
         \sqrt{d_W(F_{\mathrm{mix}}, \Phi)}
         \leq \left\{\frac{A_{\mathrm{max}}^2}{Var(\sum_{i=1}^N W_i)^{\frac{3}{2}}}\sum_{i=1}^N\mathbbm{E}[|W_i|^3] + \frac{\sqrt{28}A_{\mathrm{max}}^{\frac{3}{2}}}{\sqrt{\pi}Var(\sum_{i=1}^N W_i)}\sqrt{\sum_{i=1}^N\mathbbm{E}|W_i^4|} \right\}^{1/2},
        \vspace{-1cm}
    \end{align}
 \normalsize   
where $A_{\mathrm{max}}:=\max_{i\in[N]}\sum_{j=1}^NA_{ij}$ is the maximal degree of the network.}

\begin{proof}
Under the sufficient conditions of Theorem \ref{Asymptotic Validity}, the test statistics are computable.
 The rest of the proof follows from Lemma \ref{stein}.
\end{proof}
\subsection*{Proof of Proposition \ref{variance bias}}
\noindent \textbf{Proposition \ref{variance bias}}
  \textit{Define the uniformly weighted sample variance estimator as
$$s^2_{t}(\pi_k):=\frac{1}{(n_{tk} -1)}\sum_{i=1}^ND_i (Y_i-\bar{Y})^2,$$ where $D_i= \mathbbm{I}\{T_i=t,\Pi_i=\pi_k\},$  $\sum_{i=1}^ND_i=n_{tk},$  and $\bar{Y}=\sum_{i=1}^NY_iD_i/n_{tk}.$ The bias is   
\small
\begin{align}\label{bias main}
    \mathbbm{E}[s^2_{t}(\pi_k)-\sigma^2_t(\pi_k)] =& \frac{1}{n_{tk}} \sum_{i=1}^N \left( p_i - \frac{n_{tk}}{N} \right) D_i  Y_i(t, \pi_k)^2 \nonumber\\
    &-\frac{1}{n_{tk}(n_{tk}-1)} \sum_{i \neq j} \left( p_{ij} - \frac{n_{tk}(n_{tk}-1)}{N(N-1)} \right) Y_i(t, \pi_k) Y_j(t, \pi_k),
\end{align}
\normalsize
where $p_i=\Pr(D_i=1)$ and $p_{ij}=\Pr(D_i=1, D_j=1).$}
\begin{proof}
    We first expand the squared deviations:
\begin{align*}
\sum_{i=1}^N D_i ( Y_i-\bar{Y})^2 
&= \sum_{i=1}^N D_iY_i^2 - 2 \bar{Y} \sum_{i=1}^N D_iY_i + n_{tk} \bar{Y}^2
= \sum_{i=1}^N D_iY_i^2 - n_{tk} \bar{Y}^2.
\end{align*}
Thus:
\begin{equation*}
s^2_{t}(\pi_k) = \frac{1}{n_{tk}-1} \left( \sum_{i=1}^N D_iY_i^2 - n_{tk} \bar{Y}^2 \right) =\frac{1}{n_{tk}-1} \left( \sum_{i=1}^N D_iY_i^2 -\frac{1}{n_{tk}}\left(\sum_{i=1}^NY_iD_i \right)^2\right) .
\end{equation*}

We expand:
\[
\left( \sum_{i=1}^N D_i Y_i \right)^2 = \sum_{i=1}^N D_i Y_i^2 + \sum_{\substack{i,j=1 \\ i \neq j}}^N D_i D_j Y_i Y_j.
\]
Therefore:
\begin{align*}
s^2_{t}(\pi_k)
&= \frac{1}{n_{tk}-1} \left( \sum_{i=1}^N D_i Y_i^2 - \frac{1}{n_{tk}} \left( \sum_{i=1}^N D_i Y_i^2 + \sum_{i \neq j} D_i D_j Y_i Y_j \right) \right) \\
&= \frac{1}{n_{tk}-1} \left( \sum_{i=1}^N \left(1 - \frac{1}{n_{tk}} \right) D_i Y_i^2 - \frac{1}{n_{tk}} \sum_{i \neq j} D_i D_j Y_i Y_j \right) \\
&= \frac{1}{n_{tk}-1} \left( \frac{n_{tk}-1}{n_{tk}} \sum_{i=1}^N D_i Y_i^2 - \frac{1}{n} \sum_{i \neq j} D_i D_j Y_i Y_j \right).\\
&= \frac{1}{n_{tk}-1} \left( \frac{n_{tk}-1}{n_{tk}} \sum_{i=1}^N D_i Y_i(t,\pi_k)^2 - \frac{1}{n} \sum_{i \neq j} D_i D_j Y_i(t,\pi_k) Y_j(t,\pi_k) \right).
\end{align*}

We take expectation over the sampling design:

\begin{align*}
\mathbbm{E}[s^2_{t}(\pi_k)]
&=  \frac{1}{n_{tk}-1} \left( \frac{n_{tk}-1}{n_{tk}} \sum_{i=1}^N p_i Y_i(t,\pi_k)^2 - \frac{1}{n} \sum_{i \neq j} p_{ij} Y_i(t,\pi_k) Y_j(t,\pi_k) \right)\\
&= \frac{1}{n_{tk}} \sum_{i=1}^N p_i Y_i(t,\pi_k)^2 - \frac{1}{n_{tk}(n_{tk}-1)} \sum_{i \neq j} p_{ij} Y_i(t,\pi_k) Y_j(t,\pi_k). 
\end{align*}

The population variance is:
\begin{equation*}
\sigma^2_{t}(\pi_k) = \frac{1}{N-1} \left( \sum_{i=1}^N Y_i(t,\pi_k)^2 - \frac{1}{N} \left( \sum_{i=1}^N Y_i(t,\pi_k) \right)^2 \right).
\end{equation*}

We want $\text{Bias} = \mathbbm{E}[s^2_{t}(\pi_k)]- \sigma^2_{t}(\pi_k)$.
We rewrite:
\[
p_i = \frac{n_{tk}}{N} + \Delta_i,
\qquad
p_{ij} = \frac{n_{tk}(n_{tk}-1)}{N(N-1)} + \Delta_{ij},
\]
with $\Delta_i$ and $\Delta_{ij}$ being the deviations from simple random sampling.

Plugging into $\mathbbm{E}[s^2_{t}(\pi_k)],$ we have 

\begin{align*}
\mathbbm{E}[s^2_{t}(\pi_k)]
&=
 \frac{1}{n_{tk}} \sum_{i=1}^N \left(\frac{n_{tk}}{N} + \Delta_i\right) Y_i(t,\pi_k)^2 - \frac{1}{n_{tk}(n_{tk}-1)} \sum_{i \neq j} \left( \frac{n_{tk}(n_{tk}-1)}{N(N-1)} + \Delta_{ij} \right) Y_i(t,\pi_k) Y_j(t,\pi_k) \\
&=
\underbrace{
\left(  \frac{1}{N-1} \left( \sum_{i=1}^N Y_i(t,\pi_k)^2 - \frac{1}{N} \left( \sum_{i=1}^N Y_i(t,\pi_k) \right)^2 \right) \right)
}_{= \sigma^2_{t}(\pi_k)}\\
&+
\underbrace{
\frac{1}{n} \sum_{i=1}^N \Delta_i Y_i(t,\pi_k)^2 - \frac{1}{n(n-1)} \sum_{i \neq j} \Delta_{ij} Y_i(t,\pi_k) Y_j(t,\pi_k)
}_{\text{Bias term}}.
\end{align*}

Thus, 
\begin{align*}
    \mathbbm{E}[s^2_{t}(\pi_k)-\sigma_t(\pi_k)] =& \frac{1}{n_{tk}} \sum_{i=1}^N \left( p_i - \frac{n_{tk}}{N} \right) D_i  Y_i(t, \pi_k)^2 \nonumber\\
    &-\frac{1}{n_{tk}(n_{tk}-1)} \sum_{i \neq j} \left( p_{ij} - \frac{n_{tk}(n_{tk}-1)}{N(N-1)} \right) Y_i(t, \pi_k) Y_j(t, \pi_k),
\end{align*}
as required.
\end{proof}

\subsection*{Proof of Theorem \ref{Asymptotic validity of the CI method on CRI using the individual tests} }
\noindent \textbf{Theorem \ref{Asymptotic validity of the CI method on CRI using the individual tests} (Asymptotic Validity of the  CRI-CI Method for  $H^{\tau}_{0}$\!)} \\
\textit{Suppose the sufficient conditions of either Theorem \ref{Asymptotic Validity} or Corollary \ref{Gen Asymptotic Validity} hold.  Then, the randomization testing procedure in Procedure \ref{Alg: Ag2} is asymptotically valid  at some significant level $\alpha \in (0,1)$ i.e., 
 \vspace{-0.5cm}
\begin{equation} \label{size 3 main}
   \lim_{N\to \infty}   \Pr{}_{\mathbf{T}^{obs}}(pval_{k,\gamma}( \mathbf{Y}^{obs},\mathbf{T}^{obs}, \boldsymbol{\pi}(\mathbf{T}^{obs});\mathcal{C}(\mathbf{T}^{obs}))\leq \alpha|H^{\tau}_{0})\leq \alpha. 
\end{equation}
}
\begin{proof}
Same as in the proof of Theorem \ref{Validity of the individual unadjusted tests}, under Assumptions \ref{direct neighbors} and \ref{exp map}, and the null hypothesis,  the focal units for each focal assignment are sufficient to compute the test statistic using the conditioning procedure in Procedure \ref{Alg: Ag2}. 

Let $\tau_0$ be the true value of the nuisance parameter $\tau.$ Then, since Assumption \ref{Overlap} guarantees that the test statistic is computable for  all $\mathbf{t}^{obs} \in \mathcal{T}_0,$ 
\begin{align*}
 \Pr{}_{\mathbf{T}^{obs}}&(pval_{k,\gamma}( \mathbf{Y}^{obs},\mathbf{T}^{obs}, \boldsymbol{\pi}(\mathbf{T}^{obs});\mathcal{C}(\mathbf{T}^{obs}))\leq \alpha|H^{\tau}_{0})=\\ &  \Pr{}_{\mathbf{T}^{obs}}(pval_{k,\gamma}( \mathbf{Y}^{obs},\mathbf{T}^{obs}, \boldsymbol{\pi}(\mathbf{T}^{obs});\mathcal{C}(\mathbf{T}^{obs}))\leq \alpha, \tau_0 \in CI_{\gamma}|H^{\tau}_{0})\\  &+    \Pr{}_{\mathbf{T}^{obs}}(pval_{k,\gamma}( \mathbf{Y}^{obs},\mathbf{T}^{obs}, \boldsymbol{\pi}(\mathbf{T}^{obs});\mathcal{C}(\mathbf{T}^{obs}))\leq \alpha, \tau_0 \notin CI_{\gamma}|H^{\tau}_{0})\\
    \leq & \Pr{}_{\mathbf{T}^{obs}}(\sup_{\tau' \in CI_\gamma} pval_k( \mathbf{Y}^{obs},\mathbf{T}^{obs}, \boldsymbol{\pi}(\mathbf{T}^{obs});\mathcal{C}(\mathbf{T}^{obs}),\tau') +\gamma\leq \alpha, \tau_0 \in CI_{\gamma}|H^{\tau}_{0})\\ &+  \Pr{}_{\mathbf{T}^{obs}}(\tau_0 \notin CI_{\gamma}|H^{\tau}_{0})\\
    \leq &  \Pr{}_{\mathbf{T}^{obs}}( pval_k( \mathbf{Y}^{obs},\mathbf{T}^{obs}, \boldsymbol{\pi}(\mathbf{T}^{obs});\mathcal{C}(\mathbf{T}^{obs}), \tau') \leq \alpha-\gamma, \tau_0 \in CI_{\gamma}|H^{\tau}_{0})\\ &+ \Pr{}_{\mathbf{T}^{obs}}(\tau_0 \notin CI_{\gamma}|H^{\tau}_{0})\\
    \leq &  \alpha-\gamma + \gamma =\alpha
\end{align*}
The first equality applies the law of total probability.
The first inequality is a straightforward use of the relationship between marginal and joint probabilities for the second term. 
The second inequality  uses the fact that $\sup_{\tau' \in CI_\gamma} pval_k( \mathbf{Y}^{obs},\mathbf{T}^{obs}, \boldsymbol{\pi}(\mathbf{T}^{obs});\mathcal{C}(\mathbf{T}^{obs}), \tau')$ first-order stochastically dominates $pval_k( \mathbf{Y}^{obs},\mathbf{T}^{obs}, \boldsymbol{\pi}(\mathbf{T}^{obs});\mathcal{C}(\mathbf{T}^{obs}),\tau').$
The last inequality uses Theorem \ref{Validity of the individual unadjusted tests} (or Corollary \ref{Gen Asymptotic Validity}) and the fact that $\mathrm{CI}_\gamma$ is a valid confidence interval.
\end{proof}

\subsection*{Proof of Theorem \ref{power theorem}}

\begin{proof}
   Based on  \eqref{t2 puelz} in Lemma \ref{PL Th}, for  all $j$ such that $1\leq j\leq |\bar{\mathcal{U}}|,$ 
   we have
   $$\beta_{j,m(j)}(\tau)\geq 1-F\left(F^{-1}(1-\alpha)-\frac{\tau}{\sigma_j}\right) -\mathcal{O}({m(j)}^{-0.5+\delta}),$$
  Thus, 
   $$\beta(\tau)= \sum_{j=1}^{|\bar{\mathcal{U}}|}\beta_{j,m(j)}(\tau)\cdot \Pr\left( \mathcal{T}= \mathcal{T}_{\epsilon,j}\right)\geq 1-\sum_{j=1}^{|\bar{\mathcal{U}}|}\left\{F\left(F^{-1}(1-\alpha)-\frac{\tau}{\sigma_j}\right)\right\}\cdot\Pr\left( \mathcal{T}= \mathcal{T}_{\epsilon,j}\right) -\mathcal{O}({\bar m}^{-0.5+\delta}).$$
 
   Also, based on \eqref{t3 puelz} in Lemma \ref{PL Th}, for  all $j$ such that $1\leq j\leq |\bar{\mathcal{U}}|,$ 
   we have
   $$   \beta_{j,m(j)}(\tau) \geq \frac{1}{1 + A e^{-a \tau \sqrt{j}}} - \mathcal{O}(m(j)^{-0.5 + \delta}) - \epsilon.$$
   Thus, 
   $$ \beta(\tau)= \sum_{j=1}^{|\bar{\mathcal{U}}|}\beta_{j,m(j)}(\tau)\cdot \Pr\left( \mathcal{T}= \mathcal{T}_{\epsilon,j}\right)\geq\sum_{j=1}^{|\bar{\mathcal{U}}|}\frac{\Pr\left( \mathcal{T}= \mathcal{T}_{\epsilon,j}\right) }{1+Ae^{-\alpha\tau\sqrt{j}}}-\mathcal{O}(\bar{m}^{-0.5+\delta})-\varepsilon$$ as required.
\end{proof}
\end{alphasection}

\end{document}